\documentclass[
aps,
amsmath,amssymb,
reprint,
]{revtex4-2}

\usepackage{graphicx}
\usepackage{hyperref}
\hypersetup{colorlinks=true}
\usepackage{siunitx}
\usepackage{multirow}
\usepackage[table,dvipsnames]{xcolor}
\usepackage{braket}
\graphicspath{{./Plots/}}
\DeclareSIUnit\wn{\cm\tothe{-1}}
\DeclareSIUnit\persec{\second\tothe{-1}}
\usepackage{bm}
\definecolor{ddgreen}{RGB}{0,150,50}
\definecolor{ddcyan}{RGB}{0,150,180}
\hypersetup{colorlinks=true,
    linkcolor=blue,
    filecolor=magenta,      
    urlcolor=ddcyan,
    citecolor=ddgreen}

\newcommand{\tnu}{\tilde{\nu}}

\newcommand{\jjj}[6]{\left(\,\begin{matrix}\,#1\,&\,#2\,&\,#3\,\\\,#4\,&\,#5\,&\,#6\,\end{matrix}\,\;\right)}
\newcommand{\sixj}[6]{\left\lbrace\,\begin{matrix}\,#1\,&\,#2\,&\,#3\,\\\,#4\,&\,#5\,&\,#6\,\end{matrix}\,\;\right\rbrace}
\usepackage{pdfpages}
\usepackage{etoolbox} 

\makeatletter
\patchcmd{\@outputpage@head}{\@ifx{\LS@rot\@undefined}{}{\LS@rot}}{}{}{}
\makeatother

\begin{document}
\title{Unimolecular processes in diatomic carbon anions at high rotational excitation}
	\author{Viviane~C.~Schmidt}
	\email{viviane.schmidt@mpi-hd.mpg.de}
	\affiliation{Max-Planck-Institut f\"ur Kernphysik, 69117 Heidelberg, Germany }
	\author{Roman~\v{C}ur\'{i}k}
    \affiliation{J. Heyrovsk\'{y} Institute of Physical Chemistry, Academy of Sciences \protect\mbox{of the Czech Republic,v.v.i,} Dolej\v{s}kova 3, 18223 Prague 8, Czech Republic}
    \author{Milan~On\v{c}\'{a}k}
	\affiliation{Institut f\"ur Ionenphysik und Angewandte Physik, Leopold-Franzens-Universit\"at Innsbruck, Technikerstraße 25/3, Innsbruck 6020, Austria}
	\author{Klaus~Blaum}
	\affiliation{Max-Planck-Institut f\"ur Kernphysik, 69117 Heidelberg, Germany}
	\author{Sebastian~George}
	\affiliation{Max-Planck-Institut f\"ur Kernphysik, 69117 Heidelberg, Germany}
	\author{Jürgen~G\"ock}
	\affiliation{Max-Planck-Institut f\"ur Kernphysik, 69117 Heidelberg, Germany}
	\author{Manfred~Grieser}
	\affiliation{Max-Planck-Institut f\"ur Kernphysik, 69117 Heidelberg, Germany}
	\author{Florian~Grussie}
	\affiliation{Max-Planck-Institut f\"ur Kernphysik, 69117 Heidelberg, Germany}
	\author{Robert~von~Hahn}\altaffiliation[]{Deceased}
	\affiliation{Max-Planck-Institut f\"ur Kernphysik, 69117 Heidelberg, Germany}
	\author{Claude~Krantz}
	\affiliation{Max-Planck-Institut f\"ur Kernphysik, 69117 Heidelberg, Germany}
	\author{Holger~Kreckel}
	\affiliation{Max-Planck-Institut f\"ur Kernphysik, 69117 Heidelberg, Germany}
	\author{Old\v{r}ich~Novotn\'{y}}
	\affiliation{Max-Planck-Institut f\"ur Kernphysik, 69117 Heidelberg, Germany}
	\author{Kaija~Spruck}
	\affiliation{Max-Planck-Institut f\"ur Kernphysik, 69117 Heidelberg, Germany}
	\author{Andreas~Wolf}
	\affiliation{Max-Planck-Institut f\"ur Kernphysik, 69117 Heidelberg, Germany}

\date{\today}
	
\begin{abstract}
   On the millisecond to second time scale, stored beams of diatomic carbon anions C$_2{}^-$ from a sputter ion source feature unimolecular decay of yet unexplained origin by electron emission and fragmentation.
   To account for the magnitude and time dependence of the experimental rates, levels with high rotational and vibrational excitation are modeled for the lowest electronic states of C$_2{}^-$, also including the lowest quartet potential.
   Energies, spontaneous radiative decay rates (including spin-forbidden quartet-level decay), and tunneling dissociation rates are determined for a large number of highly excited C$_2{}^-$ levels and their population in sputter-type ion sources is considered.
   For the quartet levels, the stability against autodetachment is addressed and recently calculated rates of rotationally assisted autodetachment are applied.
   Non-adiabatic vibrational autodetachment rates of high vibrational levels in the doublet C$_2{}^-$ ground potential are also calculated.
   The results are combined to model the experimental unimolecular decay signals.
   Comparison of the modeled to the experimental rates measured at the Croygenic Storage Ring (CSR) gives strong evidence that C$_2{}^-$ ions in quasi-stable levels of the quartet electronic states are the so far unidentified source of unimolecular decay.
\end{abstract}
\maketitle

\section{Introduction}\label{sec:intro}

Observation of time-delayed, unimolecular decay of highly excited polyatomic ions, in particular with molecular and cluster ion beams in traps or storage rings, has become a powerful tool for the study of complex quantum systems \cite{hansen_observation_2001, andersen_physics_2004,toker_radiative_2007,martin_fast_2013,menk_vibrational_2014,breitenfeldt_long-term_2018,stockett_radiative_2020}.
These experiments studied the heavy-fragment split-off (fragmentation) and the electron emission (autodetachment or autoionization) driven by statistical energy exchange in systems with numerous internal degrees of freedom.  
The superposition of decays from a large number of levels with different decay rates leads to a characteristic non-exponential time dependence in the observed unimolecular fragmentation signals.
These were often found to extend over timespans of many orders of magnitude (typically milliseconds to many seconds) following the creation of the hot ions.

Surprisingly, similarly extended time-delayed autofragmentation (AF) and autodetachment (AD) were also observed for excited homonuclear diatomic anions with only a single vibrational degree of freedom \cite{fedor_2005,anderson_ag2-_2020}.
In these studies metal dimer anions (Ag$_2{}^-$, Cu$_2{}^-$) stored up to $\sim$\,$\SI{0.08}{s}$ \cite{fedor_2005} or  $\sim$\,$\SI{10}{s}$ \cite{anderson_ag2-_2020} were used.
The observed AF signal from these anions, produced in a hot (sputtering) ion source, was explained \cite{fedor_2005} by rotational tunneling fragmentation based on high rotational and vibrational excitation of the anions.
Here, relaxation by spontaneous radiative emission, as in heteronuclear diatomic molecules, cannot occur because of the absence of a dipole moment.
AD signals from these metal dimer anions observed at $>$\,$\SI{0.01}{s}$ were explained \cite{anderson_ag2-_2020,Jasik_Franz_AD_Ag_2021} by electronic--vibrational non-adiabatic coupling.

Electrostatic ion storage rings \cite{andersen_physics_2004,schmidt_first_2013, von_hahn_2015,schmidt_electrostatic_2015,nakano_design_2017} allow such decay processes to be studied for a wide range of molecular ions.
After filling the ring by a fast injection process, often using a hot (sputtering) ion source, the neutral or charged products created by AD or AF are directly observed and counted as a function of time during the beam storage time.

In the course of such studies with hot ion sources, delayed unimolecular decay signals have been reported also for the small fundamental anion C$_2{}^-$ \cite{andersen_carbon_1997,pedersen_experimental_1998,iizawa_photodetachment_2022,unimol_remark}.  
Also our experimental work, as reported here, confirms the existence of such delayed unimolecular decays of C$_2^-$.

As a homonuclear system, C$_2{}^-$ appears to be quite equivalent to the metal dimer anions studied previously.
However, the interpretation of its delayed AD and AF signals is much more complex because of the intrinsic properties of the system.
The C$_2{}^-$ anion has excited doublet states lying above the doublet ground state (X$^2\Sigma_g^+$) by $\sim$\,$\SI{2.3}{\electronvolt}$ (B$^2\Sigma_u^+$ \cite{Herzberg_1968,lineberger_two_1972}) and by only $\sim$\,$\SI{0.5}{\electronvolt}$ (A$^2\Pi_u$ \cite{mead,rehfuss}).
Radiative transitions between molecular levels in these electronic states are dipole allowed and have been studied in various experiments and calculations \cite{Herzberg_1968,lineberger_two_1972,mead,rehfuss,rosmus_werner,pedersen_experimental_1998,iida_state-selective_2020,*iida_correction_2021,quartet_theo_shi,da_silva_transition_2024}.
As also addressed by previous storage-ring studies \cite{pedersen_experimental_1998,iida_state-selective_2020,*iida_correction_2021} this is known to lead to rapid radiative decay of many excited vibrational levels and thus seems to make the occurrence of AD and AF after long beam-storage times less compatible with the interpretation of the previous metal-dimer signals, calling for a more detailed understanding of the system.

In the present work, we summarize the observations of delayed AF and AD signals from  C$_2{}^-$.
Then, as the main objective, we consider the various relaxation and unimolecular decay pathways of this diatomic anion in order to model the time dependence of the observed AF and AD signals from the expected molecular properties.
In addition to the excited electronic states of C$_2{}^-$ with doublet spin symmetry, we also consider the lowest C$_2{}^-$ state with quartet spin symmetry.
Speculations about a possible role of the C$_2{}^-$ quartet levels were already brought forward in previous work \cite{pedersen_experimental_1998,iizawa_photodetachment_2022}, while the results of our study give strong evidence that they play a significant role in explaining the observed AD and AF signals.

With an emphasis on highly excited ro-vibrational levels, this study hopes to complement the extensive earlier work on this anion, which mainly focused on situations with smaller internal excitation.
Since the discovery of bound electronically excited states in C$_2{}^-$ \cite{Herzberg_1968,lineberger_two_1972} -- the first ones for any negatively charged molecule -- this system became arguably the most studied molecular anion in history.
Next to its potential for laser spectroscopic experiments \cite{lineberger_two_1972,rehfuss,mead,jones_photodetachment_1980,hefter_ultrahigh_1983,pedersen_experimental_1998,
shan-shan_study_2003,
iida_state-selective_2020,*iida_correction_2021,iizawa_photodetachment_2022}
C$_2{}^-$ is also present in a wide range of low-temperature plasma reactions \cite{weltner_carbon_1989}.
Furthermore, its astrophysical importance was investigated by searches in the interstellar medium and in the atmospheres of carbon stars \cite{weltner_carbon_1989,carbon_stars_vardya,carbon_stars_wallerstein,civis_search_2005}.

Additionally, C$_2{}^-$ has long been named as an anion for which laser-cooling to sub-millikelvin temperatures may be possible in an ion trap \cite{lasercooling_c2-_yzombard,lasercooling_c2-_gerber}.  
Hence, it is a candidate for sympathetic cooling of antiprotons to similar temperatures, since these particles can be co-trapped with laser-cooled C$_2{}^-$. 
Furthermore, studies on ro-vibrational quenching with different buffer gases \cite{rovibrational_quenching_c2-}, stimulated Raman pumping \cite{raman_pumping} and electron collision experiments \cite{electron_collision_c2-} have been performed for this anion.
Recent studies addressed laser excitation also with C$_2{}^-$ beams from a hot sputtering ion source \cite{pedersen_experimental_1998,iizawa_photodetachment_2022}, and  the related observations will be summarized below.

The outline for the remainder of this paper is as follows.
In Sec.\ \ref{sec:exp}, we present the various experimental observations on delayed unimolecular decay of internally excited C$_2{}^-$ anions for storage times up to $\sim$\,$\SI{1}{s}$.
Moreover, we characterize the conditions of high rotational and vibrational excitation expected for the applied C$_2{}^-$ ions.

In Sec.\ \ref{sec:the_levels} dedicated ab-initio calculations for low-lying electronic states of C$_2$ and C$_2{}^-$ with different spin symmetries are presented.
We consider the molecular energy levels allowing for strong nuclear rotation and derive the vibrational eigenstates.
We then discuss the resulting energetic level structure of the system and its consequences for the expected AD and AF signals.
The energetic structure suggests that C$_2{}^-$ levels of quartet symmetry become quasi-stable against AD at high rotation and can be efficiently produced also for quasi-thermal ion-source conditions.

In Sec.\ \ref{sec:the_rates} we then specify the methods to calculate the unimolecular and radiative decay rates, which will allow us to model the experimental AD and AF signals, and summarize the results for the various decay rates.
The mechanisms of the AD process and the respective importance of vibrational and rotational excitation turn out to be different between the doublet and quartet levels of C$_2{}^-$.

In Sec.\ \ref{sec:model} we model the AD and AF signals by combining the decay rates of the individual levels with their estimated populations.
These results, separately considering the spin symmetries, are compared to the experimental findings.
The time dependence of the observed AD signals turns out to be better modeled by the decay rates obtained for strongly rotating quartet levels of C$_2{}^-$.
This suggests that the quartet levels of C$_2{}^-$ play a significant role for the observed unimolecular decay, which is further discussed in the conclusion (Sec.\ \ref{sec:conclusion}).

\section{Experiment}\label{sec:exp}

\subsection{Time-dependent unimolecular decay signals}
\label{sec:exp_signals}

Studies with stored  C$_2{}^-$  ion beams, using a sputter ion source for producing the initial ion sample, appear to have been first reported for the magnetic storage ring ASTRID (Aarhus, Denmark) \cite{andersen_carbon_1997,pedersen_experimental_1998}.
These were later continued \cite{naaman_metastable_2000} at the electrostatic ion beam trap in Rehovot, Israel, and at electrostatic storage rings, including measurements at the TMU E-ring and RICE in Japan \cite{takao_storage_2007,iizawa_photodetachment_2022} and preliminary reports from the DESIREE storage ring in Sweden \cite{unimol_remark}.
From unimolecular decay of excited  C$_2{}^-$, neutral fragments are produced by AD, 
\begin{equation}
(\text{C}_2{}^-)^* \to \text{C}_2 + e ,
\label{eq:ADreact}
\end{equation}
and AF,
\begin{equation}
(\text{C}_2{}^-)^*  \to \text{C}^- + \text{C}.
\label{eq:AFreact}
\end{equation}
The product neutrals exit the ion storage device at the same high laboratory-frame velocities as the stored ions, and are therefore detected with high efficiency by the charge pulses they release when they impinge on the surfaces of electron-multiplier-type particle detectors.
Typically $10^6$--$10^8$ stored ions were found to yield a few ($\sim$\,$10$) up to $\sim$\,$10^3$ neutral events per second.
Thus the experiments are sensitive to small fractions of anions that are sufficiently excited to produce spontaneous unimolecular decay.  
However, neutral fragments from collisions of the stored anions (irrespective of their excitation) with residual gas molecules in the storage device can generate a background signal that may overlay the experimental signal from Eqs.\ (\ref{eq:ADreact}) and (\ref{eq:AFreact}), requiring low residual gas density in the storage device.
While some of the experiments \cite{pedersen_experimental_1998,iizawa_photodetachment_2022} also analyzed laser-induced electron detachment with the same products as Eq.\ (\ref{eq:ADreact}), only spontaneous unimolecular decay is considered here.

Our measurements on this system were carried out during the early operation period (2015) of the electrostatic Cryogenic Storage Ring (CSR) at the Max-Planck-Institut f\"ur Kernphysik in Heidelberg, Germany, described in detail earlier \cite{von_hahn_2015}.
A continuous beam of mass-selected $(^{12}\text{C})_2{}^-$ ions was produced in a Cs sputter source and accelerated to $\SI{60}{\kilo\electronvolt}$ from the large CSR high-voltage ion-source platform.
The  Cs sputter source used a solid graphite target which was bombarded by Cs$^+$ ions in the intense discharge of the ion source.
The typical beam current amounted to several \SI{0.1}{\micro\ampere}.
The ions were guided through an injection beamline, where the beam was chopped by an electrostatic deflector before entering the storage ring.
An ion pulse of $\SI{35}{\micro\second}$ duration was injected.
Finally, approximately $\SI{4e7}{}$ ions were stored in the ring for up to $\SI{20}{\second}$ in a beam of typically \SI{20}{\mm} transverse diameter.

\begin{figure}[tb]
    \includegraphics[width=0.45\textwidth]{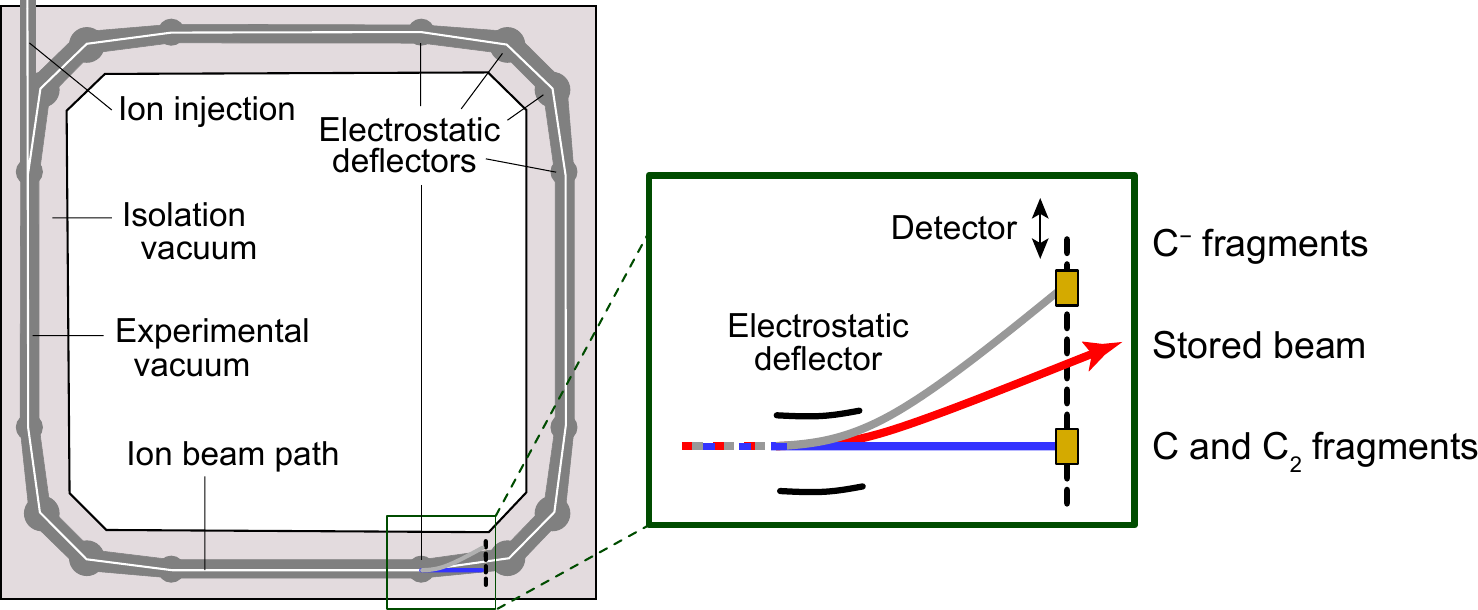}
    \caption{Schematic of the cryogenic storage ring CSR \cite{von_hahn_2015} and the detection of unimolecular reaction products using a moveable single particle detector \cite{spruck_compact_2015}.}
    \label{fig:CSR}
\end{figure}

A schematic drawing of the set-up is shown in Fig.\ \ref{fig:CSR}. 
The square-shaped storage ring has a circumference of $C_0=\SI{35}{\meter}$ (revolution time $\sim\SI{50}{\micro\second}$ at $\SI{60}{\kilo\electronvolt}$ beam energy) with sets of electrostatic bending dipole and focussing quadrupole deflectors situated in its four corners.  
With four field-free straight sections in-between the ion-optics sectors, the CSR is used for various experiments including interaction studies of the stored ions with photons \cite{meyer_oh-_2017}, electrons \cite{novotny_heh+_2019} and neutral atoms \cite{kreckel_astrostudies_2019}.  
As a result of cryogenic cooling \cite{von_hahn_2015} the residual gas density corresponds to a pressure $\lesssim\SI{1e-13}{\milli\bar}$ at room-temperature and leads to essentially background-free conditions for observing the signals of Eqs.\ (\ref{eq:ADreact}) and (\ref{eq:AFreact}).

Considering the low particle densities in the ion beam of $\lesssim\SI{1e4}{\per\cubic\cm}$ and the residual gas ($\lesssim\SI{2e3}{\per\cubic\cm}$ at the CSR) we assume that external influences on the radiative relaxation of the stored anions and on their unimolecular decay can be neglected.
With similar conditions (partly with somewhat higher densities of the residual gas likely reaching up to $\sim\SI{1e6}{\per\cubic\cm}$) this also applies to the other experimental studies quoted 
\cite{andersen_carbon_1997,pedersen_experimental_1998,takao_storage_2007,iizawa_photodetachment_2022,unimol_remark}.

Expected reaction products from unimolecular decay are neutral C$_2$ from AD, Eq.\ (\ref{eq:ADreact}), and neutral C as well as C$^-$ from AF, Eq.\ (\ref{eq:AFreact}).
At the CSR, the products created along one of the field-free straight sections ($l_0=\SI{4.5}{\meter}$) travel with the stored ion beam until they reach the electrostatic deflector terminating that section.
While the deflector bends the stored ion (C$_2{}^-$) trajectory by nominally 6$^\circ$, it deflects the C$^-$ by twice that angle while letting the neutral fragments pass straight.
As shown in Fig.\ \ref{fig:CSR}, a single MCP-based single-particle detector \cite{spruck_compact_2015,krantz_single-particle_2017}, mounted on a $\sim\SI{0.3}{\meter}$ translation stage in the cryogenic region of the CSR, could be positioned to detect either the neutral (C and C$_2$) or the charged (C$^-$) products.
The two cases will be referred to below as the `neutral' and `charged' detector positions.
The electronic detection efficiency is $\gtrsim0.5$ \cite{krantz_single-particle_2017} and the geometrical efficiency considering the transverse beam size at the detector positions is estimated to $\gtrsim0.3$.
As an estimate for the fraction counted by the detector against all decay events in the storage ring, we use $0.5\times0.3\times l_0/C_0=0.02$.

\begin{figure}
    \centering
    \includegraphics[width=0.45\textwidth]{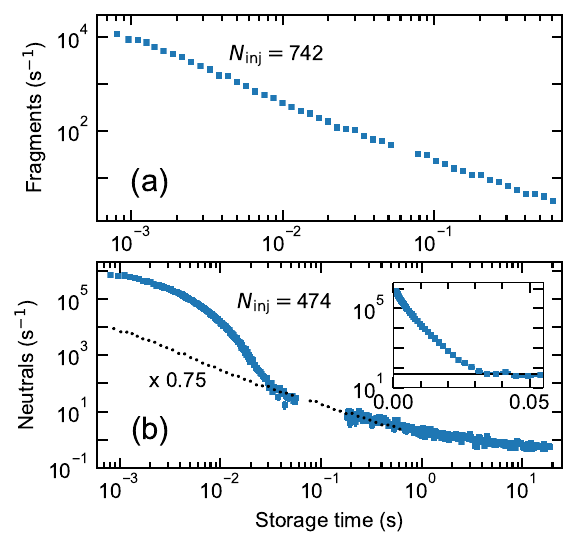}
    \caption{Unimolecular decay signals from C$_2{}^-$ stored in the CSR, accumulated over the given numbers $N_{\rm inj}$ of injection cycles. 
    (a) `Charged' signal $s_-(t)$ of C$^-$ attributed to AF.
    (b) `Neutral' signal $s_0(t)$ of C$_2$ and C produced by AD and AF.  
    Small black symbols in (b) show the signal $\eta s_-(t)$ from (a) with $\eta$ as indicated, chosen to fit $s_0(t)$ at $t \gtrsim \SI{30}{\milli\second}$.  
    Statistical uncertainties are mostly smaller than the data points (blue symbols).
    Inset: Difference signal $s_d(t)$ from Eq.\ (\ref{eq:signal_sd}), interpreted as the AD rate, using a logarithmic signal scale.
    To mitigate the effect of sign changes on this presentation, an arbitrary offset of $\SI{50}{\per\second}$ (horizontal line) was added to $s_d(t)$.
    \label{fig:measured_data}}
\end{figure}

Unimolecular decay signals were accumulated as a function of the time after the injection, denoted as the storage time $t$.
At a given detector position a series of many injection cycles was performed.
Each cycle consisted of a single injection followed by extended storage and final dumping of the ion beam.
The measurements with the `charged' and `neutral' detector positions were carried out on different days with similar ion-source and beam-storage conditions, but without more precise comparisons of the stored beam intensities.
The storage time was derived from the pulsing of the injection chopper such that $t=0$ represents the time when the ions left the acceleration section of the high-voltage platform.
Within each injection cycle of the CSR measurement, the data flow of the event counting system had to be shortly interrupted at $t\sim\SI{50}{\milli\second}$ to process the events arriving at a high initial rate just after the injection.
The event data included the detector pulse heights which were carefully analyzed to ensure a well-characterized detection efficiency~\cite{krantz_single-particle_2017}.
The small effect of the regular break in the data flow is included in the uncertainties of the data shown.

The signal $s_-(t)$ of C$^-$, attributed to AF, is shown in Fig.\ \ref{fig:measured_data}(a).
Roughly, $s_-(t)$ corresponds to a power-law decay ($\propto t^{-\kappa}$ with $\kappa\sim1$) similar to that observed for other homonuclear diatomic anions by \citet{fedor_2005}.
The signal $s_0(t)$ of neutral events, measured over independent injection cycles with the adequate detector position,  is shown in Fig.\ \ref{fig:measured_data}(b).
From a higher initial rate than $s_-(t)$, it decreases in a near-exponential dependence until $t\sim\SI{30}{\milli\second}$ and then resembles $s_-(t)$ at least over the time span recorded for this signal.
The scaling factor $\eta=0.75$ applied in Fig.\ \ref{fig:measured_data}(b) to match $s_0(t)$ and $s_-(t)$ at $t \gtrsim \SI{30}{\milli\second}$ is compatible with estimated variations of the injection and storage conditions.

We interpret the measured `charged' signal to directly represent AF,
\begin{equation}
     s_f(t)=s_-(t).
    \label{eq:signal_sf}
\end{equation}
In contrast, considering that the `neutral' signal reflects both AF and AD and that its decay appears to feature an additional component, we assign the time dependence of the AD process to
\begin{equation}
      s_d(t)=s_0(t)-\eta s_-(t).
      \label{eq:signal_sd}
\end{equation}
This difference is displayed in the inset of Fig. \ref{fig:measured_data}(b).

For analysis, each signal representing a unimolecular process $x$ is assumed to be composed of contributions from several excited levels $j$. Estimates of the state-specific unimolecular decay rates $k_{x,j}$ (with $x=f$ or $d$) will be discussed in Sec.\ \ref{sec:the_rates}.
Moreover, for C$_ 2{}^-$ it is appropriate to also consider the rates $A_j$ of spontaneous radiative decay to energetically lower levels in different electronic states of the anion, further discussed in Sec.\ \ref{sec:calculation_rad}.
Hence, each level $j$ will be characterized by its total decay rate
\begin{equation}
      k'_j=k_{f,j}+k_{d,j}+A_j.
      \label{eq:total_decay_rate}
\end{equation}
From knowledge of the excited-level decay rates, the time dependence of the observed unimolecular processes can be modeled.
Each level $j$ being distinguished by rotational, vibrational, and electronic quantum numbers, the quantity of relevant excited levels can become substantial and the introduction of suitable limitations is important in practice.

Here, we model the signal $s_x$ from the decay process $x$, observed in a stored ion beam as a function of the storage time, by
\begin{equation}
      s_x(t)=\epsilon_x N_i\sum_j k_{x,j} P_j e^{-k'_j t}
\label{eq:signal_simulate}
\end{equation}
where $P_j$ is the relative population of level $j$ in the ion beam at $t=0$, $N_i$ the ion number assumed to be constant, and $\epsilon_x$ the detection efficiency for the products of process $x$.
The sum over $j$ runs over all levels considered in the model.
Clearly, Eq.\ (\ref{eq:signal_simulate}) only accounts for decay processes out of these levels.
Since the levels in C$_2{}^-$ also undergo a radiative decay, which is not connected with ion loss from the stored beam, 
radiative cascades from energetically even higher levels may influence the population in a level $j$ and modify the time dependence compared to Eq.\ (\ref{eq:signal_simulate}).
We estimate such re-population effects in Sec.\ \ref{sec:model}.
Finding that they remain small, we use in our final comparison a slightly augmented form of the time dependence to include a small correction for re-population via the next-higher radiative cascade step. 

Interestingly, via the stored ion number and the counting efficiencies, the absolute size of the observed count rates $s_x$  for a unimolecular process $x$ is linked to the relative populations $P_j$ of the underlying highly excited levels $j$.
In particular, as the observation in the time interval 
$t_1\leq t\leq t_2$ covers most of the temporal decay in the event rate, the integral $S_x$ over this time interval can be expressed as
\begin{eqnarray}
      S_x&=&\int_{t_1}^{t_2} \!\! s_x(t)dt \nonumber\\
      &\approx&\int_{t_1}^\infty \!\! s_x(t)dt=
      \epsilon_x N_i\sum_{(j)} Y_{x,j}P_j,
      \label{eq:eventintegral}
\end{eqnarray}
where the quantity
\begin{equation}
    Y_{x,j}=e^{-k'_j t_1} \frac{k_{x,j}}{k'_j}\leq 1
    \label{eq:yield}
\end{equation}
describes the {\em yield} at which a level $j$ can contribute to the integrated decay signal for process $x$.
These yield values represent the combined effects of the decay rates for level $j$ according to Eq.\ (\ref{eq:total_decay_rate}) as calculated below.
For example, fast radiative decay of a level can strongly reduce its effect on the unimolecular decay signals.
Because of the finite end time $t_2$ of observation, the reasoning of Eq.\ (\ref{eq:eventintegral}) only extends to the set of sufficiently unstable levels with decay rates $k_j'\gg t_2^{-1}$, marked as $(j)$ in the final sum.

Since the yields are limited to $\leq 1$, the {\em minimum} relative population of excited levels compatible with the observed integral count rate can be estimated as
\begin{equation}
    \sum_{(j)} P_j\geq S_x/\epsilon_x N_i
\label{eq:minipop}
\end{equation}
where the lowest value is obtained if all levels contribute with unity yield.
The results in Sec.\ \ref{sec:model} show that for a number of levels with total decay rates in the range  $t_2^{-1} \leq k_j'\leq t_1^{-1}$, the yields become close to 1.
Hence, relative populations summed over levels contributing to AF and AD can be estimated using the integrated count rates $S_f$ and $S_d$ obtained in the experiment.
From the signals in Fig.\ \ref{fig:measured_data}, with the limits $t_1=\SI{0.8}{\milli\second}$ and $t_2=\SI{800}{\milli\second}$,
the integrals of $S_f=\SI{2.9e1}{}$ and $S_d=\SI{1.57e3}{}$ are derived.
For $N_i=\SI{4e7}{}$ and $\epsilon_f=\epsilon_d=0.02$ this yields for AF
\begin{equation}
    \sum_j P_j\gtrsim 
    \SI{4e-5}{}
\end{equation}
and for AD
\begin{equation}
    \sum_j P_j\gtrsim
    \SI{2e-3}{}.
\end{equation}
Hence, the strong AD signal appears to be caused by excited C$_2{}^-$ ions whose total relative population, likely distributed over several levels, represents a few per mille of the stored ions.
When the same beam intensities are assumed, the population at the origin of the AF signal is estimated to be about two orders of magnitude lower.

\subsection{Internal excitation of diatomic anions from sputter ion sources}
\label{sec:exp_population}

In the described experiments, $(\text{C}_2{}^-)^*$ anions were produced using cesium sputter ion sources common for the production of atomic and molecular anions in accelerator applications \cite{middleton_close_1974,middleton_versatile_1983}, where an intense Cs$^+$ beam impinges on a solid (in the present case graphite) surface.
As was already discussed in some of the experimental work \cite{iizawa_photodetachment_2022} and can be seen as a common observation of various studies \cite{snowdon_ZPA_1984,wucher_JCP_1996,corderman_laser_1980,anders_NIM_2015}, molecules arising from sputtering on a surface can be rotationally, vibrationally, and electronically highly excited with temperatures (likely not uniform over the degrees of freedom) up to several thousand kelvin.  
In addition to the formation of neutral molecular sputter products, also the electron transfer to them from ground-state (Cs) and excited (Cs$^*$) neutral cesium atoms at rather high density \cite{vogel_anion_2015} has to be considered to understand the internal excitation of sputtered C$_2{}^-$ anions.
Thus, C$_2{}^-$ may be formed by charge exchange of Cs or Cs$^*$ on sputtered C$_2$ molecules.
Since the energy defect in the reaction C$_2$ + Cs$^*$ $\to$ C$_2{}^-$ + Cs$^+$ may be relevant, the formation yield is likely to depend on the internal (and, in particular, rotational) excitation of the C$_2$ molecules.
Together with the fact that the ions are removed from the Cs-ion sputtered surface by a directly applied extraction field, the excitation-dependent production yields may lead to non-thermal internal energy distributions with different temperatures for different degrees of freedom and with significant high-energy tails.

Excitation distributions of diatomic anions from a Cs-ion sputter ion source were measured for diatomic oxygen and metal oxide anions \cite{corderman_laser_1980}, where a vibrational temperature $T_v$ of the order of $\SI{5000}{\K}$ was found, with smaller or similar values of the rotational temperature $T_r$. 
Later, molecular dynamics studies of neutral products sputtered by ion impact on a surface \cite{wucher_JCP_1996} yielded $T_v\sim\SI{3100}{\K}$ and $T_r\sim\SI{5600}{\K}$ for sputtered neutral Ag$_2$, but the transfer of these results to anions is unclear.
Studies \cite{menk_vibrational_2014} on vibrational AD of hot SF$_6{}^-$ anions from a Cs-ion sputter source yielded  $T_v$ in the range of $1000\ldots\SI{3500}{\K}$ and $T_r \sim 3000\ldots\SI{5500}{\K}$ depending on the operating parameters of the ion source.   
A recent measurement on resonant inner-shell photodetachment of  C$_2{}^-$ from a Cs-ion sputter ion source, showing vibrational structure, could be well modeled using $T_v$ and $T_r$ near $\SI{1100}{\K}$ \cite{schippers_CPC_2023}.  

An interpretation of the present experiments, where AD and AF signals are observed, should take into account that these signals are expected to be caused only by the highly excited anions.
In contrast, a spectroscopic signal such as from resonant photodetachment \cite{schippers_CPC_2023} probes all parts of the energy distribution, including the ground state and lower excitation.
For a consistent description of both types of experiments, it seems appropriate to consider a bi-modal internal energy distribution with smaller temperatures relevant for the low excitation and with larger temperatures representing a high-energy tail.
In the AF and AD signals, the relative contributions of different highly excited levels may then be weighted by the larger temperatures in the energetic tail, while for spectroscopic experiments such as photodetachment the relative contributions of the contributing states may still be adequately described by smaller temperatures.
Levels in the higher excited tail of a bi-modal distribution only negligibly contribute to the integral photodetachment signal, while they become dominant in the AD and AF measurement.
The parameters assumed to describe the population of excited levels in the stored anion beam are further discussed in Sec.\ \ref{sec:the_vibration}.

\section{\boldmath Energy levels of strongly rotating C$_2{}^-$}
\label{sec:the_levels}

\subsection{Rotational-electronic potentials}
\label{sec:the_potentials} 

\begin{figure*}[!t]
    \includegraphics[width=1\textwidth] {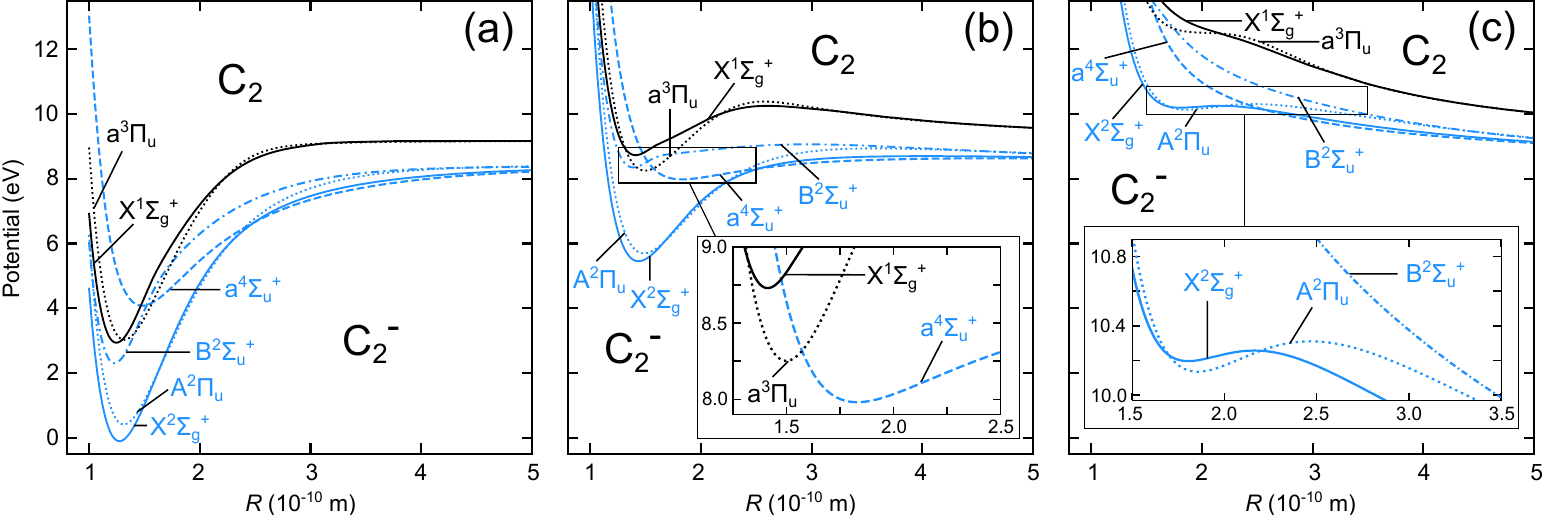}
    \caption{Potential curves of the electronic states X$^2\Sigma_g^+$, A$^2\Pi_u$, B$^2\Sigma_u^+$, and a$^4\Sigma_u^+$
    in C$_2{}^-$ (blue curves) and X$^1\Sigma_g^+$ and a$^3\Pi_u$ in 
    C$_2$ (black curves).  
    (a) Potential energy functions $V_\gamma(R)$ from ab-initio calculations described in the text. 
    (b) Rotational-electronic potentials $V_{\gamma N}(R)$ for $N=171$ ($\Pi_u$ and $\Sigma^+_u$ states) or $N=170$ ($\Sigma^+_g$ states), showing the stabilization of the a$^4\Sigma_u^+$ electronic state.
    (c) $V_{\gamma N}(R)$ for $N=251$ ($\Pi_u$ and $\Sigma^+_u$ states) or $N=250$ ($\Sigma^+_g$ states), showing the change in energetic sequence for the  X$^2\Sigma_g^+$ and A$^2\Pi_u$ states of the anion.
    The zero of the energy scale corresponds to the lowest vibrational level in the $V_\gamma(R)$ potential for X$^2\Sigma_g^+$ and the  asymptotic value of the X$^2\Sigma_g^+$ curve [fitting an $R^{-4}$ long-range part to $V_\gamma(R)$] corresponds to a dissociation limit of $D_0=\SI{8.409}{\electronvolt}$.}
    \label{fig:rearrange}
\end{figure*}

Motivated by the early optical and photodetachment spectroscopy \cite{Herzberg_1968,lineberger_two_1972}, a number of potential energy curves of C$_ 2{}^-$ were theoretically determined by ab-initio calculations since 1974 \cite{barsuhn_nonempirical_1974,thulstrup_theoretical_1974}.  These calculations were refined in various works \cite{zeitz_theoretical_1979,rosmus_werner,watts_coupled_1992,quartet_theo_shi,da_silva_transition_2024} some of which also addressed radiative transition probabilities.  

Since the main interest of the present studies are the strongly excited states of C$_2{}^-$ as well as their electron detachment,  electronic potentials are required for the anionic and the neutral carbon dimer over a large range of the internuclear distance $R$. 
We have obtained the electronic potentials in non-rotating C$_ 2{}^-$ and C$_ 2$ from state-averaged multi-configuration self-consistent field (MCSCF) calculations, combined with the multi-reference configuration interaction (MRCI) method with the active space of 8 and 9 electrons in 8 orbitals for C$_ 2$ and C$_ 2{}^-$, respectively, using the {\sc molpro} package \cite{MOLPRO12}.
The augmented correlation-consistent basis of quadruple-zeta quality, aug-cc-pVQZ \cite{Dunning_ccpVXZ}, was applied.
As functions of $R$ we obtain the potential energy curves $V_\gamma(R)$ for the X$^2\Sigma_g^+$, A$^2\Pi_u$, B$^2\Sigma_u^+$, and a$^4\Sigma_u^+$ states of C$_2{}^-$ as well as for the X$^1\Sigma_g^+$ and a$^3\Pi_u$ states of C$_ 2$.
We use the symbol $\gamma$ to label these electronic states.
The potential energy functions in the low-rotation limit are shown in Fig.\ \ref{fig:rearrange}(a).

All  $V_\gamma(R)$ directly represent the {\sc molpro} results except for the a$^4\Sigma_u^+$ electronic state, where $V_\gamma(R)$ was slightly modified.
The purely electronic potential of the a$^4\Sigma_u^+$ state lies above the potential curves of neutral C$_2$ for $R\lesssim\SI{1.6}{\AA}$.
As discussed in the companion paper \cite{PRL}, very fast spin-allowed AD into C$_2$(a$^3\Pi_u$) can at low rotation take place from all its vibrational levels.
The electronic state is unstable for $R\lesssim\SI{1.6}{\AA}$ and in fact a resonance with $\sim\SI{0.5}{\electronvolt}$ width at $R=\SI{1.2}{\AA}$.
To derive an effective complex potential for the AD calculations, a parametrized a$^4\Sigma_u^+$ potential curve was determined \cite{PRL} as a fit to (a) previously calculated \cite{Halmova_JPB_2006} resonance data in the unstable region ($R\lesssim\SI{1.6}{\AA}$) and to (b) the present ab-initio results at larger $R$.
This parametrized curve is applied as the purely electronic potential $V_\gamma(R)$ for the  a$^4\Sigma_u^+$ state in the present work and presented in Fig.\ \ref{fig:rearrange}(a).
From the expected fast AD decay of the a$^4\Sigma_u^+$ and, likely, also the higher quartet states we conclude that processes originating from C$_2{}^-$ quartet levels  can hardly be observed at the experimental time scales relevant unless rotation becomes strong.

As discussed in the following, the set of these potentials allows us to predict the approximate energies and wave functions of the vibrational levels for a wide range of rotational excitation.
Combining the vibrational level results with matrix elements of electric dipole transitions and spin--orbit coupling, radiative transitions within C$_2{}^-$ can be predicted at high rotation.
Also the relative positions of anionic and neutral levels can be found in order to assess the stability of anionic levels against AD when the rotational excitation is high.

The Supplemental Material \cite{suppl}
includes a comparison of the spectroscopic data for the present potentials in the low-rotation limit with other theory and experiments.
We accept somewhat larger deviations from spectroscopic data in the low-rotation regime than in other recent calculations \cite{quartet_theo_shi,da_silva_transition_2024} since we cannot predict the possible influence of further optimization of the ab-initio calculation on the results at $R$ values far from the low-rotation equilibrium.
More significantly, the detachment energy for ground-state C$_2{}^-$ obtained with the present ab-initio potentials is underestimated by $\sim$\,$\SI{0.22}{\electronvolt}$.
Hence, for calculating the AD rates of C$_2{}^-$ doublet states at low rotation (see Sec.\ \ref{sssec:doubletsigmag}), the vibrational energies and wave functions were obtained from analytic potential functions fitted to the spectroscopic data.

To describe rotating C$_ 2{}^-$ and C$_ 2$ we consider \cite{bernath_spectra_2005,hougen_calculation_1970} the rotational Hamiltonian $H_r=B(R) (\vec{N}-\vec{L})^2_\perp/\hbar^2$, where $B(R)=\hbar^2/2m_rR^2$ is the rotational variable for the reduced mass $m_r$.  
With the conventional angular momentum symbols and related coordinate definitions (see Appendix \ref{app:definitions}) the rotational Hamiltonian can be written as \cite{bernath_spectra_2005,hougen_calculation_1970}
\begin{eqnarray}
    H_r&=&B(R)\,[(N_x-L_x)^2+(N_y-L_y)^2]/\hbar^2\nonumber\\
    &=&B(R)\,[(\vec{N}^2-N_z^2)+(\vec{L}^2-L_z^2)\nonumber\\
    &&\hphantom{B(R)\,[}{}-(N_+L_-+N_-L_+)]/\hbar^2
    \label{eq:hrot}
\end{eqnarray}
using $N_\pm=N_x\pm {\rm i} N_y$ and $L_\pm=L_x\pm {\rm i} L_y$. 

For the fixed-$R$ rotational-electronic eigenstates we consider the quantum numbers  $J$, $N$, $S$, $\Lambda$, and $\Omega$ in the conventional definition summarized in Appendix \ref{app:definitions}.
In the sub-space for a given electronic state $\gamma$, the matrix of the terms in the first parenthesis of the second line in Eq.\ (\ref{eq:hrot}) is diagonal and yields the centrifugal energy
\begin{equation}
    E_{N}(R)=B(R)\left[N(N+1)-\Lambda^2\right]
    \label{eq:erot}
\end{equation}
where $N=|\Lambda|,|\Lambda|+1,\ldots$ and which uses the fact that $N_z=L_z$. The matrix elements resulting from the second term ($\vec{L}^2-L_z^2$) are usually \cite{bernath_spectra_2005,hougen_calculation_1970} treated as a small correction to the electronic energy.  In our case of strong rotation (large $N$) they are clearly much smaller than $E_{N}$, so that their actual value is irrelevant and masked by the larger expected inaccuracies of the ab-initio potentials $V_\gamma(R)$.

Consequently, we consider the rotational-electronic potential curves defined as
\begin{equation}
    V_{\gamma N}(R) = V_\gamma(R) + E_N(R)
    \label{eq:pot}
\end{equation}
in the range of strong nuclear rotation.  
Vibrational levels with energies $E_{\gamma Nv}$, distinguished by the vibrational quantum number $v$, were determined (see Sec.\ \ref{sec:the_vibration}) in the rotational-electronic states $\gamma N$ with potentials $V_{\gamma N}(R)$.
Their parity is expressed (see Appendix \ref{app:definitions}) by $\epsilon_N(-1)^{N}$ using the parity label $\epsilon_N=\pm1$ adapted to Hund's case (b) basis states \cite{watson_honllondon_2008}.
There are two near-degenerate eigenstates with $\epsilon_N=\pm1$, except for $\Sigma$ states, which have a single parity with label $\epsilon_N=1$ for $\Sigma^+$ ($\epsilon_N=-1$ for $\Sigma^-$).  

Panels (b) and (c) of Fig.\ \ref{fig:rearrange} show the rotational-electronic potentials $V_{\gamma N}(R)$ for two specific large values of $N$ to demonstrate the significant differences from $V_{\gamma}(R)$.
The landscape of potential curves within the C$_2{}^-$ anion itself as well as their placement relative to the neutral C$_2$ potentials change substantially.  
Vibrational states within the $V_{\gamma N}(R)$ potentials are in general localized at higher $R$ and less deeply bound as $N$ increases.  
Moreover, several or all of the vibrational levels are only quasi-bound resonant states which can fragment by tunneling dissociation (AF).  
Also, the potentials $V_{\gamma N}(R)$ have limiting $N$ values of $\sim200\ldots260$ above which they no longer support even such quasi-bound levels.

By the choice of $N$-values for Fig.\ \ref{fig:rearrange}(b) and (c) we focus on two cases where this re-arrangement of potential curves significantly affects the unimolecular processes.  
Regarding the AD of C$_2{}^-$ vibrational levels in the a$^4\Sigma_u^+$ potential, the situation for low rotation, described above, changes at $N\sim155$, when the potential $V_{\gamma N}(R)$ for the anionic a$^4\Sigma_u^+$ state moves below that of the neutral a$^3\Pi_u$ state.
At $N=171$, as shown in Fig.\ \ref{fig:rearrange}(b), a considerable part of the a$^4\Sigma_u^+$ potential lies below that of C$_2$(a$^3\Pi_u$).
The stability of a$^4\Sigma_u^+$ vibrational levels against AD at high rotation is considered in the companion paper \cite{PRL}.
While vibrational levels in the potential $V_{\gamma N}(R)$ of the C$_2{}^-$ quartet state lie below the lowest level in the C$_2$(a$^3\Pi_u$) potential for the {\em same} $N$ at sufficiently high rotation ($N\geq155$ for our model), such states can still undergo AD into neutral levels with a more or less reduced angular momentum $N$.  
We have developed a method \cite{PRL} to calculate the AD rates which require changes in $N$ by up to 8 units.
These rates turn out to decrease strongly with the required change $\Delta N$.
As also summarized in Sec.\ \ref{sssec:quartetsigmau}, the AD rates for $\Delta N=5$ and 6 lie close to the decay rates observed in the near-exponential autodetachment signal presented in Sec.\ \ref{sec:exp}.  
As the C$_2{}^-$ quartet levels become quasi-stable at high $N$, also their radiative lifetimes against spin-forbidden electric dipole decay to the  X$^2\Sigma_g^+$ state become relevant and, hence, are addressed below (Sec.\ \ref{sec:calculation_rad_quartet}).
  
At even higher $N$, the minima of the two lowest electronic levels of C$_2{}^-$ lie significantly above the dissociation threshold to C$^-(2p^3\,^4S)$ $+$  C($2p^2\,^3P$).  
In this range, demonstrated in Fig.\ \ref{fig:rearrange}(c) for $N$ near 250, levels in the X$^2\Sigma_g^+$ and A$^2\Pi_u$ potentials can undergo AF by tunneling dissociation, while being stable against AD considering the much higher energies of the C$_2$ potentials at this $N$.
Depending on their energetic positions in the potential $V_{\gamma N}(R)$, vibrational levels in these two states show tunneling rates which we estimate below (Sec.\ \ref{sec:calculation_diss}).
While C$_2{}^-$ is homonuclear and thus, on first sight, similar to systems for which rotational tunneling was studied before, the vibrational levels will still show radiative decay driven by the electronic dipole moment between the X$^2\Sigma_g^+$ and A$^2\Pi_u$ states.  
Remarkably, at high $N$ the centrifugal energy $E_{N}(R)$ shifts the minimum of the X$^2\Sigma_g^+$ state energetically above that of the A$^2\Pi_u$ state, thus inverting their energetic arrangement as compared to low rotation.  
Correspondingly, also the radiative lifetimes are expected to be strongly modified by the rotation.  Hence, before addressing the AF rates, we also consider the radiative lifetimes of levels in the  X$^2\Sigma_g^+$ and A$^2\Pi_u$ states at high $N$ (Sec.\ \ref{sec:calculation_rad_doublet}).

Before closing this Section, we discuss two further aspects of rotational-electronic potentials and the related eigenstates.  In the homonuclear and isotope-symmetric molecules $(^{12}\text{C})_2{}^-$ and $(^{12}\text{C})_2$ (as fixed by the ionic mass selection in the experiments) the total wavefunction of the dimer must be symmetric against exchange of the nuclei, which are indistinguishable bosons.
Hence, a rotational-electronic eigenstate must have positive nuclear exchange parity.
The latter can be found by multiplying the total parity of the state with its electronic inversion parity \cite{bernath_spectra_2005}.
The total parity being given by $\epsilon_N(-1)^{N}$ and the electronic inversion parity by the $g$ or $u$ label of the electronic states, we find that for $g$-states, even-$N$ states must have $\epsilon_N=1$ and odd-$N$ states $\epsilon_N=-1$.
In agreement with earlier presentations \cite{rehfuss} this restricts $N$ to even values for all $\Sigma^+_g$ states (which all have $\epsilon_N=1$).
For $u$-states (in particular the $\Pi_u$ states) even-$N$ states must have $\epsilon_N=-1$ and odd-$N$ states $\epsilon_N=+1$.  Hence, only single parity-sublevels, with $\epsilon_N$-values alternating for varying $N$, exist for the $\Pi_u$ states, while $N$ must be odd for $\Sigma^+_u$ states (which again all have $\epsilon_N=1$).
The nuclear-symmetry suppression of some levels does not influence the radiative lifetimes as the radiative transitions only connect levels with the same nuclear exchange symmetry.

\begin{figure}[t]
    \centering
    \includegraphics[width=\columnwidth]{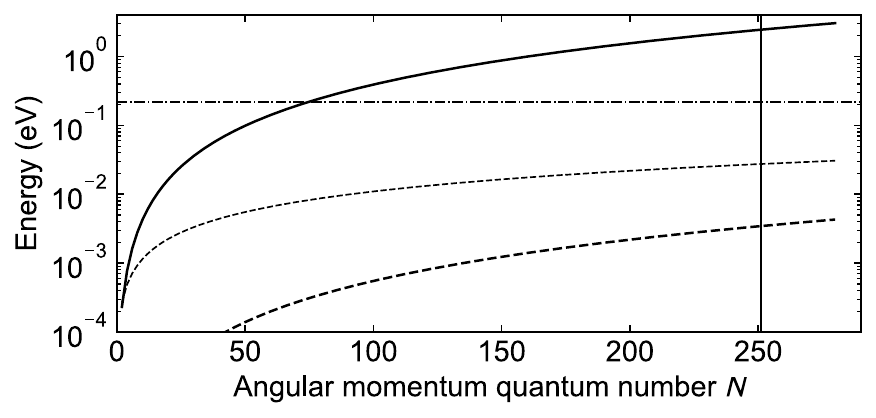}
    \caption{Terms of the rotational Hamiltionian $H_r$ at fixed internuclear distance $R=R_b=3.0$\,\AA.
    Full line: Centrifugal energy $E_r=B(R_b)\,N(N+1)$. Thin short-dashed line: Estimated $\Lambda$-coupling term $E_{\Lambda\Lambda'}=2B(R_b)\sqrt{2N(N+1)}$ between the A$^2\Pi_u$ and B$^2\Sigma^+_u$ electronic states.
    Dash-dotted line:   Potential difference $\Delta E_{AB}=V_B(R_b)-V_A(R_b)$ between these states. 
    Thick dashed line: Second-order perturbation $(E_{\Lambda\Lambda'})^2/\Delta E_{AB}$ of the A$^2\Pi_u$ and B$^2\Sigma^+_u$ level energies.  
    At $N=251$ (vertical line) this energy perturbation is $3.5\times10^{-3}$ eV.}
    \label{fig:rotenergy}
\end{figure}

We have also verified the effect of the terms $(N_+L_-+N_-L_+)$ in the rotational Hamiltonian of Eq.\ (\ref{eq:hrot}).  They can couple potentials of different $|\Lambda|$ according to the occurrence of non-zero matrix elements for $L_+$ and $L_-$.
At the larger internuclear distances becoming relevant for very high $N$, the energetic distance between electronic states of the same symmetry but with different $|\Lambda|$ becomes small, such that they may be efficiently coupled by matrix elements of $L_+$ and $L_-$.  
As most relevant, we consider the A$^2\Pi_u$ ($\Lambda=\pm1$) and B$^2\Sigma_u^+$ ($\Lambda=0$) states at $R\geq2$\,\AA, near the rotational barrier of the A$^2\Pi_u$ potential in Fig.\ \ref{fig:rearrange}(c).  
The energies related to the rotational Hamiltonian $H_r$ at the large internuclear distance $R_b=\SI{3}{\AA}$ are shown in Fig.\ \ref{fig:rotenergy} as a function of $N$.
For $N=250$, near the highest value to be considered, the centrifugal energy amounts to several eV and the $\Lambda$-coupling matrix element \cite{kovacs_1969} reaches $\sim\SI{30}{\milli\electronvolt}$.
This energy is still significantly below the potential difference of the A$^2\Pi_u$ and B$^2\Sigma^+_u$ states [see also Fig.\ \ref{fig:rearrange}(c)] such that the related energy shift can be derived as a second-order perturbation.
This results in $\sim\SI{3}{\milli\electronvolt}$ which we consider still small enough to be neglected.

\subsection{Vibrational levels}
\label{sec:the_vibration}

For modeling radiative and unimolecular decay with an emphasis on highly rotating levels, we require energies and wave functions for a large number of bound and quasibound vibrational eigenstates in the potentials $V_{\gamma N}(R)$.
Inspired by Ref.\ \cite{fedor_2005}, we apply an approximate method to describe the quasibound levels between the dissociation energy $E_{\gamma N}^d=V_{\gamma N}(R\to\infty)$ and the rotational barrier.
At the higher $N$ in the focus of our work, a typical potential $V_{\gamma N}(R)$ has a local minimum at moderate internuclear distance $R_1$ and rises again up to the rotational barrier $V_{\gamma N}(R_2)$ reached at $R_2\lesssim3R_1$. 
We chose to obtain the vibrational energies and wave functions in an exactly bound potential $V'_{\gamma N}(R)$ which keeps the value at the rotational barrier for $R>R_2$, i.e., $V'_{\gamma N}(R>R_2)=V_{\gamma N}(R_2)$.
We expect this modification of the potential to have only little effect on levels energetically well below the barrier.
For such levels, the amplitude of their wave function in the classically forbidden region becomes very low, while also their tunneling dissociation rates decrease strongly.
Previous work on vibrational levels in local potential minima of molecular di-cations \cite{chen_mean-lifetime_1994} used this modification of the potential to predict vibrational level energies and, when comparing to more exact solutions, reached about 4-digit accuracy in this prediction for the highest vibrational level, having tunneling dissociation rates of the order of \SI{e8}{\persec}.
Since our focus is on much smaller tunneling dissociation rates, we expect the method using $V'_{\gamma N}(R)$, largely facilitating the solution of the eigenvalue problem, to be sufficient.

\begin{figure}[t]
    \centering
    \includegraphics[width=\columnwidth]{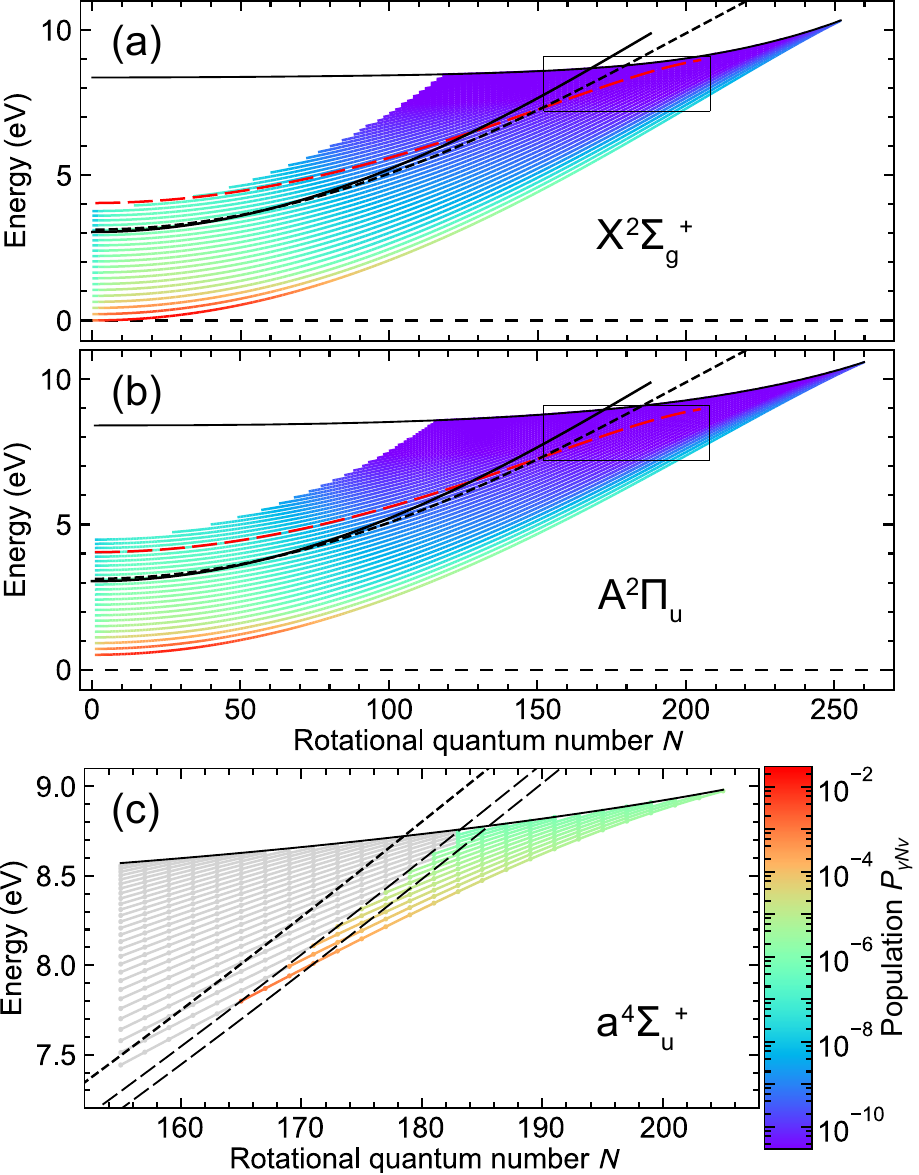}
    \caption{Energies $E_{\gamma Nv}$ and indicative populations $P_{\gamma Nv}$ for vibrational levels in the rotational-electronic potentials $V_{\gamma N}(R)$.
    For the indicated electronic states $\gamma$ of  C$_2{}^-$, energy points $E_{\gamma Nv}$ are shown as functions of $N$ for all levels covered by the present calculations.
    Colored lines connect points for the same $v$ and varying $N$ values (lowest colored lines for $v=0$). 
    The color of a line segment starting at $E_{\gamma Nv}$ encodes, according to the color bar, the estimated level population $P_{\gamma Nv}$, assuming the conditions of Scenario~I discussed in the text.
    Thin full lines indicate the rotational barrier height $V^b_{\gamma N}$.
    Panels (a) and (b) also show the energies of the $v=0$ levels in rotating neutral C$_2$, using $V_{\gamma N}(R)$ for the states $\gamma$ of X$^1\Sigma_g^+$ (thicker full line) and a$^3\Pi_u$ (dotted line).
    Dashed red line: Energy of the $v=0$ level in the C$_2{}^-$ quartet state, using $V_{\gamma N}(R)$ for $\gamma=\text{a$^4\Sigma_u^+$}$.
    Rectangles: Range of energy and of $N$ covered by panel (c).
    Panel (c) also shows the $v=0$ level energy in neutral C$_2$ in the a$^3\Pi_u$ state for angular momentum $N_n=N$ (short-dashed line) as well as for $N_n=N-4$ and $N-6$ (long-dashed lines). a$^4\Sigma_u^+$ levels above the energy for $N_n=N-4$ (shown in grey) are highly unstable as discussed in the text and thus not present in the stored beam. }
    \label{fig:vib_lev_pop}
\end{figure}

The ensemble of vibrational levels calculated in the rotational-electronic potentials $V_{\gamma N}(R)$ is shown in Fig.\ \ref{fig:vib_lev_pop}.
Vibrational levels for the C$_2{}^-$ electronic states X$^2\Sigma_g^+$ and A$^2\Pi_u$ were calculated up to the barrier energy $V^b_{\gamma N}$ for $N\gtrsim110$ and for a limited range of excitation energies ($\lesssim\SI{4.5}{\electronvolt}$) at smaller $N$.
Moreover, the $v=0$ levels in the neutral C$_2$ electronic states X$^1\Sigma_g^+$ and a$^3\Pi_u$ were determined at all $N$ and are included as continuous lines.
Levels of the C$_2{}^-$ doublet states which are energetically allowed to undergo AD into C$_2$ can thus be identified  by being positioned above the corresponding lines in Fig.\ \ref{fig:vib_lev_pop} (a) and (b).

The energy of the $v=0$ level in the a$^4\Sigma_u^+$ state of  C$_2{}^-$  is indicated as a function of $N$ by another (dashed red) line in Fig.\ \ref{fig:vib_lev_pop} (a) and (b).
As long as $N\lesssim150$, this line lies above that indicating the $v=0$ level energy for the neutral C$_2$(a$^3\Pi_u$), reflecting the fact that all quartet levels of C$_2{}^-$ then efficiently decay by spin-allowed AD.
For $N\geq155$, however, several levels in the C$_2{}^-$ quartet state move below the C$_2$(a$^3\Pi_u$) $v=0$ level with the same $N$.
In this $N$-range, vibrational levels up to the barrier energy were therefore also calculated for the a$^4\Sigma_u^+$ state of C$_2{}^-$, as shown in Fig.\ \ref{fig:vib_lev_pop}(c).

Below the (short-dashed) line indicating the $v=0$ level energy of neutral C$_2$(a$^3\Pi_u$) for an angular momentum of $N_n=N$, all C$_2{}^-$ (a$^4\Sigma_u^+$) $Nv$ levels shown in Fig.\ \ref{fig:vib_lev_pop}(c) are energetically stable against AD into a$^3\Pi_u$ neutral levels {\em with the same} $N$ (i.e., $N_n=N$).
Yet, C$_2$(a$^3\Pi_u$) levels with lower $N$ remain energetically accessible if the AD process is allowed to involve a decrease of the nuclear angular momentum.
Such rotationally assisted AD processes were considered in the companion paper \cite{PRL} and will be summarised in Sec.\ \ref{sssec:quartetsigmau}.
To account for this, lines indicating the $v=0$ level energies of neutral C$_2$(a$^3\Pi_u$) for angular momenta $N_n=N-4$ and $N_n=N-6$ are also included in Fig.\ \ref{fig:vib_lev_pop}(c).
Decay rates for rotationally assisted AD are found  \cite{PRL} to lie in the range of $10^{2}\ldots10^{4}$\,s$^{-1}$ relevant for the present experimental conditions when the energies $E_{Nv}$ of anionic quartet levels lie below the C$_2$(a$^3\Pi_u$) $v=0$ level for $N_n=N-4$, but above that for $N_n=N-6$.
The decaying C$_2{}^-$(a$^4\Sigma_u^+$) $Nv$ levels then have energies in the band between the $N_n=N-4$ and $N_n=N-6$ lines in Fig.\ \ref{fig:vib_lev_pop}(c).

For the many excited $\gamma Nv$ levels presented in Fig.\ \ref{fig:vib_lev_pop}, we also considered their relative populations $P_j=P_{\gamma Nv}$.
These quantities directly enter the calculation of the unimolecular signals according to Eq.\ (\ref{eq:signal_simulate}). 
Therefore, their size is relevant for the comparison of the model result to the experimental data, addressed in Sec.\ \ref{sec:comparison_data_rate_model}. 

Since accurate predictions for them are not available, we choose to define for the level populations two Scenarios (see Table \ref{tab:pop}) which are primarily distinguished by the amount of anions in the quartet level allowed to be present in the stored beam.

\begin{table}[tb]
\caption{\label{tab:pop} 
Population scenarios assumed in the model.
For doublet levels ($\gamma={}$X$^2\Sigma_g^+$ and A$^2\Pi_u$) and quartet levels ($\gamma={}$a$^4\Sigma_u^+$), the electronic population fractions $P^e_\gamma$, and the temperatures (in K) of rotational and vibrational excitation are given for the core ($T_r^c$, $T_v^c$) and the tail ($T_r^t$, $T_v^t$) together with the tail fraction $\alpha$, as applied in Eqs.\ (\ref{eq:prot}) and (\ref{eq:pvib}).}
\centering
\begin{ruledtabular}
\begin{tabular}{clcclcc}
&\multicolumn{3}{c}{Doublet}& \multicolumn{3}{c}{Quartet}\\
\cline{2-4} \cline{5-7} \vphantom{$\hat{A}$}
  &\multicolumn{1}{c}{$P^e_\gamma$} & 
  \begin{tabular}{c}$T_r^c$\\$T_r^t$\end{tabular} & 
  \begin{tabular}{c}$T_v^c$\\$T_v^t$\end{tabular} & 
  \multicolumn{1}{c}{$P^e_\gamma$} & 
  \begin{tabular}{c}$T_r^c$\\$T_r^t$\end{tabular} & 
  \begin{tabular}{c}$T_v^c$\\$T_v^t$\end{tabular}\\
\hline
 \multicolumn{7}{l}{Scenario I}\\
  &0.497 
  & \begin{tabular}{c} 2100\\
  8000\footnote{Tail fraction $\alpha=0.01$}\end{tabular} & 
  \begin{tabular}{c} 1200\\
  8000\footnotemark[1]
  \end{tabular} & 
  0.006 
  & \begin{tabular}{c} 2100\\
  ---\footnote{Tail fraction $\alpha=0$}\end{tabular} & 
  \begin{tabular}{c} 1200\\
  ---\footnotemark[2]
  \end{tabular}\\
 \multicolumn{7}{l}{Scenario II}\\
   &0.5 
  & \begin{tabular}{c} 2100\\
  30000\footnote{Tail fraction $\alpha=0.035$}\end{tabular} & 
  \begin{tabular}{c} 1200\\
  30000\footnote{Tail fraction $\alpha=0.05$}
  \end{tabular} & 0 & & 
\end{tabular}
\end{ruledtabular}
\end{table}
The population probabilities are obtained as
\begin{equation}
    P_{\gamma Nv}=P^e_\gamma\, P^{\rm rot}_N(\gamma)\,P^{\rm vib}_v(\gamma N)
    \label{eq:provib}
\end{equation}
with electronic, rotational, and vibrational population fractions $P^e_\gamma$, $P^{\rm rot}_N(\gamma)$, and $P^{\rm vib}_v(\gamma N)$, respectively.
While $P^e_\gamma$ directly follows from Table \ref{tab:pop}, bi-modal thermal distributions are used for $P^{\rm rot}_N(\gamma)$ and $P^{\rm vib}_v(\gamma N)$.
These distributions are obtained as
\begin{equation}
    P^{\rm rot}_N(\gamma)=(1-\alpha) p^{\rm th}_N(\gamma;T^c_r) + \alpha p^{\rm th}_N(\gamma;T^t_r)
    \label{eq:prot} 
\end{equation}
and
\begin{equation}
    P^{\rm vib}_v(\gamma N)=(1-\alpha) p^{\rm th}_v(\gamma N;T^c_v) + \alpha p^{\rm th}_v(\gamma N;T^t_v).
    \label{eq:pvib}    
\end{equation}
Using the normalised thermal single-temperature distributions $p^{\rm th}_m(n;T)$, they define two-component thermal distributions characterized by the core and tail temperatures $T^c_x$ and $T^t_x$ ($x=r$ or $v$ for rotations and vibrations, respectively) and by the tail fraction $\alpha$.
The single-temperature distributions follow the general definition
\begin{equation}
    p^{\rm th}_m(\Gamma;T)=\frac{w_me^{{-E_m(\Gamma)/k_BT}}}
    {\sum_{m'}w_{m'} e^{{-E_{m'}(\Gamma)/k_BT}}}
\end{equation}
where, for rotations ($m=N$, $\Gamma=\gamma$),  $w_m=w_N=2N+1$, the energy is given by
\begin{equation}
    E_N(\gamma)=E_{\gamma N\,v=0}-E_{\gamma\,0\,v=0},
\end{equation}
and the sum runs over the quantum numbers of all 
rotational levels for which stable or quasi-stable vibrational levels exist in the electronic state $\gamma$.
Similarly, for vibrations ($m=v$, $\Gamma=\gamma N$), $w_m=w_v\equiv 1$, the energy is given by
\begin{equation}
    E_v(\gamma N)=E_{\gamma N\,v}-E_{\gamma N\,v=0},
    \label{eq:evibpvib}
\end{equation}
and the sum runs over the quantum numbers of all stable or quasi-stable vibrational states in the rotational-electronic state $\gamma N$.
For the high tail temperature in Scenario II, the high vibrational levels not included for small $N$ in Fig.\ \ref{fig:vib_lev_pop}(a) and (b) have a (yet very limited) influence on the normalization and  for this purpose their energies were estimated by a simplified method \cite{suppl}.

Scenario I assumes that (at the beginning of observation, $t=t_1$) a small fraction $P^e_\gamma$ of the anions in the stored beam is in quartet levels ($\gamma=\text{a$^4\Sigma_u^+$}$) at quantum numbers $Nv$ that ensure quasi-stability against AD.  
According to the discussion above and in Sec.\ \ref{sssec:quartetsigmau}, this stabilization occurs for all levels whose energy lies below that of the a$^3\Pi_u,N-4,v\!=\!0$ level of C$_2$, cf.\ Fig.\ \ref{fig:vib_lev_pop}(c).
The size of this fraction (Table \ref{tab:pop}) is set as large as required to match the experimental amplitude of the AD signal in the comparison with the model results (Sec.\ \ref{sec:comparison_data_rate_model}).
Scenario I also puts priority on keeping the rotational and vibrational temperatures within a range compatible with the current evidence about the experimentally relevant sputter-type of ion sources (see Sec.\ \ref{sec:exp_population}).
Thus, the values of $T_r^t,T_ v^t,$ and $\alpha$, listed in Table \ref{tab:pop}, are set as high as can be reconciled with these expected ion source properties.
Figure \ref{fig:vib_lev_pop}(a) and (b) then reveal that, for the doublet levels of C$_2{}^-$, the population of high-$N$ levels which might contribute to tunneling dissociation (AF) is still exceedingly small, apparently contradicting the experimental result.
However, the comparison with experiment (Sec.\ \ref{sec:comparison_data_rate_model}) shows that a model including tunneling dissociation from the quartet levels can explain also the AF signal, choosing the set of rotational and vibrational temperatures as listed in Table \ref{tab:pop} for Scenario~I.

Scenario II assumes that the C$_2{}^-$ quartet levels are not populated at all in the stored beam.
As further discussed in Sec.\ \ref{sec:comparison_data_rate_model}, some agreement between the model result and the observations can be obtained for elevated values of the rotational and vibrational temperatures in the tail and vibrational and rotational tail fractions further increased from Scenario I to several percent instead of only 1\%.
Such extremely strong tails in rotational and vibrational distributions are not well supported by the available evidence about the applied sputter ion source.
Nevertheless, we keep this Scenario in order to verify how well the observations might be explained without considering population of the  C$_2{}^-$ quartet levels.

\subsection{Dipole moments and spin--orbit coupling}
\label{sec:the_matelements}

\begin{figure}[b]
    \centering
    \includegraphics[width=\columnwidth]{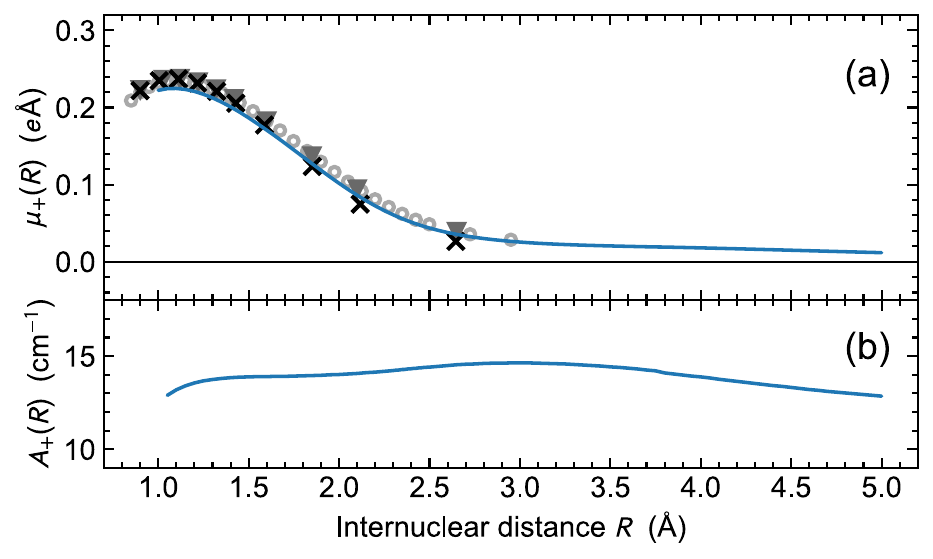}
    \caption{Coupling functions derived from the ab-initio calculations. (a) Dipole moment function $\mu_+(R)$ between the states X$^2\Sigma_g^+$ and A$^2\Pi_u$, compared with previous calculations shown by symbols: 
    $\blacktriangledown$ \citet{quartet_theo_shi}, $\times$ \citet{rosmus_werner},  $\circ$ \citet{sedivcova_potential_2006}.
    (b) Spin--orbit mixing function $\hat{A}_+(R)$ between a$^4\Sigma_u^+$ ($\Lambda=0,\Sigma=3/2$) and
    A$^2\Pi_u$ ($\Lambda=1,\Sigma=1/2$).}
    \label{fig:dipole_transition_moment}
\end{figure}

For the decay rate calculations below, further properties of the electronic states were extracted from the presented series of ab-initio calculations.
We focus on the matrix elements for electric dipole transitions between the X$^2\Sigma^+_g$ and A$^2\Pi_u$ states as well as on the spin--orbit mixing of the A$^2\Pi_u$ and a$^4\Sigma_u^+$ states.

As for the transition dipole moment, the spinless orbital calculations using {\sc molpro} yield the matrix elements of the cartesian components $\hat{\mu}_x,\hat{\mu}_y$ of the dipole moment operator with the electronic eigenstates of angular character $\ket{z}$ for X$^2\Sigma^+_g$ and of characters $\ket{x}$ and $\ket{y}$ for A$^2\Pi_u$.
At any fixed $R$, the cartesian dipole matrix elements obtained by {\sc molpro} are
\begin{equation}
   \bra{\text{X$^2\Sigma^+_g$}, z}\hat{\mu}_{x}\ket{\text{A$^2\Pi_u$}, x}=
   \bra{\text{X$^2\Sigma^+_g$}, z}\hat{\mu}_{y}\ket{\text{A$^2\Pi_u$}, y} = d
\end{equation}
and, thus, given by a single $R$-dependent value $d$, resulting as a negative real number.
Expressing the cartesian components of $\hat{\mu}$ as well as the eigenfunctions by spherical tensor components $\ket{\Lambda}$ and $\hat{\mu}_{q}$, respectively, results in
\begin{eqnarray}
   \mu^{\text{A-X}}_+
   &=&\bra{\text{X$^2\Sigma^+_g$},\Lambda\!=\!0}
   \hat{\mu}_{1}\ket{\text{A$^2\Pi_u$},\Lambda\!=\!-1} \nonumber\\
   \mu^{\text{A-X}}_-&=&\bra{\text{X$^2\Sigma^+_g$},\Lambda\!=\!0}
   \hat{\mu}_{-1}\ket{\text{A$^2\Pi_u$},\Lambda\!=\!+1}
\end{eqnarray}
with 
\begin{eqnarray}
   \mu^{\text{A-X}}_+=\mu^{\text{A-X}}_-=-d.
\end{eqnarray}
Interpolating the series of fixed-$R$ calculations, we obtain the dipole moment function $\mu^{\text{A-X}}_+(R)$ shown in Fig.\ \ref{fig:dipole_transition_moment}(a), which is in good agreement with other calculations for this transition in C$_2{}^-$.

The vibrational averaging of $\mu^{\text{A-X}}_+(R)$ is performed with vibrational wavefunctions in the rotational-electronic potentials.
This yields for the higher level in electronic state $a$ the wavefunction $\ket{a,N'v'}$ in the potential $V_{aN'}(R)$, while for the  lower level in state $b$ the wavefunction 
$\ket{b,N''v''}$ is obtained in the potential $V_{bN''}(R)$. In the resulting dipole overlap integral
\begin{equation}
    \mu^{a\text{-}b}\bm{(}a,N'v';b,N''v''\bm{)} = \bra{a,N'v'}
    \mu^{\text{A-X}}_{+}(R) 
    \ket{b,N''v''}
    \label{eq:mu_vibint}
\end{equation}
the same dipole moment function $\mu^{\text{A-X}}_+(R)$ can be used for A$^2\Pi_u$-X$^2\Sigma^+_g$ and X$^2\Sigma^+_g$-A$^2\Pi_u$ transitions.

Also the spin--orbit Hamiltonian $\hat{H}_{\rm SO}$ was obtained in the {\sc molpro} calculations.
This involves the component $\Sigma$ of the total electron spin along the internuclear axis as an additional quantum number.
Moreover, the component $\Omega=\Sigma+\Lambda$ of the total electronic angular momentum along the internuclear axis is a conserved quantity.
It is directly given by $\Omega=\Sigma$ for the a$^4\Sigma_u^+$ electronic state.
Only states of the same $\Omega$ are coupled by $\hat{H}_{\rm SO}$.
In terms of the cartesian eigenstates used by {\sc molpro}, the non-zero matrix elements which couple the a$^4\Sigma_u^+$ and A$^2\Pi_u$ states are found for $\Omega=\pm3/2$ as
\begin{eqnarray}
  \bra{\text{a$^4\Sigma_u^+$},z,
  \Sigma\!=\!\pm\textstyle\frac{3}{2}}
   \hat{H}_{\rm SO}\ket{\text{A$^2\Pi_u$},x,
  \Sigma\!=\!\pm\textstyle\frac{1}{2}} &=& \alpha
   \nonumber\\
  \bra{\text{a$^4\Sigma_u^+$},z,
  \Sigma\!=\!\pm\textstyle\frac{3}{2}}
   \hat{H}_{\rm SO}\ket{\text{A$^2\Pi_u$},y,
  \Sigma\!=\!\pm\textstyle\frac{1}{2}} &=& \mp{\rm i}\alpha,
  \hspace{1cm}
\end{eqnarray}
where only the upper or the lower sign should be used in all occurrences of $\pm$ or $\mp$ in each line.
In the same notation,  the non-zero matrix elements for $\Omega=\mp1/2$ are found as
\begin{eqnarray}
  \bra{\text{a$^4\Sigma_u^+$},z,
  \Sigma\!=\!\mp\textstyle\frac{1}{2}}
   \hat{H}_{\rm SO}\ket{\text{A$^2\Pi_u$},x,
  \Sigma\!=\!\pm\textstyle\frac{1}{2}} &=& \beta
   \nonumber\\
  \bra{\text{a$^4\Sigma_u^+$},z,
  \Sigma\!=\!\mp\textstyle\frac{1}{2}}
   \hat{H}_{\rm SO}\ket{\text{A$^2\Pi_u$},y,
  \Sigma\!=\!\pm\textstyle\frac{1}{2}} &=& \pm{\rm i} \beta.
  \hspace{1cm}
\end{eqnarray}
Hence, the eight cartesian non-zero spin--orbit mixing matrix elements can be expressed at any fixed $R$ by the two $R$-dependent values $\alpha$ and $\beta$, resulting as negative real numbers.
Inserting the tensorial definition of the $\ket{\Lambda}$ states in terms of the cartesian eigenfunctions (and omitting the state labels) this yields
\begin{eqnarray}
\hat{A}_{\pm3/2}&\!=\!&
\bra{\Lambda\!=\!0,\Sigma\!=\!
{\pm\textstyle\frac{3}{2}}}
\hat{H}_{\text{SO}}
\ket{\Lambda\!=\!\pm1,\Sigma\!=\!
{\pm\textstyle\frac{1}{2}}}
= \mp\sqrt{2}\alpha
\nonumber\\
\hat{A}_{\mp1/2}&\!=\!&
\bra{\Lambda\!=\!0,\Sigma\!=\!
{\mp\textstyle\frac{1}{2}}}
\hat{H}_{\text{SO}}
\ket{\Lambda\!=\!\mp1,\Sigma\!=\!
{\pm\textstyle\frac{1}{2}}}
= \pm\sqrt{2}\beta.
\nonumber\\
\label{eq:afunc_val}
\end{eqnarray}
At all $R$, we find $\hat{A}_{-1/2}=-\hat{A}_{3/2}/\sqrt{3}$.
Hence, it is sufficient to use the series of fixed-$R$ results for $\hat{A}_{3/2}$ to interpolate the coupling function $\hat{A}_+(R)$ for the $\Omega=3/2$ states.
The function $\hat{A}_+(R)$ is shown in Fig. \ref{fig:dipole_transition_moment} (b).
The coupling functions for the states with other $\Omega$ result from $A_-(R)=-A_+(R)/\sqrt{3}$ for $\Omega=-1/2$ and observing the sign change $\hat{A}_{-\Omega}=-\hat{A}_{\Omega}$ implied by Eq.\ (\ref{eq:afunc_val}).

The ab-initio spin--orbit calculations also yield the fine-structure splitting function $A(R)$ between the levels $|\Omega|=3/2$ and $1/2$ of the A$^2\Pi_u$ electronic state.
Vibrational averaging over the $v=0$, $N=1$ vibrational wave function using the ab-initio potential yields $A=\SI{22.6}{\wn}$, which agrees within $\sim$10\% with the experimental A$^2\Pi_u$ fine-structure splitting constant of $A=\SI{25.009(15)}{\wn}$ \cite{rehfuss}.

\section{Radiative and unimolecular decay rates}\label{sec:the_rates}

\subsection{Radiative decay}\label{sec:calculation_rad}

The procedures of the radiative rate calculations are described in Appendix \ref{app:rad} and the obtained expressions and results are given below.

\subsubsection{Doublet levels}\label{sec:calculation_rad_doublet}
Regarding the doublet levels, we consider the spin-allowed electric dipole transitions A$^2\Pi_u$--X$^2\Sigma^+_g$ (decay of A$^2\Pi_u$) and X$^2\Sigma^+_g$--A$^2\Pi_u$ (decay of X$^2\Sigma^+_g$).  
For each $N$, the considered doublet terms can have levels with $J=N+1/2$ (the F1 term, $i=1$) and $J=N-1/2$ (the F2 term, $i=2$).  
The two A$^2\Pi_u$ levels are spaced by spin--orbit interaction described by the splitting constant $A(R)$ of order \SI{25}{\wn}.  
This is significant compared to the rotational energy for low $N$ values.
Considering the ratio $Y$ of the values $A(R)$ and $B(R)$ (see Sec.\ \ref{sec:the_potentials}) averaged over the vibrational wave function, its value is found to be $\lesssim 60$ which leads to typical fine-structure-related correction terms of $<0.1$ for $N\gtrsim70$ [cf.\ Eq.\ (\ref{eq:erot})].
Hence, we can neglect spin--orbit effects and use pure Hund's case (b) states for A$^2\Pi_u$ in this high-$N$ range.
    
The decay rates for transitions originating in the A$^2\Pi_u(N'v')$ levels are based on Eq.\ (\ref{eq:resultAX1}).
The states A$^2\Pi_u(N'v',\epsilon'_N=+1)$ decay by $\Delta N = \pm1$ with the rates
    \begin{eqnarray}
      A^{{\rm A},N'v'}_ {{\rm X},(N'\mp1)v''}&=&
      2\,C_0 \left(\tnu^{{\rm A},N'v'}_{{\rm X},(N'\mp1)v''}\right)^3
      \hspace{15mm}\nonumber\\ &&{}\times
      \left[\mu^{\text{A-X}}\bm{(}{\rm A},N'v';{\rm X},(N'\mp1)v''\bm{)}\right]^2
      \nonumber\\ &&{}\times
                     \left\lbrace
          \begin{array}{lr}\frac{N'+1}{2(2N'+1)}&\text{for $\Delta N=+1$}\\
                    \frac{N'}{2(2N'+1)}&\text{for $\Delta N=-1$}
                  \end{array}\right. 
      \label{eq:resultAXplus}
    \end{eqnarray}
which for $N'\gg1$ is close to
\begin{eqnarray}
  A^{{\rm A},N'v'}_ {{\rm X},(N'\mp1)v''} = 
  \frac{C_0}{2} \left(\tnu^{{\rm A},N'v'}_{{\rm X},(N'\mp1)v''}\right)^3
  \hspace{15mm}\nonumber\\ {}\times
  \left[\mu^{\text{A-X}}\bm{(}{\rm A},N'v';{\rm X},(N'\mp1)v''\bm{)}\right]^2.
  \label{eq:resultAXlimplus}
\end{eqnarray}
The states A$^2\Pi_u(N'v',\epsilon'_N=-1)$ decay by $\Delta N = 0$ with the rates
    \begin{eqnarray}
       A^{{\rm A},N'v'}_{{\rm X},N'v''}&=&
      C_0 \left(\tnu^{{\rm A},N'v'}_{{\rm X},N'v''}\right)^3
      \nonumber\\ &&{}\times
      \left[\mu^{\text{A-X}}\bm{(}{\rm A},N'v';{\rm X},N'v''\bm{)}\right]^2
      \label{eq:resultAXminus}
    \end{eqnarray}
as the squared 3$j$ symbol in Eq.\ (\ref{eq:resultAX1}) for this case exactly equals $[2(2N''+1)]^{-1}$.
$C_0$ is given in Eq.\ (\ref{eq:c0}).
With this, the total radiative decay rate of A$^2\Pi_u(N'v')$ levels is
\begin{eqnarray}
    A^{{\rm A},N'v'}=\sum_{v'',\Delta N}
    A^{{\rm A},N'v'}_{{\rm X},(N'-\Delta N)v''},
    \label{eq:atotal_A}
\end{eqnarray}
where $\Delta N=\pm 1$ for those with $\epsilon'_N=+1$ and $\Delta N=0$ for the $\epsilon'_N=-1$ levels.  
For isotopically symmetric C$_2{}^-$, $\epsilon'_N=+1$ corresponds to odd and $\epsilon'_N=-1$ to even $N'$, so that the resulting transitions into the X$^2\Sigma^+_g$ state lead only into the levels of even $N''$ which are allowed by spin statistics.

The partial decay rates from  X$^2\Sigma^+_g(N'v')$  levels are based on Eq.\ (\ref{eq:resultXA1}).
The X$^2\Sigma^+_g$ states have an $\epsilon'_N=+1$ component only and decay to
A$^2\Pi_u(\epsilon''_N=+1)$ by $\Delta N = \pm1$ transitions with the rates
\begin{eqnarray}
  A^{{\rm X},N'v'}_ {{\rm A},(N'\mp1)v''}
  &=&2\,C_0 \left(\tnu^{{\rm X},N'v'}_ {{\rm A},(N'\mp1)v''}\right)^3
      \hspace{15mm}\nonumber\\
  &&{}\times
    \left[\mu^{\text{A-X}}\bm{(}{\rm X},N'v';{\rm A},(N'\mp1)v''\bm{)}\right]^2
  \nonumber\\ &&{}\times
                 \left\lbrace
      \begin{array}{lr}\frac{N'-1}{2(2N'+1)}&\text{for $\Delta N=+1$}\\
                \frac{N'+2}{2(2N'+1)}&\text{for $\Delta N=-1$}
              \end{array}\right.         
   \label{eq:resultXApm}
\end{eqnarray}                            
which for $N'\gg1$ is close to
    \begin{eqnarray}
      A^{{\rm X},N'v'}_ {{\rm A},(N'\mp1)v''}
      =\frac{C_0}{2} \left(\tnu^{{\rm X},N'v'}_ {{\rm A},(N'\mp1)v''}\right)^3
      \hspace{15mm}\nonumber\\ {}\times
      \left[\mu^{\text{A-X}}\bm{(}{\rm X},N'v';{\rm A},(N'\mp1)v''\bm{)}\right]^2.
      \label{eq:resultXAlimpm}
    \end{eqnarray}
They also decay to A$^2\Pi_u(\epsilon''_N=-1)$ by $\Delta N = 0$ transitions with  the rates
    \begin{eqnarray}
      A^{{\rm X},N'v'}_ {{\rm A},N'v''}
      &=& C_0 \left(\tnu^{{\rm X},N'v'}_ {{\rm A},N'v''}\right)^3  \hspace{15mm}\nonumber\\ &&{}\times
      \left[\mu^{\text{A-X}}\bm{(}{\rm X},N'v';{\rm A},N'v''\bm{)}\right]^2
      \label{eq:resultXAzero}
    \end{eqnarray}
as the relevant squared 3$j$ symbol again equals $[2(2N''+1)]^{-1}$.
With this, the total radiative decay rate of X$^2\Sigma^+_g$ levels is
\begin{eqnarray}
    A^{{\rm X},N'v'}=\sum_{v'',\Delta N}
    A^{{\rm X},N'v'}_{{\rm A},(N'-\Delta N)v''},
    \label{eq:atotal_X}
\end{eqnarray}
the $\Delta N$ sum extending over $0$ and $\pm 1$. 

\begin{figure}[t]
    \centering
    \includegraphics[width=\columnwidth]{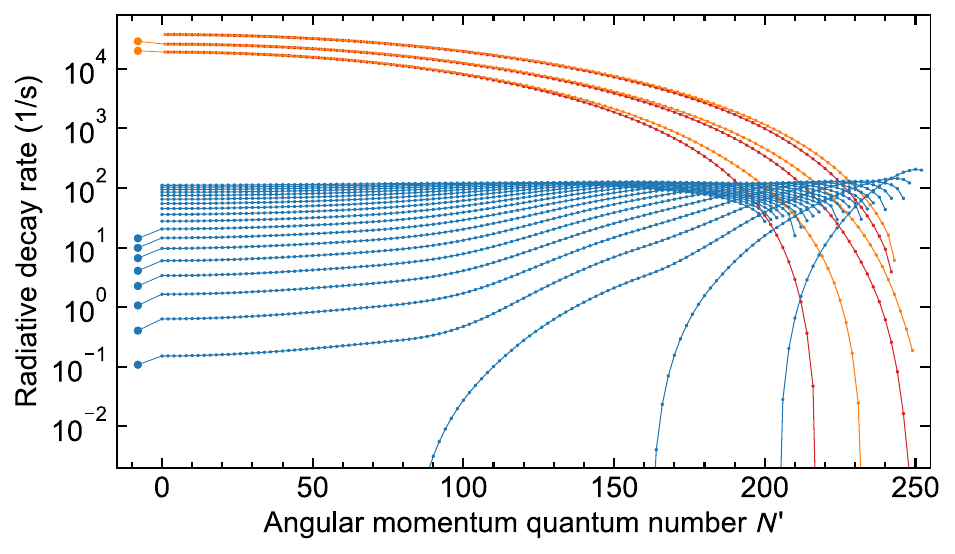}
    \caption{Total rates of spin-allowed radiative decay as functions of the upper-state angular momentum $N=N'$.  
    Upper group of line pairs: $A^{A,N'v'}$ from Eqs.\ (\ref{eq:resultAXplus}),  (\ref{eq:resultAXminus}), and (\ref{eq:atotal_A}) for levels A$^2\Pi_u(N'v')$ with $v'=0,2,6$.
    In the line pairs, the upper (orange) line is for the $\epsilon_N=+1$ levels (odd $N'$ for isotopically symmetric C$_2{}^-$, decay by $\Delta N=\pm1$) and the lower (red) one for $\epsilon_N=-1$ levels (even $N'$ for isotopical symmetry, decay by $\Delta N=0$).
    Lower group of lines: $A^{X,N'v'}$ from Eqs.\ (\ref{eq:resultXApm}),  (\ref{eq:resultXAzero}), and (\ref{eq:atotal_X}) for the X$^2\Sigma^+_g(N'v')$ levels with $v'\leq20$.
    The small symbols along all lines indicate the allowed $N'$ values assuming isotopical symmetry.
    The larger symbols connected to some of the data for the lowest $N'$ are earlier theoretical results \cite{rosmus_werner}, partly re-scaled as discussed in the text.}
    \label{fig:atotal_AX}
\end{figure}

The total decay rates from Eqs.\ (\ref{eq:atotal_A}) and (\ref{eq:atotal_X}) are shown in Fig.\ \ref{fig:atotal_AX}. 
In earlier studies \cite{iida_state-selective_2020,iizawa_photodetachment_2022} it was pointed out that levels X$^2\Sigma^+_g(v')$ have a large preference to decay into the highest energetically open level A$^2\Pi_u(v'')$, corresponding to $v'\to v''$ transitions of (for the lowest values) $3\to0$, $4\to1$, $5\to2$, etc., with wavenumbers around \SI{1000}{\wn}. 
This, together with unfavorable vibrational overlap, leads to rather small decay rates. 
In contrast, the A$^2\Pi_u(v')$ levels decay mostly with $v''\sim v'$, leading to transition wavenumbers larger by at least a factor of 4 and much larger vibrational overlap. 
This is clearly reflected in our results, except for the highest $N'$. 

The nature of the decay channels makes the level decay rates depend very critically on the energies and wave functions of excited vibrational levels.
The results of Fig.\ \ref{fig:atotal_AX} for A$^2\Pi_u(N'=1)$ agree with other calculations \cite{rosmus_werner,quartet_theo_shi,da_silva_transition_2024} within a factor of $\sim$2.
Our X$^2\Sigma^+_g$ decay rates turn out to differ even more strongly (factor of $\sim$3) in a direct comparison with earlier results \cite{rosmus_werner}.
However, using the spectroscopic parameters derived from the earlier calculation (Table III of Ref.\ \cite{rosmus_werner}), the energies of the most relevant X$^2\Sigma^+_g(v')$-A$^2\Pi_u(v'')$ transitions in this calculation are found to be larger than our transition energies by the substantial factor of $\sim$1.65 (all cases up to $v'=10$, $v''=7$) which enters to the third power in the total decay rates.
Scaling the earlier calculated rates \cite{rosmus_werner} by the ratio of the transition-energy factors, our results turn out to be in more reasonable agreement, now being larger by a factor of $\sim$1.4.
We also compared our transition energies for the dominant $v'\to v''$ cases with the ones obtained from the experimental \cite{rehfuss} spectroscopic parameters, finding agreement within about 10\% up to the case of X$^2\Sigma^+_g(v'\!=\!6)$, corresponding to the $6\to3$ vibrational transition.
Thus, the deviations from other low-$N$ theoretical results probably reflect uncertainties of the ab-initio calculations and in spite of these imperfections we apply the data of Fig.\ \ref{fig:atotal_AX} in our modeling studies.

Although this is not pursued in the present work, the large set of partial decay rates obtained here may also be used to model a full radiative cascade of rotational de-excitation.  
As long as the small fine-structure shifts of the transition energies are neglected, the total decay rates of the fine-structure components should be the same within the multiplets and independent of the coupling scheme \cite{watson_honllondon_2008}.

\subsubsection{Quartet levels}\label{sec:calculation_rad_quartet}

Considering the possibility of long-lived levels in the potential a$^4\Sigma_u^+$ at high rotation ({$N\gtrsim160$}), it is of interest to estimate their spontaneous radiative decay rates.  
Radiative decay is expected to be caused by spin--orbit perturbation of levels in the a$^4\Sigma_u^+$ potential by doublet ungerade ($u$) levels, in particular in the A$^2\Pi_u$ potential.  
The resulting doublet admixtures in the coupled ungerade eigenstates will then lead to a dipole moment with X$^2\Sigma_g^+$ levels causing spontaneous radiative decay.

For each $N'$, the a$^4\Sigma_u^+$ excited levels of the spontaneuous transitions are fine structure terms with the sublevels F1$\ldots$F4 with $J'=N'+3/2$, $J'=N'+1/2$, $J'=N'-1/2$, and $J'=N'-3/2$, respectively.   Also, the X$^2\Sigma_g^+$ lower levels of the spontaneous transitions are fine structure terms with the sublevels F1 and F2 with $J''=N''+1/2$ and $J''=N''-1/2$, respectively.  

Important for the doublet admixtures in the quartet states, giving rise to the spin-forbidden decay, is the difference of the energies $E_{\text{a},N'v'}$ and $E_{\text{A},\bar{N}\bar{v}}$ between the vibrational levels a$,N'v'i'$ and A$,\bar{N}\bar{v}$.  
As discussed in Appendix \ref{app:rad_quart}, the spin--orbit perturbation by the A$^2\Pi_u$ electronic state is restricted to the sublevels of this state with $\bar{J}=J'$, which belong to quantum numbers $\bar{N}=J'\pm1/2$.
The strength of the perturbation is given by the vibrational averages $\hat{A}_+(N'v',\bar{N}\bar{v})$ of the spin--orbit mixing function shown in Fig.\ \ref{fig:dipole_transition_moment}.

\begin{figure}[t]
    \centering
    \includegraphics[width=\columnwidth]{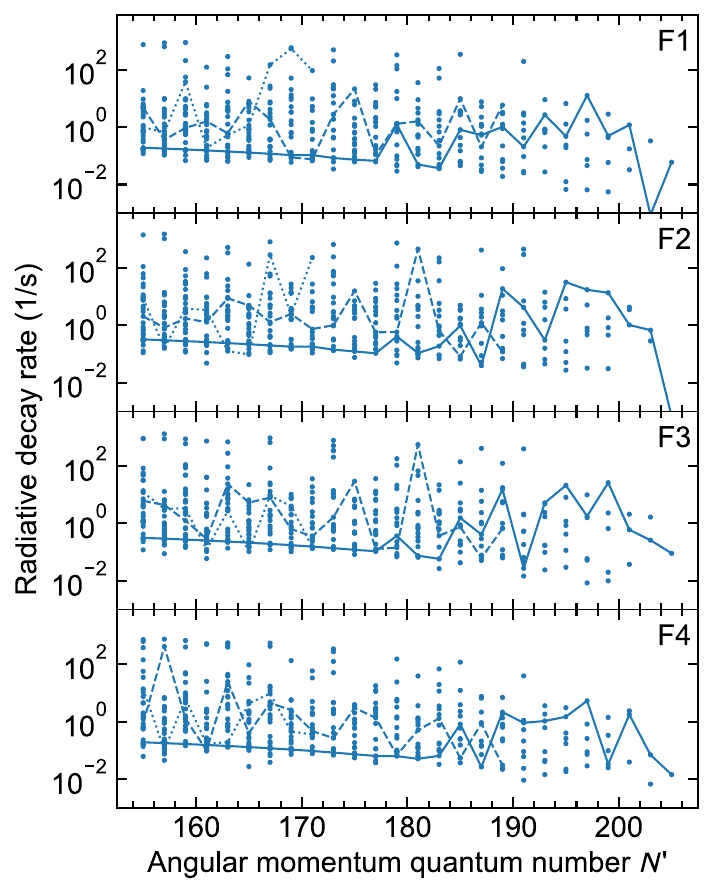}
    \caption{Total rates of spin-forbidden radiative decay of all vibrational levels of the a$^4\Sigma_u^+$ electronic state with upper-state angular momentum $N=N'\geq155$ and in the fine-structure sublevels F1$\ldots$F4.
    Interconnected are the levels with $v'=0$ (full lines), $v'=10$ (dashed lines) and $v'=20$ (dotted lines).}
    \label{fig:rdec_quartet}
\end{figure}

Starting in an initial fine-structure sub-level a$,N'v'i'$, the partial decay rates into individual final sub-levels labeled by X$,N''v''i''$ are given by
\begin{equation}
    A_{{\rm a},N'v'i'}^{{\rm X},N''v''i''} = C_0\left(\tnu_{{\rm a},N'v'}^{{\rm X},N''v''}\right)^3\; \frac{S_{N'v'J'}^{N''v''J''}}{2J'+1}
\end{equation}
according to Eq.\ (\ref{eq:arate}) using the line strength from Eq.\ (\ref{eq:s_dq_p}) with $J'=N'+5/2-i'$ for terms F$(i')$ and $J''=N''+3/2-i''$ for terms F$(i'')$.
As described in Appendix \ref{app:rad_quart}, the final angular momentum quantum numbers are $N''=N'-\Delta N$ with odd $\Delta N$ where $|\Delta N|\leq3$.
An example of partial decay rates summed over final fine-structure levels $i''$ and over the final angular momenta $N''$, still dependent on the final vibrational level $v''$, is given in Fig.\ \ref{fig:quartet_partial} of  Appendix \ref{app:rad_quart}.
As expected since the perturber levels A$,\bar{N}\bar{v}$ are densely spaced, the spin--orbit perturbations lead to significant doublet admixture in certain $N'v'$ initial levels and, depending also on the A$,\bar{N}\bar{v}$-X$,N''v''$ matrix elements, the decay into certain $v''$ is enhanced.
However, the effect scatters significantly, including also significant differences between the initial fine-structure levels $i'$.

The total radiative decay rates by spin-forbidden transitions into the X$^2\Sigma_g^+$ electronic state,
\begin{equation}
    A_{{\rm a},N'v'i'}^{{\rm X}} = \sum_{N''v''i''} A_{{\rm a},N'v'i'}^{{\rm X},N''v''i''}
\end{equation}
($E_{\text{X},N''v''}<E_{\text{a},N'v'}$) for fine-structure levels a$,N'v'i'$  are shown in Fig.\ \ref{fig:rdec_quartet}.
The radiative decay rates range down to $\sim$ \SI{e-2}{\persec}, so that many a$^4\Sigma_u^+$ levels can live for several seconds unless other decay channels are relevant. 
For many other levels, however, the radiative decay rates increase up to $\sim$ \SI{e3}{\persec} and thus shorten their possible lifetimes down into the millisecond range.
The scatter of the radiative decay rates between the different initial levels $N'v'$ and also their fine-structure terms F($i'$) is strong and mostly irregular.

\subsection{Tunneling dissociation} \label{sec:calculation_diss}

A fit of the asymptotic energy to the X$^2\Sigma_g^+$ ab-initio potential curve at $N=0$ and the calculation of the lowest vibrational level in this potential (see Sec.\ \ref{sec:the_vibration}) yield a dissociation energy of $D_0=\SI{8.409}{\electronvolt}$ for our C$_2{}^-$ model.
Rotational-vibrational levels with energies $E_{\gamma Nv}$ above the dissociation limit $D_0$ can undergo tunneling dissociation through the rotational barrier, described by the rotational-electronic potential $V_{\gamma N}(R)$ with a local maximum $V^b_{\gamma N}$ (see Sec.\ \ref{sec:the_vibration}).

\begin{figure}
    \centering
    \includegraphics[width=0.45\textwidth]{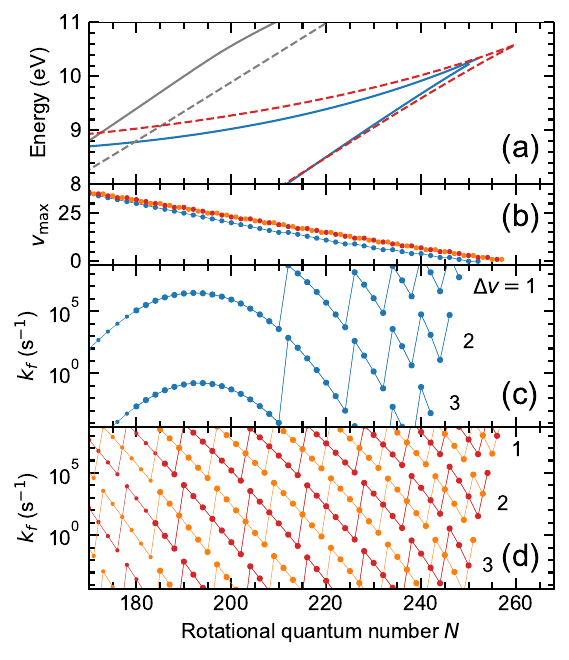}
    \caption{Tunneling dissociation rates $k_f$ of levels $N\, v$, where $v=v_{\rm max}{-\Delta v}$, with $\Delta v={1}\ldots{3}$ in the doublet states X$^2\Sigma_g^+$ and A$^2\Pi_u$.
    (a) Rotational barrier energy $V^b_{\gamma N}$ and lowest vibrational energy $E_{\gamma N\,v=0}$ for X$^2\Sigma_g^+$ (blue full lines) and A$^2\Pi_u$ (red dashed lines) shown with the lowest energies $E_{\gamma_n N\,v=0}$ of neutral C$_2$ in the X$^1\Sigma_g^+$ and a$^3\Pi_u$ states (grey full and dashed lines, respectively).
    (b) Highest vibrational excitation $v_{\rm max}$ in the potentials $V_{\gamma N}(R)$ for X$^2\Sigma_g^+$ (blue symbols) and A$^2\Pi_u$ (red and orange symbols for even and odd $N$, respectively).
    (c) Dissociation rates $k_f$ for levels in the  X$^2\Sigma_g^+$ state with $\Delta v$ as indicated.
    (d) Rates $k_f$ for levels in the  A$^2\Pi_u$ state with indicated $\Delta v$, orange and red symbols as in (b).
    Larger symbols in (c) and (d): energy levels below the energy $E_{\gamma_n N\,v=0}$ of neutral C$_2$ (a$^3\Pi_u$) and hence vanishing AD rates.
    Straight interconnecting lines in (b)-(d) are to guide the eye.}
\label{fig:kd_doublet}
\end{figure}

We estimate the dissociation rates by the semiclassical Wentzel--Kramers--Brillouin (WKB) approximation, following in this step the procedure outlined by \citet{fedor_2005}.
From the level energy $E_{\gamma Nv}<V^b_{\gamma N}$ but above the asymptotic energy, the classical turning points $R_{1,2}$ of the barrier are determined using $V_{\gamma N}(R_1)=V_{\gamma N}(R_2)=E_{\gamma Nv}$.
The semiclassical transmission coefficient through the barrier for outward moving nuclei is then given by
\begin{equation}
    T=\exp\left[-2\int_{R_1}^{R_2}\!\!\sqrt{\frac{2m_r}{\hbar^2}\left(
    \vphantom{\sum}V_{\gamma N}(R)-E_{\gamma Nv}\right)}\;dR\right].
    \label{eq:dissociation_p}
\end{equation}
This yields an estimate for the fragmentation rate $k_f$ by tunneling dissociation of the level $\gamma Nv$ by multiplication with the semiclassical oscillation frequency $\nu$, 
\begin{equation}
    k_f=\nu T,
    \label{eq:dissociation}
\end{equation}
where we determine the classical attempt-frequency $\nu$ from the differences of the vibrational energies for levels adjacent to the level $\gamma Nv$.
For levels $v<v_{\rm max}$, we used the average energy difference to adjacent levels, $\nu=(E_{\gamma N\,v+1}-E_{\gamma N\,v-1})/2h$, while in the less relevant cases of $v=v_{\rm max}$ only the energy difference to the next lower level was employed.
(In the exceptional, even less relevant cases of $v=v_{\rm max}=0$, $\nu$ was replaced by the harmonic frequency found for this level.)

\begin{figure}
        \centering
        \includegraphics[width=0.45\textwidth]{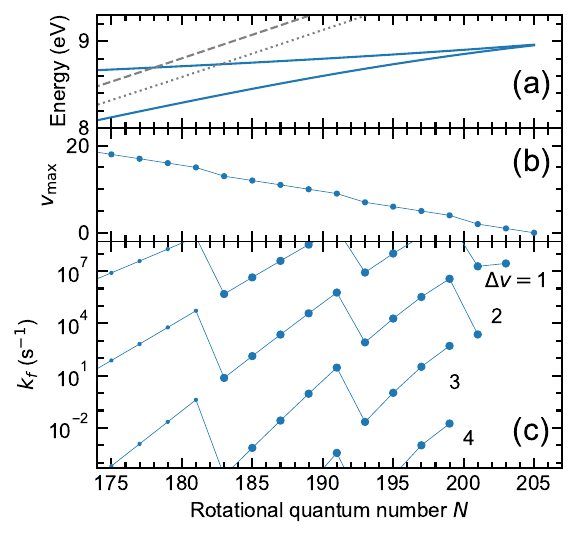}
    \caption{Tunneling dissociation rates $k_f$ of  levels $N\, v$, where $v=v_{\rm max}{-\Delta v}$, with $\Delta v={1}\ldots{4}$ in the quartet state a$^4\Sigma_u^+$.
    (a) Rotational barrier energy $V^b_{\gamma N}$ and lowest vibrational energy $E_{\gamma N\,v=0}$ for a$^4\Sigma_u^+$ (blue full lines) and the lowest energies $E_{\gamma_n N_n\,v=0}$ of neutral C$_2$ in the a$^3\Pi_u$ state at same rotation as in the anion ($N_n=N$, grey dashed line) and at smaller rotation $N_n=N-4$ (grey dotted line).
    (b) Highest vibrational excitation $v_{\rm max}$ in the potential $V_{\gamma N}(R)$ for a$^4\Sigma_u^+$.
    (c) Dissociation rates $k_f$ for levels in the a$^4\Sigma_u^+$ state with $\Delta v$ as indicated.
    Large symbols: energy levels below $E_{\gamma_n\, (N-4)\,v=0}$ with correspondingly small AD rates; 
    small symbols: energy levels above $E_{\gamma_n\, (N-4)\,v=0}$ with faster AD (see Sec.\ \ref{sssec:quartetsigmau}).
    Straight interconnecting lines in (b) and (c) are to guide the eye.}
        \label{fig:kd_quartet}
    \end{figure}

Tunneling dissociation rates of strongly rotating levels in the doublet states X$^2\Sigma_g^+$ and A$^2\Pi_u$, calculated with Eqs.\ (\ref{eq:dissociation_p}) and (\ref{eq:dissociation}), are shown in Fig.\ \ref{fig:kd_doublet}.
The energies of the rotational barrier $V^b_{\gamma N}$ and, for very high $N$, of the lowest vibrational level, $E_{\gamma N\,v=0}$, are shown for the relevant $N$ range in Fig.\ \ref{fig:kd_doublet}(a).  
Also shown are the lowest energy levels $E_{\gamma N\,v=0}$ of neutral C$_2$ in the X$^1\Sigma_g^+$ and a$^3\Pi_u$ states.
At sufficiently low $N$, the $v=0$ level energy in the neutral a$^3\Pi_u$ state becomes lower than the rotational barrier and competing decay by AD can be expected for high-$v$ anionic levels.
Decay rates in the range of interest for our experimental case occur for the levels $N\,(v_{\rm max}{-\Delta v)}$ with $\Delta v =1\ldots3$ and are indicated in Fig.\ \ref{fig:kd_doublet}(c) and (d).
The values of $v_{\rm max}$ for the two considered anionic states are shown in Fig.\ \ref{fig:kd_doublet}(b).

Similarly, the tunneling dissociation rates of levels in the quartet state a$^4\Sigma_u^+$ are shown in Fig.\ \ref{fig:kd_quartet}.
The opening of the AD decay paths for some of these levels at lower $N$ is illustrated in Fig.\ \ref{fig:kd_quartet}(a).
Here, the dissociation rates of levels $Nv$ with $v_{\rm max}-1$ down to $v_{\rm max}-4$ turn out to be of interest for our experimental case.

The widely varying rates for tunneling dissociation (AF), extending far below $\sim \SI{e4}{\persec}$ especially for levels with vibrational quantum numbers lying  below $v_{\rm max}$ by $\Delta v\geq2$ units, are likely to explain the non-exponentially decaying fragmentation signal observed over four orders of magnitude of C$_2{}^-$ storage time ($\sim \SI{e-3}\ldots\SI{e1}{\second}$).
Suitable AF rates are estimated for highly excited levels of  C$_2{}^-$ in doublet states as well as in the lowest quartet states.
However, radiative decay and electron detachment, as discussed in Sec.\ \ref{sec:calculation_rad} and \ref{sec:the_detach}, must be included before modeling the AF signal in more detail (see Sec.\ \ref{sec:model}).

\subsection{Electron detachment}
\label{sec:the_detach} 

The electron detachment from doublet and quartet states is considered in the present study. 
We first summarize the results of our calculations of the rates of rotationally assisted AD from levels of the a$^4\Sigma_u^+$ state, presented in the companion paper \cite{PRL}.
In the further subsections we then present our calculations on AD induced by non-adiabatic coupling for the X$^2\Sigma_g^+$ and B$^2\Sigma_u^+$ states.

\subsubsection{\label{sssec:quartetsigmau}AD from the quartet levels}

Levels $\gamma Nv$ of the $\gamma={}$a$^4\Sigma_u^+$ state of C$_2{}^-$ undergo AD on the pico- to femtosecond time scale when their energy $E_{\gamma Nv}$ is higher than that of levels $\gamma_n N_n v_n$ of the neutral state $\gamma_n={}$a$^3\Pi_u$ of C$_2$ with the same $N$.
At an energy $E_{\gamma Nv}<E_{\gamma_n, N_n,v_n=0}$ they would be excluded from AD by energy conservation as long as changes of $N$ are disregarded ($N_n=N$).  
However, AD into neutral levels with lower $N$ may still be energetically open, corresponding to $E_{\gamma Nv}>E_{\gamma_n, N_n,v_n=0}$ for $N_n<N$.
As only the change in $N$ makes AD possible for these levels, such processes are referred to as rotationally assisted AD.
The calculation presented in Ref.\ \cite{PRL} shows that their rates are reduced in large steps as $N_n$ is allowed to decrease.
Thus, when channels are energetically open only if $N-4\leq N_n< N-2$ , their AD rates lie in the range $10^{7}$--$10^{9}$\,s$^{-1}$.
Quartet C$_2{}^-$ ions in such levels will decay before leaving the accelerating platform of the CSR and thus cannot be present in the stored beam.
The next step occurs when levels become energetically open only if $N-6\leq N_n< N-4$.
The AD rates then are reduced to values in the range $10^{2}$--$10^{4}$\,s$^{-1}$ which lie in the interval of sensitivity defined by the observed storage times.

\begin{figure}
    \centering
    \includegraphics[width=0.5\textwidth]{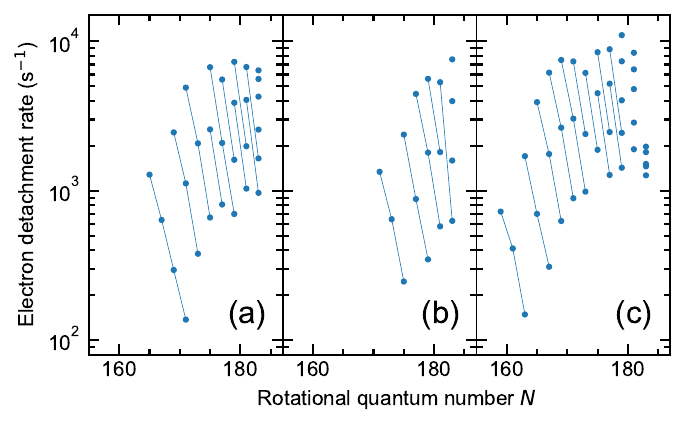}
    \caption{Rates of rotationally assisted AD calculated \cite{PRL} for levels $Nv$ of the lowest C$_2{}^-$ quartet state whose AD decay becomes energetically open only if $N-6\leq N_n< N-4$.
    The lines connect the results for levels with the same $v$.
    Results are shown for three cases within the uncertainty of the a$^4\Sigma_u^+$ potential energy curve.
    (a) Potential Q0,  lines (left to right) for $v=0$--9. 
    (b) Potential Qp, lines for $v=0$--4. (c) Potential Qm, lines for $v=0$--10.
    The $v$ quantum numbers of the highest points not identified by connecting lines follow from the Tables in the Supplemental Material \cite{suppl}.
    \label{fig:AD_rates_quartet}}
\end{figure}

The calculations \cite{PRL} apply a parametrized a$^4\Sigma_u^+$ potential curve.
As summarized in Sec.\ \ref{sec:the_potentials}, this parametrization is for internuclear distances $R\gtrsim \SI{1.6}{\AA}$ based on the present ab-initio results.
The ab-initio calculations have an estimated inaccuracy of $\pm\SI{0.17}{\electronvolt}$ at $R=\SI{2.64}{\AA}$ ($=5a_B$ with the Bohr radius $a_B$). To account for this uncertainty, two modified parametrized potentials were also obtained which reproduce the shift corresponding to the given upper and lower deviations.
The curves based on the original ab-initio result and on the up- and down-shifted modifications are denoted as Q0, Qp, and Qm potentials, respectively.

For levels whose decay is energetically open when $N-6\leq N_n< N-4$, the calculated rates of rotationally assisted AD are shown in Fig.\ \ref{fig:AD_rates_quartet}.
The sets of AD rates shown in the three panels are all considered to be allowed within the accuracy of the ab-initio bound potential calculation.
Results for all three sets will be used in the model calculations.

\subsubsection{AD from the doublet levels\label{sssec:doubletsigmau}
\label{sssec:doubletsigmag}}

\paragraph{Method.}
The doublet states examined here are X$^2\Sigma_g^+$ and B$^2\Sigma_u^+$.
The autodetachment lifetimes from B$^2\Sigma_u^+$ were determined experimentally \cite{hefter_ultrahigh_1983} to be of the order of nanoseconds and hence they are not relevant for the present experiment.
However, this initial state is included as a benchmark that allows us to estimate the plausibility and accuracy of the methods employed.

The procedure used here closely follows that described in Ref.~\cite{Douguet_Slava_JCP_2015}. 
The probability rate $P$ of a transition from an initial vibronic state $\ket{i}$ of C$_2{}^-$ to a final state $\ket{f}$ of the $e^- + \mathrm{C}_2$ system, is given by Fermi’s golden rule \cite{hefter_ultrahigh_1983,Douguet_Slava_JCP_2015,Berry_JCP_1966,Jasik_Franz_AD_Ag_2021}
\begin{equation}
\label{eq-Fermgold}
P = 2 \pi \left| \bra{f} \Lambda \ket{i} \right|^2 \rho_f \;,
\end{equation}
where $\Lambda$ is the operator of the non-adiabatic coupling and $\rho_f$ the density of final states. 
The operator $\Lambda$ represents all the terms neglected in the
the Born--Oppenheimer approximation.
It consists of the electronic--vibrational and the electronic--rotational non-adiabatic couplings. 
As in the above mentioned studies, the first-order electronic--vibrational term
\begin{equation}
\label{eq-vibcoupl}
\Lambda = \frac{1}{\mu} \left. \frac{\partial}{\partial R} \right|_\mathrm{ele} 
\left. \frac{\partial}{\partial R} \right|_\mathrm{vib}
\end{equation}
is considered to be dominant. 
The second-order term is neglected.
As a result, the matrix elements $\Lambda_{fi}$ between the initial and final states are obtained by a non-adiabatic Franck--Condon average of the electronic non-adiabatic couplings
\begin{equation}
\label{eq-NAFC}
\Lambda_{fi} = \frac{1}{\mu} \bra{ \chi_f(R) }
\Lambda_{fi}(R,\varepsilon) \frac{\partial}{\partial R} \ket{\chi_i(R)}_R \;,
\end{equation}
where the electronic non-adiabatic couplings can be expressed as
\begin{equation}
\label{eq-NAel}
\Lambda_{fi}(R,\varepsilon) = \bra{\psi_f(\vec{r};R)} \frac{\partial}{\partial R}
\ket{\psi_i(\vec{r};R)}_{\vec{r}} \;.
\end{equation}
Symbols $\ket{\psi_i(\vec{r};R)}$ and $\ket{\psi_f(\vec{r};R)}$ represent the initial and final electronic Born--Oppenheimer states, respectively.
Similarly, $\ket{\chi_i(R)}$ and $\ket{\chi_f(R)}$ are the initial and final vibrational states. 
The scalar products $\braket{\cdot}_R$ and $\braket{\cdot}_{\vec{r}}$ are carried out on the nuclear and electronic spaces, respectively.
Note that $\psi_f(\vec{r};R)$ is a continuous state labeled by the symbol set $f \equiv \{\Gamma\gamma l\}$.
The asymptotic kinetic energy of the ejected electron is denoted as $\varepsilon$ and only a linear energy dependence for the l.h.s.\ of Eq.~(\ref{eq-NAel}) will be assumed here.
In this Subsection,  $\Gamma$ stands for the final electronic state of the neutral molecule, $\gamma$ for the symmetry of the ejected electron,  and $l$ for the angular momentum of the ejected electron in the symmetry $\gamma$.
The reduced mass (in atomic units) is denoted as $\mu$.
Eqs.\ (\ref{eq-Fermgold})--(\ref{eq-MorseSpec}) use atomic units.

\paragraph{\label{sssec:potcurves}Potential curves.}
Born--Oppenheimer potential energy curves (PECs) of the relevant electronic states are required to compute the vibrational states $\chi_i(R)$ and $\chi_f(R)$ in Eq.~(\ref{eq-NAel}). 
Furthermore, corresponding vibronic energies of the initial and final states define the asymptotic kinetic energy $\varepsilon$ of the released electron.
We have observed that the autodetachment rates are very sensitive to the quality of the employed PECs.
Especially, the relative position of neutral and anionic curves, defined by the vertical electron affinity, plays an important role. 
While PECs are available from the ab-initio calculations (Sec.\ \ref{sec:the_potentials}), such calculations cannot reproduce the experimental electron affinity of C$_2$ \cite{Ervin_Lineberger_JPC_1991} with sufficient accuracy.
Therefore, in all calculations of non-adiabatic electronic--vibrational AD, we adjust the relative energy between the neutral and anionic PECs to the electron affinity of C$_2$ (3.269~eV \cite{Ervin_Lineberger_JPC_1991}) applying a suitable overall energy shift to the curves.
Moreover, we also implement spectroscopically parametrized PECs, defined as Morse potentials
\begin{equation}
\label{eq-Morse}
V(R) = D_e\left[ e^{-2a(R-R_e)} - 2 e^{-a(R-R_e)} \right].
\end{equation}
The parameters of these ``spectroscopic PECs'' are
\begin{eqnarray}
\nonumber
a &=& \sqrt{2 \mu\; \omega_e x_e} \;, \\
D_e &=& \left( \frac{\omega_e}{a}\right)^2 \frac{\mu}{2}.
\label{eq-MorseSpec}
\end{eqnarray}
Literature values \cite{Ervin_Lineberger_JPC_1991} are used for the equilibrium vibrational frequency 
and the anharmonicity where $\omega_e$ and $\omega_e x_e$ here represent the corresponding energy values converted to atomic units.
When calculating the B$^2\Sigma_u^+$ AD rates (for a test of the method), we use the spectroscopic PECs for the anionic as well as for the X$^1\Sigma_g^+$ and a$^3\Pi_u^+$ neutral states.
When producing the AD rates for the X$^2\Sigma_g^+$ levels, we use for the anionic state always the ab-initio result (Sec.\ \ref{sec:the_potentials}).
Our motivation for this choice is that the relevant levels in the  X$^2\Sigma_g^+$ state are highly excited and therefore probe parts of the PEC which are less verified by the spectroscopic data.
For the neutral states, however, we consider either the spectroscopic or the ab-initio PECs and give results for both cases.
The comparison between these cases illustrates the sensitivity of the results on the detailed shape of the PECs, which enters via the vibrational averaging and the ejected electron energy $\varepsilon$.

\paragraph{Electronic non-adiabatic coupling curves.}

The electronic non-adiabatic coupling functions $\Lambda_{fi}(R,\varepsilon)$ were computed by using Eq.~(\ref{eq-NAel}) and the quantum chemistry software {\sc molpro} 2012 \cite{MOLPRO12}. 
The MRCI method was employed with the aug-cc-pVTZ basis set \cite{Dunning_ccpVXZ}. 
The technique that allows to represent the electronic continuum was essentially taken from the Ref.~\cite{Douguet_Slava_JCP_2015}.
The final continuum states were represented as a set of bound discrete states. 
This discretized continuum was generated by a set of five additional diffused functions that were added to the original aug-cc-pVTZ set. 
These were either the $s$-type or $p$-type Gaussians depending which symmetry of the continuum was in the question as summarized in Table \ref{tab-syms}.
\begin{table}[t]
\caption{\label{tab-syms}
Symmetries of the initial anion and the final neutral states studied here. 
The symmetry of the continuum electron and the lowest partial-wave contribution are shown in the third and fourth columns.}
\centering
\begin{ruledtabular}
\begin{tabular}{@{\quad}cccc@{\quad}}

Anion & Neutral & Electron & Lowest $l$ \\
\hline
\multirow{2}{*}{$^2\Sigma_u^+$} & $^1\Sigma_g^+$ & $^2\Sigma_u^+$ & $p$ \\
                                & $^3\Pi_u$ & $^2\Pi_g$ & $d$ \\
\hline
\multirow{2}{*}{$^2\Sigma_g^+$} & $^1\Sigma_g^+$ & $^2\Sigma_g^+$ & $s$ \\
                                & $^3\Pi_u$ & $^2\Pi_u$ & $p$ \\
\end{tabular}
\end{ruledtabular}
\end{table}
\begin{figure}[b]
\begin{center}
\includegraphics[width=0.49\textwidth]{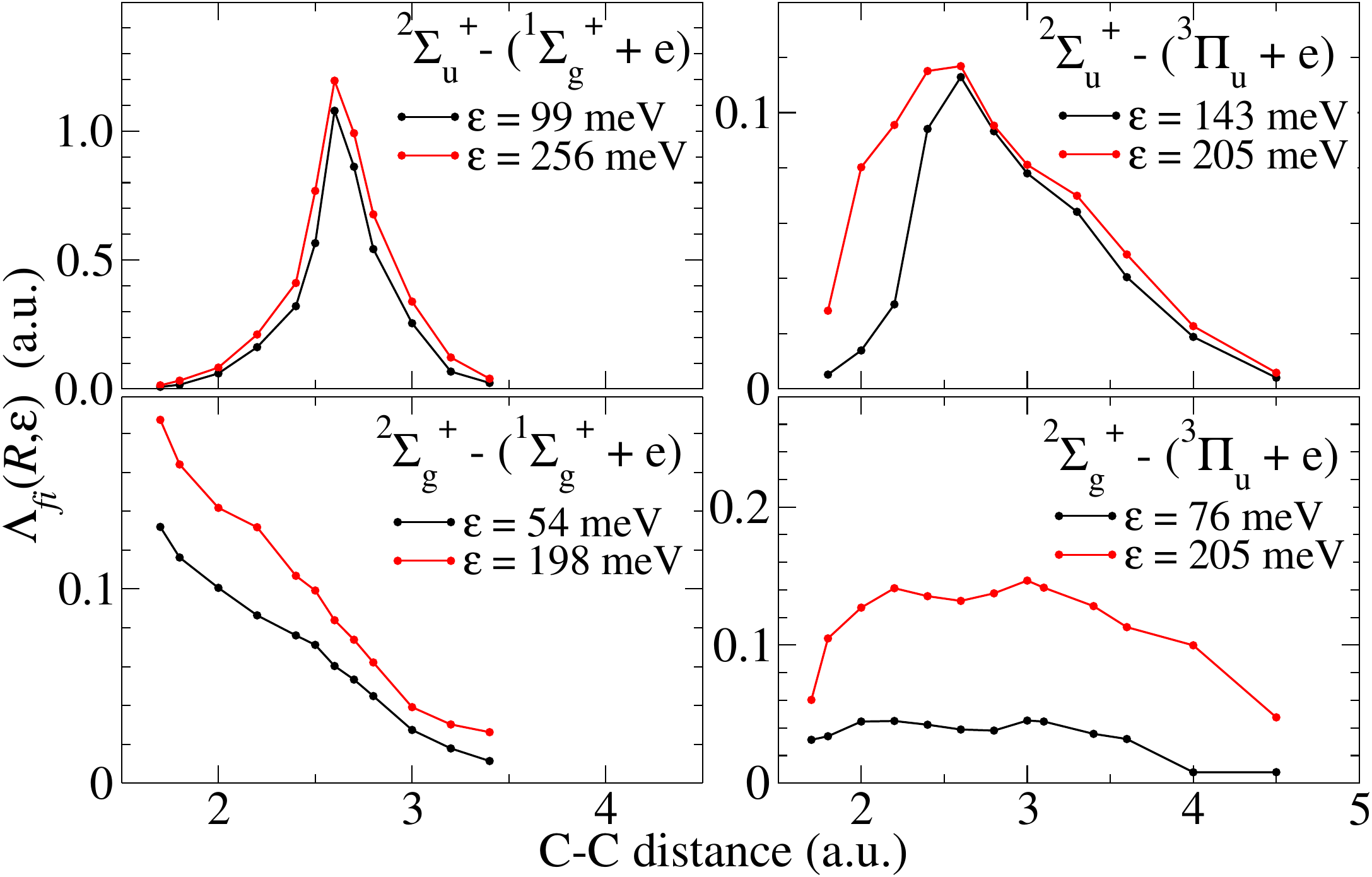}
\caption{\label{fig-nace}
The matrix elements $\Lambda_{fi}(R,\varepsilon)$ of the non-adiabatic coupling operator $\Lambda$ computed for the two different discretized continuum states with two different continuum energies $\varepsilon$ and two different initial states (using atomic units, a.u.).
}
\end{center}
\end{figure}

The coupling elements, Eq.\ (\ref{eq-NAel}), resulting from the {\sc molpro} software \cite{MOLPRO12} are volume normalized. 
Therefore, their value depends significantly on the extent of the state representing the discretized continuum, since very diffused continuum states will offer less density in the non-adiabatic overlap as described by Eq.\ (\ref{eq-NAel}). 
In order to check the stability of the results with respect to the basis, the discretized continuum state was renormalized by $(\Delta \varepsilon)^{-1/2}$ to achieve the energy normalization of the final state in Eq.~(\ref{eq-NAel}) \cite{Douguet_Slava_JCP_2015}.
Incidentally, such a renormalization of the continuum states leads to $\rho_f=1$ in the expression of Fermi's golden rule, Eq.\ (\ref{eq-Fermgold}).

All the four electronic non-adiabatic coupling elements necessary for the studied combinations of initial and final states (listed in Table \ref{tab-syms}) are shown in Fig.~\ref{fig-nace}. 
It can be seen that the electronic non-adiabatic couplings are fairly weak around the Franck--Condon region of the final vibrational state, except for the case of an initial $^2\Sigma_u^+$ and a final $^1\Sigma_g^+ + e(l=1)$ state. 
These two states undergo an avoided crossing at $R=2.6\,a_B$ with a strong coupling peak shown in the top-left panel of Fig.~\ref{fig-nace}.

\paragraph{Results for initial anion B$^2\Sigma_u^+$.}

Computed autodetachment rates from the initial B$^2\Sigma_u^+$ state from vibrational levels $v'$ are summarized in Table \ref{tab-ratesSu}. 
These results contradict the mechanism for the sudden rate jump for $v' > 5$ proposed in Refs.~\cite{hefter_ultrahigh_1983,mead}. 
The suggested mechanism \cite{hefter_ultrahigh_1983,mead} is an opening of the new final state a$^3\Pi_u$ for $v' > 5$ that strongly couples to the initial B$^2\Sigma_u^+$ state.
However, the electronic non-adiabatic coupling to this state, shown in Fig.~\ref{fig-nace}(top-right panel) is weak, probably due to the $d$-wave nature of the ejected electron (see Table \ref{tab-syms}). 
Present calculations show that the steep increase of the AD rates for $v'>5$ happens already for the final X$^1\Sigma_g^+$ state. 
Furthermore, in agreement with these computational findings, the proposed necessity of the new a$^3\Pi_u$ states, to facilitate the increased rates, was later disproven experimentally \cite{Ervin_Lineberger_JPC_1991}.

\begin{table}[t]
\caption{\label{tab-ratesSu} Computed rates from the initial B$^2\Sigma_u^+$ state and the two final neutral states. The rates are given in units of s$^{-1}$ and they are compared with the spectroscopic data of \citet{hefter_ultrahigh_1983}}
\centering
\begin{ruledtabular}
\begin{tabular}{@{\quad}r@{\quad\quad}lll@{\quad}}
$v'$ & final X$^1\Sigma_g^+$ \quad\quad\quad & final a$^3\Pi_u$ \quad\quad\quad & \citet{hefter_ultrahigh_1983} \\
\hline
5 & 2.1 $\times 10^6$ & 2.2 $\times 10^6$ & $\leq$ 1.0 $\times 10^7$ \\
6 & 4.3 $\times 10^8$ & 8.8 $\times 10^7$ & \quad\,3.8 $\times 10^8$ \\
7 & 4.1 $\times 10^9$ & 2.1 $\times 10^8$ & \quad\,1.3 $\times 10^9$ \\
8 & 8.4 $\times 10^9$ & 3.2 $\times 10^8$ & \quad\,3.2 $\times 10^9$ \\
9 & 1.0 $\times 10^{10}$ & 9.8 $\times 10^8$ & \quad\,8.7 $\times 10^9$ \\
10 & 3.3 $\times 10^{10}$ & 1.0 $\times 10^9$ & $\geq$ 3.0 $\times 10^{10}$ \\
\end{tabular}
\end{ruledtabular}
\end{table}

While the quantitative agreement between the present calculations and the experimental data of \citet{hefter_ultrahigh_1983} is not perfect, the order of magnitude and the increase with $v'$ agrees with the experiment. 
These results also set a level of confidence that can be expected in the following calculations for which experimental data are not available.

\paragraph{Results for initial anion X$^2\Sigma_g^+$.}

\begin{figure}[t]
    \centering
    \includegraphics[width=0.45\textwidth]{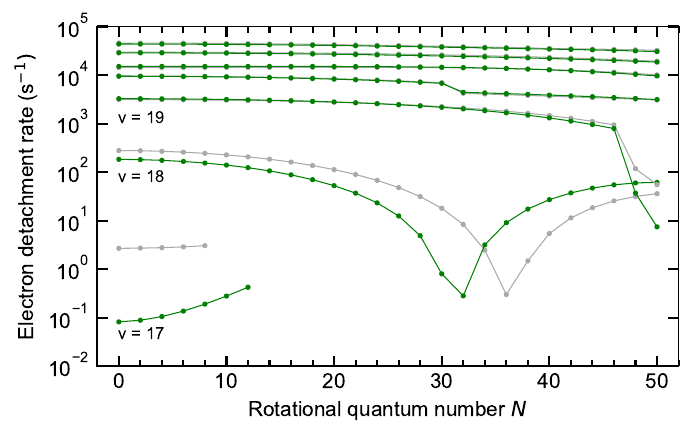}
    \caption{Calculated total AD rates for levels in X$^2\Sigma_g^+$ versus the angular momentum $N=N'$ of the initial level. 
    The rates from levels of given $v$ (as indicated) are connected with a thin line.
    The unlabeled lines are for continuously increasing $v$ up to $v=23$.
    Rates using the spectroscopic and the ab-initio potentials for the neutral states (see the text) are drawn in green and grey, respectively.}
    \label{fig:AD_rates_doublet}
\end{figure}

Computed autodetachment rates from the initial X$^2\Sigma_g^+$ state are summarized in Fig.\ \ref{fig:AD_rates_doublet}.
The AD process is found to dominantly lead into the  X$^1\Sigma_g^+$ neutral state, with the rates into 
a$^3\Pi_u$ smaller by typically one to two orders of magnitude.
These rates are much smaller than those for the initial B$^2\Sigma_u^+$ state for two reasons: 
\begin{itemize}
\item[(i)] Electronic non-adiabatic couplings shown in the lower panels of Fig.~\ref{fig-nace} are an order of magnitude smaller than the dominant $^2\Sigma_u^+$ -- $(^1\Sigma_g^+ + e)$ coupling.
\item[(ii)] Initial states with $v' \geq 17$ possess less favorable non-adiabatic Franck--Condon overlap (\ref{eq-NAFC}) when compared with states with $v' \geq 5$ discussed in the previous section.
\end{itemize}

The rotational dependence of the computed AD rates was obtained by an addition of the centrifugal term to the PECs described in Sec.~\ref{sssec:potcurves}.
Such modification changes the shapes of vibrational wave functions as well as the energy levels. 
However, the direct non-adiabatic term that couples molecular rotations with the electronic angular momenta \cite{hefter_ultrahigh_1983} is not considered here. 
This rotational non-adiabatic coupling term becomes stronger for higher rotational levels \cite{hefter_ultrahigh_1983} and, by providing another AD pathway, it would further increase the AD rates for the higher $N'$. 
At present we report only the contribution generated by the vibrational non-adiabatic coupling term of Eq.\ (\ref{eq-vibcoupl}), leading to the dependence of the AD rates on the rotational quantum number $N$ of the initial state in Fig. \ref{fig:AD_rates_doublet}.

As discussed in Sec.\ \ref{sssec:potcurves}, the  X$^2\Sigma_g^+$ AD rates were obtained  for two choices of the neutral-state PECs.
As shown in Fig.\ \ref{fig:AD_rates_doublet}, deviations between the two cases are found mainly for the $v=17$ and 18 initial levels.
In the results below, AD decay signals obtained with both data sets are included to roughly characterize the uncertainty of the model.

\section{Model results and comparison to experiment}\label{sec:model}

\subsection{Level-specific signal yields for AF and AD}

\begin{figure}[b]
    \centering \includegraphics[width=0.49\textwidth]{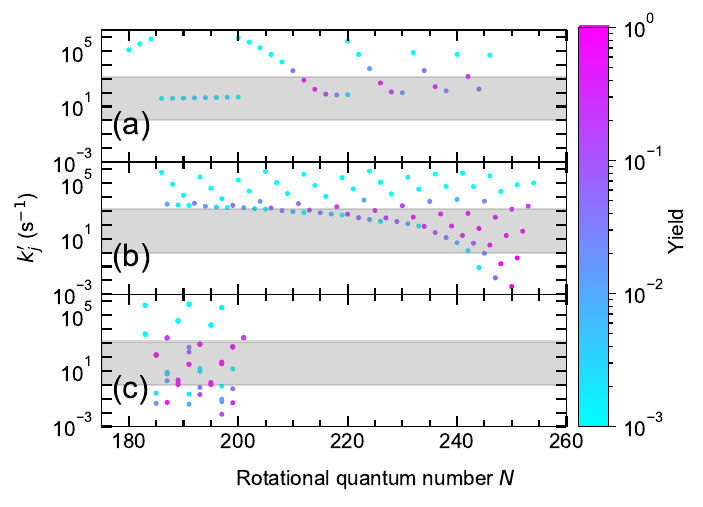}
    \caption{Total decay rates $k'_j$ of levels relevant for the measured AF signal in electronic state (a) X$^2\Sigma_g^+$, (b) A$^2\Pi_u$, and (c) a$^4\Sigma_u^+$ versus $N$. The points are color-coded according to the relative contribution to the AF channel $Y_{AF,j}$ (Eq. (\ref{eq:yield})). The grey shaded area marks the region of decay rates which are visible in the measured experiment.}
    \label{fig:yield_AF}
\end{figure}

With the various decay rates derived in Sec.\ \ref{sec:the_rates} the total decay rate $k_j'$ and the signal yields $Y_{x,j}$ can be calculated for the excited levels $j$ and the processes $x={}$AD and AF, using Eqs.\ (\ref{eq:total_decay_rate}) and  (\ref{eq:yield}), respectively.
The relative yield determines the importance of a level for the observed decay signals, while the total decay rate describes the temporal dependence of the level contribution.
The window of total decay rates relevant for the measurement time window (see Sec.\ \ref{sec:exp_signals}) is considered to be $1$\,s$^{-1} \leq k_j' \leq 10^3$\,s$^{-1}$.

For the AF signal, the level yields approach the limit of 1 for a number of levels $\gamma Nv$ with high rotation $N$.
The decay rates $k_j'$ and the yields $Y_{{\rm AF},j}$ are mapped in Fig.\ \ref{fig:yield_AF}.
For $\gamma={}\text{X}$$^2\Sigma_g^+$ in Fig.\ \ref{fig:yield_AF}(a), the dots reflect the tunneling fragmentation rates $k_f$ in Fig.\ \ref{fig:kd_doublet}(c) as well as the radiative decay rates of high-$N$ 
X$^2\Sigma_g^+$ levels in Fig.\ \ref{fig:atotal_AX}.
Substantial yields ($>0.1$) arise at $N\geq210$ for tunneling AF of the $v=v_{\rm max}(N)-2$ levels.
When the AF rates become small, radiative decay causes a minimal value of the signal decay rate with a rather flat $N$ dependence and also limits the yield.
Tunneling AF from $v=v_{\rm max}(N)-1$ levels becomes too fast to be observed with good experimental yield as $t_1=0.8$\,ms.

Likewise, the dots in Fig.\ \ref{fig:yield_AF}(b) for $\gamma={}$A$^2\Pi_u$ reflect the tunneling AF rates in Fig.\ \ref{fig:kd_doublet}(d) and the A$^2\Pi_u$ radiative decay rates in Fig.\ \ref{fig:atotal_AX}.
The radiatative decay rates become similar to those of X$^2\Sigma_g^+$ levels at high $N$.
The levels with high AF yield mostly have $v=v_{\rm max}(N)-2$.
The increase of the  A$^2\Pi_u$ radiative decay rates toward lower $N$ is reflected in the minimal values of the decay time.
For this electronic state, the total decay rates scatter over several orders of magnitude.
Since the A$^2\Pi_u$ radiative decay rates become very small at the highest $N$, the signal yield remains high also for levels with $v_{\rm max}(N)-3$ or even lower, so that very slowly decaying AF signals become observable.

Finally, Fig.\ \ref{fig:yield_AF}(c) shows the total decay rates and the AF yields for the quasi-stable quartet levels, $\gamma={}$a$^4\Sigma_u^+$, at $N\gtrsim185$.
These results reveal strong competition between the radiative decay rates (Fig.\ \ref{fig:rdec_quartet}) and those of tunneling AF in Fig.\ \ref{fig:kd_quartet}(c).
The signal decay rates from levels with a high yield scatter over several orders of magnitude.

\begin{figure}[b]
    \centering
    \includegraphics[width=0.49\textwidth]{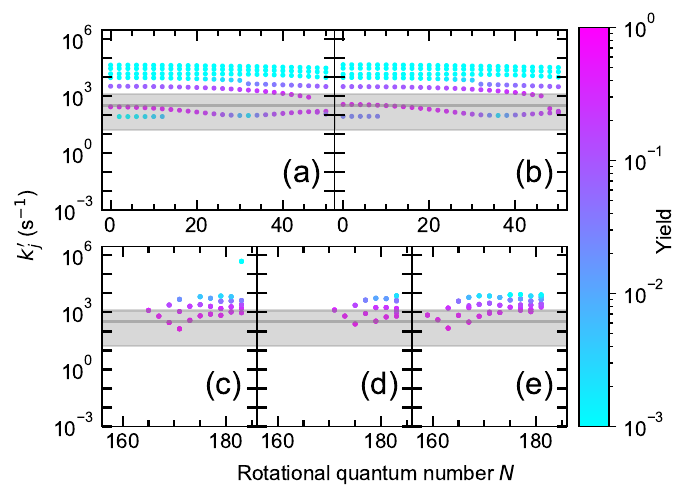}
    \caption{Total decay rates $k'_j$ of levels relevant for the measured AD signal versus $N$. Panel (a) and (b) show levels in electronic state X$^2\Sigma_g^+$ using the spectroscopic and ab-initio neutral potentials, respectively. Panels (c), (d), and (e) correspond to levels in a$^4\Sigma_u^+$, where the unshifted, upwards shifted and downwards shifted potentials are used, respectively. The points are color-coded according to the relative contribution to the AD channel $Y_{AD,j}$ [Eq.\ (\ref{eq:yield})]. The horizontal grey bar marks the main decay component observed in the experiment of $\sim 300\,$s$^{-1}$.}
    \label{fig:yield_AD}
\end{figure}

For the AD signal, the total decay rates and the yields $Y_{{\rm AD},j}$ are mapped in Fig.\ \ref{fig:yield_AD}.
The AD from levels $Nv$ of the X$^2\Sigma_g^+$ state with high $v$ and moderate $N$ is considered in Fig.\ \ref{fig:yield_AD}(a)--(b).
The dots reflect the AD rates from Fig.\ \ref{fig:AD_rates_doublet} and the radiative decay rates from Fig.\ \ref{fig:atotal_AX}.
For the high-yield contributions, the signal decay rates cluster in ranges corresponding to the AD rates of $v=18$ and 19.
AD decay of X$^2\Sigma_g^+$ levels with $v\geq20$   becomes too fast to be observed with good experimental yield at the given starting time $t_1$. Similar rate distributions are obtained for the two sets which use the spectroscopic and ab-initio neutral potential curves.

The rates of rotationally assisted AD from a$^4\Sigma_u^+$ levels with $160\lesssim N\lesssim185$ (Fig.\ \ref{fig:AD_rates_quartet}) result in the total decay rates and yields given in Fig.\ \ref{fig:yield_AD}(c)--(e).
Compared to the AD rates, the intercombination radiative decay rates (Fig.\ \ref{fig:rdec_quartet}) are small for most of these levels.
Correspondingly, AD signal yields are high and the contributions of the vibrational levels, starting with $v=0$ for the smallest decay rates, can easily be retrieved using Fig.\ \ref{fig:AD_rates_quartet}.
Somewhat different distributions of the decay rates result for the three cases within the uncertainty range of the a$^4\Sigma_u^+$ potential curve (see Sec.\ \ref{sssec:quartetsigmau}).

These results show that, at small $N$, levels  $v=18,19$ of the X$^2\Sigma_g^+$ state may cause an AD signal with decay rates in the experimentally relevant window.
The small AD rates of these $v$ levels just above the detachment threshold can be observed because the radiative decay rates of higher-$v$, X$^2\Sigma_g^+$ levels are relatively small.
Much faster radiative decay is expected for the corresponding A$^2\Pi_u$ levels.
Hence, AD signals with decay rates in the experimental window are not expected from the 
A$^2\Pi_u$ state.
The AD or radiative decay rates of B$^2\Sigma_u^+$ levels will be of similar size or even higher ($\gtrsim10^6$\,s$^{-1}$) so that also for this state signals with decay rates in the experimental window are not expected.

At higher $N$, AD signals decaying within the experimental window can be expected from low-$v$ levels of the a$^4\Sigma_u^+$ state.
Levels of the a$^4\Sigma_u^+$ state are expected to contribute also to AF as $N$ increases further.
Finally, at even higher $N$, an AF contribution may also arise from the X$^2\Sigma_g^+$ and A$^2\Pi_u$ states.
Here, the number of levels contributing with a high yield is larger for the  A$^2\Pi_u$ state.
In this $N$-range, X$^2\Sigma_g^+$ levels have higher radiative decay rates than A$^2\Pi_u$ levels.
Hence, a small amount of radiative re-population occurs from X$^2\Sigma_g^+$ levels with slow AF into A$^2\Pi_u$ levels.

\subsection{Rate model}\label{sec:comparison_data_rate_model}

Taking into account a single radiative cascade step from higher levels, the time dependent population of level $j$ can be expressed as
\begin{equation} 
    \label{eq:population_feeding}
    P_j(t) = \left[P_j + \sum_i
    P_i A_i^j \frac{
    1-e^{-(k'_i-k'_j)t}}
    {k'_i-k'_j}\right]
    \,e^{-k'_j t}.
\end{equation}
As represented by the second term in the brackets, re-population contributions can be included for higher levels $i$, where $A_i^j$ is the radiative decay rate from $i$ into $j$. 
All these contributions are positive-valued and add a further time dependence. $P_j$ and $P_i$ are the populations at $t=0$.

Re-population contributions are taken into account in the AF signals, allowing for transitions between high-$N$ X$^2\Sigma_g^+$ and A$^2\Pi_u$ levels.
We neglect re-population for all other cases.
Regarding the AD and AF signals from the a$^4\Sigma_u^+$ levels, radiative cascades are neglected because any higher quartet levels are likely to be very short-lived, while spin-forbidden radiative transitions from X$^2\Sigma_g^+$ levels into a$^4\Sigma_u^+$ levels represent only a weak decay branch compared to decay into A$^2\Pi_u$.
Cascade population from above is also neglected for the AD signal from X$^2\Sigma_g^+$, because all higher X$^2\Sigma_g^+$ or A$^2\Pi_u$ show fast radiative or AD decay.

The corrections due to cascade effects in the AF signal observed at $t>t_1$ are found to be small relative to the remaining deviations in the comparison of the model and the experimental time dependence.
For this comparison, the parameters defining the initial populations $P_j$ of the model were chosen to reproduce the overall signal amplitudes.
While cascade processes with rates $\gg t_1^{-1}$ do not influence the temporal dependence of the signals observed at $t\geq t_1$, they still enter the relation between the strength of the observed signals [effectively given by the values $P_j(t_1)$] and the initial populations $P_j$ [see Eq.\ (\ref{eq:population_feeding})].
Such fast cascades may occur in the early high-$v$ X$^2\Sigma_g^+$ and A$^2\Pi_u$ level populations.
However, the AD decay rates of these levels are mostly higher than the radiative decay rates and we believe that remaining effects of the radiative cascade are not large enough to significantly affect our conclusions on the level populations. 

The AD and AF signals $s_d(t)$ and $s_f(t)$, respectively, are modeled by 
\begin{equation}
      s_x(t)=\epsilon_x N_i\sum_j k_{x,j} P_j(t)
\label{eq:signal_model}
\end{equation}
($x=d,f$) with $P_j(t)$ from Eq.\ (\ref{eq:population_feeding}), including cascade effects to the extent discussed.
Using all levels $\gamma Nv$ contained in the model, separate results are kept for the electronic states $\gamma$.
The values of the ion number $N_i$ and the detection efficiency $\epsilon_x$ are given in Sec.\ \ref{sec:exp_signals}.
The charged and neutral signals are then reconstructed as
\begin{eqnarray}
      s_-(t)&=&s_f(t)\nonumber\\
      s_0(t)&=&s_d(t)+\eta s_f(t)
\label{eq:signal_model_cn}
\end{eqnarray}
reversing Eq.\ (\ref{eq:signal_sf}) and (\ref{eq:signal_sd}) ($\eta=0.75$).

\subsection{\boldmath Model including quartet C$_2^-$ levels \label{sec:comp_s1}}

\begin{figure}[t]
    \centering
    \includegraphics[width=1\columnwidth]{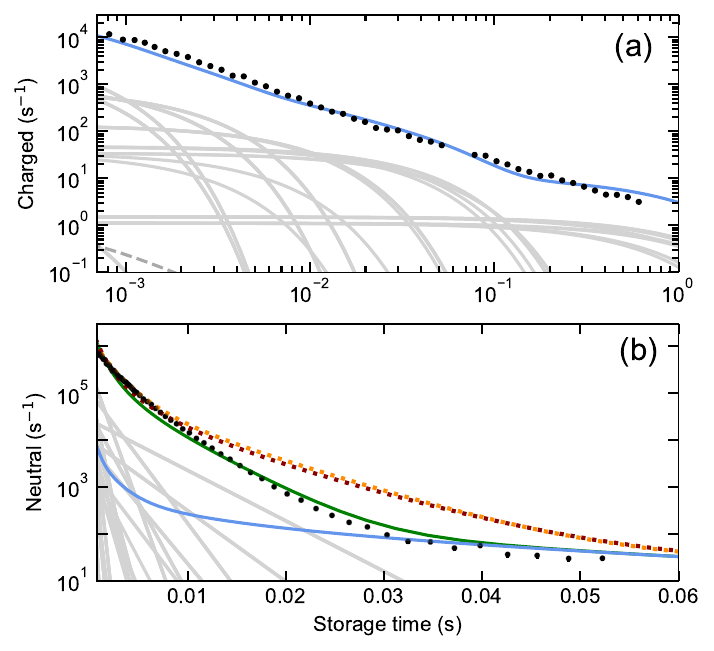}
    \caption{Experimental count rates of C$^-$ fragments (a) and neutral fragments (b) compared to the modeled signals using Scenario I.
    (a) Signal $s_-$ as measured (dots) and modeled by the total AF rate (solid blue line).
    The individual contributions (grey lines) are from levels of the a$^4\Sigma_u^+$ state (solid line style) and, in the lower left corner, the total contribution of both the X$^2\Sigma_g^+$ and A$^2\Pi_u$ state (dashed line style).
    (b) Signal $s_0$ as measured (dots) and modeled by the sum of the AD rate (mainly due to a$^4\Sigma_u^+$) and the scaled AF rate.  Solid green, dotted orange, and dotted dark-red lines correspond to AD rates using the quartet potentials Qp, Qm, and Q0, respectively.
    The scaled AF rate [Eq.\ (\ref{eq:signal_model_cn})] is shown by the solid blue line.
    Individual contributions (grey lines) are depicted for the levels in the a$^4\Sigma_u^+$ state (Qp potential).
    \label{fig:scenario_i}}
\end{figure}

In Fig.\ \ref{fig:scenario_i} the measured charged and neutral rates are compared to the model calculations using Scenario I (Table \ref{tab:pop}) which defines the level populations $P_{\gamma Nv}$ according to Eq.\ (\ref{eq:provib}).
The population of all quasi-stable quartet levels in the C$_2{}^-$ beam is set to $P_\gamma^e=6\times10^{-3}$, while the remaining population is assumed to be in  the X$^2\Sigma_g^+$ and A$^2\Pi_u$ electronic states with equal shares.
With this value the model, using the calculated rates of rotationally assisted AD, reproduces the amplitude of the neutral signal in Fig.\ \ref{fig:scenario_i}(b).
The core vibrational temperature ($T_v^c = 1200$\,K) is set according to the current evidence on the applied ion-source type (Sec.\ \ref{sec:exp_population}).
The modeled AD rate does only weakly depend on the vibrational temperature as it is caused by low-$v$ levels.
When a core rotational temperature $T_r^c=2100$\,K is chosen, the modeled AF rates from the rotationally stabilized C$_2^-$ quartet levels can explain the signal of charged fragments shown in Fig.\ \ref{fig:scenario_i}(a) as well as the long-time asymptote of the neutral fragment signal in Fig.\ \ref{fig:scenario_i}(b).

\begin{figure}[t]
    \centering
    \includegraphics[width=1\columnwidth]{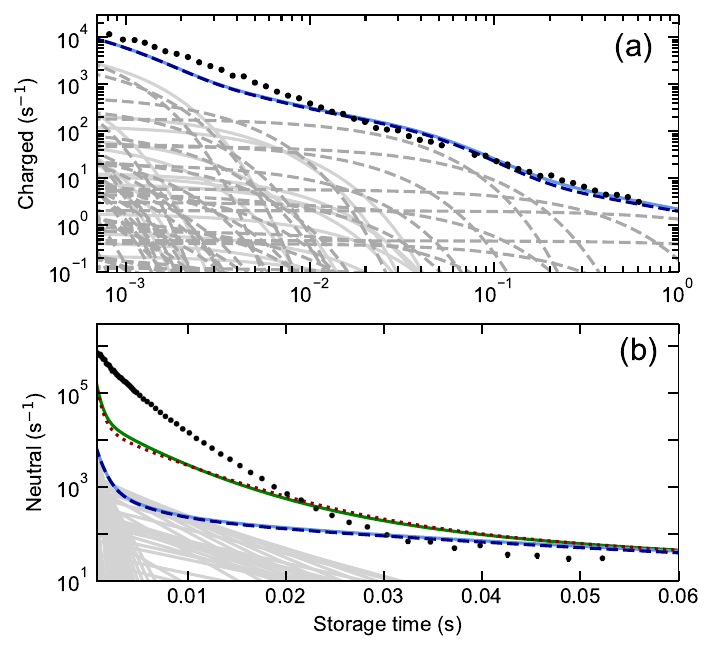}
    \caption{Experimental count rates of C$^-$ fragments (a) and neutral fragments (b) compared to the modeled signals using Scenario II.
    (a) Signal $s_-$ as measured (dots) and modeled by the total AF rate (solid blue line).
    The individual contributions (grey lines) are from levels of the  X$^2\Sigma_g^+$ state (full line style) and the A$^2\Pi_u$ state (dashed line style).
    (b) Signal $s_0$ as measured (dots) and modeled by the sum of the AD rate (due to X$^2\Sigma_g^+$) and the scaled AF rate.
    Solid green line and dotted dark-red line correspond to AD rates using the neutral potentials in the spectroscopic parametrization or from the ab-initio calculation, respectively.
    The scaled AF rate [Eq.\ (\ref{eq:signal_model_cn})] is shown by the solid blue line.
    Individual contributions (grey lines) are depicted for the levels in the X$^2\Sigma_g^+$ state (spectroscopic potential).
    In both frames, lines representing the  total AF signal are also shown without any radiative cascade contributions (dashed, dark blue).
    \label{fig:scenario_ii}}
\end{figure}

The AD and AF signals from the doublet levels of C$_2{}^-$ are contained in the presented model results.
For these electronic states, AD requires high vibrational and AF high rotational as well as vibrational excitation. 
In order to favor these doublet-C$_2{}^-$ contributions in the model, a thermal tail is added to the distributions in both degrees of freedom; however, the temperatures and the tail fractions (Table \ref{tab:pop}) are kept in a range consistent with the ion-source knowledge (Sec.\ \ref{sec:exp_population}).
In spite of this effort, the modeled AF and AD signals remain extremely small and mostly too weak to be visible in 
Fig.\ \ref{fig:scenario_i}.

\subsection{\boldmath Model including doublet C$_2^-$ only \label{sec:comp_s2}}

Scenario II defined in Sec.\ \ref{sec:the_vibration}, Table \ref{tab:pop}, sets the X$^2\Sigma_g^+$ and A$^2\Pi_u$ electronic states to equally share all population between them, while quartet levels of C$_2{}^-$ are not allowed to be populated. 
The high excitation required for the doublet AF and AD signals to occur at the observed count rates is made possible by increasing the temperatures of the tails in the  rotational and vibrational distributions.
For the neutral signal, the comparison in Fig.\ \ref{fig:scenario_ii}(b) shows that a vibrational tail temperature $T^t_v=30000$\,K, at a tail fraction of 0.05, brings the modeled AD signal to the observed intensity.
The modeled charged signal in Fig.\ \ref{fig:scenario_ii}(b) can be brought in agreement with experiment when a rotational tail temperature $T^t_r=30000$\,K is set and in addition the tail fraction for this degree of freedom is increased to 0.035.

\subsection{Discussion}

As shown in Sec.\ \ref{sec:comp_s1}, the level populations of Scenario I yield a reasonable agreement of the modeled AD and AF signals with experiment.
Assumptions that must be made about the rotational and vibrational excitation in the applied sputter ion source remain in rough agreement with other evidence about this ion source type (Sec.\ \ref{sec:exp_population}).
However, the Scenario requires that C$_2{}^-$ ions in the lowest quartet electronic state, which become quasi-stable for high rotation, accumulate in the ion source and reach a fraction of $\sim6\times 10^{-3}$.
The production of these anions probably involves neutral precursors.
Larger compounds containing more than two C atoms may create C$_2{}^-$ in dissociative electron attachment, and these processes may well lead to electronic as well as vibrational and rotational excitation of the fragments.
Another effect that may favor the production of quasi-stable C$_2{}^-$ ions is that the total energies of a$^3\Pi_u$ levels in C$_2$ and  a$^4\Sigma_u^+$ levels in C$_2{}^-$ become nearly equal at rotational energies of $\sim$\,3\,eV, as visible in Fig.\ \ref{fig:vib_lev_pop}(a) and (b).
Hence, low-energy electrons can be resonantly (and therefore possibly quite efficiently) captured on a$^3\Pi_u$ C$_2$ molecules to form a$^4\Sigma_u^+$ C$_2{}^-$.
While details of possible formation processes must remain speculative, their relevance is suggested by the good agreement with experiments reached by the model results for Scenario I.

Since the absolute count rates of the experiment quite directly reflect the population of the relevant excited states (see Sec.\ \ref{sec:exp_signals}), the assumption of very high ion source temperatures for Scenario II can hardly be avoided.
However, such assumption can also hardly be reconciled with available knowledge on the properties of sputter ion sources (Sec.\ \ref{sec:exp_population}).

Instead, the comparison with the model strongly supports the significance of C$_2{}^-$ ions in the a$^4\Sigma_u^+$ electronic state in the storage ring experiments and the production of such ions in the applied sputter ion source.
As mentioned in Sec.\ \ref{sec:intro}, the possible role of quasi-stable quartet C$_2{}^-$ ions was already discussed in connection with recent spectroscopic measurements on a stored  C$_2{}^-$ beam \cite{iizawa_photodetachment_2022}, produced from the same type of ion source.
Many optical transitions observed in this experiment could not be assigned to X$^2\Sigma_g^+$ and A$^2\Pi_u$ level energies.
Moreover, the effective lifetimes of the optical line signals (Fig.\ 9 of Ref.\ \cite{iizawa_photodetachment_2022}) were found to show a large scatter not unlike the behavior of the spin-forbidden radiative decay rates in Fig.\ \ref{fig:rdec_quartet} in Sec.\ \ref{sec:calculation_rad_quartet}.

\section{Conclusion}\label{sec:conclusion}

A detailed study of highly excited rotational and vibrational levels in the electronic states of C$_2{}^-$ allowed us to model the time dependence and intensity of fragment signals occurring in stored beams of this anion, as observed in previous experiments  \cite{andersen_carbon_1997,pedersen_experimental_1998,iizawa_photodetachment_2022,unimol_remark} and also reported here.
In particular, quasi-stable ions exist in the lowest quartet state at high rotation.
Calculated rates of rotationally assisted electron autodetachment \cite{PRL} showed that some of these quartet C$_2{}^-$ ions decay toward neutral C$_2$ at rates consistent with the experimental signal decay rates. 
The near-exponential time dependence of the experimental autodetachment signal can be reproduced within the precision of the model.
Considering the excited-level populations, the model also showed that the observed fragmentation signal can be explained by tunneling fragmentation from the quartet electronic state.
Hence, the results underline the key role quasi-stable quartet C$_2{}^-$ ions can play in diatomic carbon anion beams.

For future studies, spectroscopic measurements on stored C$_2{}^-$ ions, focusing on the unassigned transitions found recently \cite{iizawa_photodetachment_2022}, appear promising.
They would help to further validate the present results.
Moreover, an assignment of observed transitions to highly excited levels could be used to verify or possibly fit the C$_2{}^-$ electronic potential curves at larger internuclear distances, presently only covered by ab-initio calculations.
They might in particular also yield some spectroscopic information about the a$^4\Sigma_u^+$ state, which proved experimentally elusive so far.

\begin{acknowledgements}
This article comprises parts of the doctoral thesis of V.C.S. submitted to the Ruprecht-Karls-Universität Heidelberg, Germany. 
The work of R.\v{C}. has been supported by the Czech Science Foundation (Grant No.\ GACR 21-12598S).
Financial support by the Max Planck Society is acknowledged. The computational results presented have been in part achieved using the HPC infrastructure LEO of the University of Innsbruck.

\end{acknowledgements}
\appendix
\section{Symbol definitions and quantum numbers}
\label{app:definitions}
We use the molecular-frame coordinates $x,y,z$, where $z$ is the internuclear axis and ${}_\perp$ denotes the vector component perpendicular to it.  In addition to the conserved total angular momentum $\vec{J}$, we use the conventional angular momentum variables $\vec{L}$ and  $\vec{S}$ (total electronic orbital and spin angular momentum, respectively) and the total angular momentum excluding spin, $\vec{N}=\vec{J}-\vec{S}$.

For the rotational-electronic states we use the quantum numbers  $J$, $N$, $S$, $\Lambda$, and $\Omega$, where $J$, $N$, and $S$ are the quantum numbers for the magnitudes of the angular momenta $\vec{J}$, $\vec{N}$, and $\vec{S}$, while $\Lambda$ and $\Omega$ are quantum numbers for the $z$ components of $\vec{L}$ and $\vec{J}$, respectively.  We also use the quantum number $\Sigma=\Omega-\Lambda$ for the $z$-component of $\vec{S}$.  The states for $\Lambda=0$ ($\Sigma$ electronic states) are parity eigenstates with the total parity $(-1)^{s+N}$, where  $s=0$ for $\Sigma^+$ and $s=1$ for  $\Sigma^-$.  For states with $|\Lambda|>0$ parity eigenstates are constructed by linear combination of parity-inverted states and distinguished by a parity label $\epsilon_N=\pm1$ such that the total parity is $\epsilon_N(-1)^{N}$ \cite{watson_honllondon_2008}.  The parity label is included in the state quantum numbers and this definition remains valid for $\Sigma$ states, which have $\epsilon_N=(-1)^{s}$.  We note that the more commonly used parity label $\epsilon$, defining the total parity as $\epsilon(-1)^{J}$, follows from the symmetry properties of the eigenstates \cite{watson_honllondon_2008} as $\epsilon=\epsilon_N(-1)^{S+\text{int}(S)-(J-N)}$.  

\section{Calculation of radiative transition rates}
\label{app:rad}
The calculation of radiative transition rates for diatomic molecules was discussed frequently \cite{bernath_spectra_2005} and we refer here to the most recent critical summaries \cite{hansson_comment_2005,watson_honllondon_2008,western_pgopher_2017}.
We consider Hund's case (b) upper and lower levels 
$aN'v'J'$ and $bN''v''J''$, respectively, with a transition wavenumber $\tnu_{aN'v'}^{bN''v''}$.
The letters $a,b$ in the state labels represent a number of other state properties, in particular, for any state $\alpha=a, b$, its quantum numbers $S$, $|\Lambda|$, and $s$ and its parity label $\epsilon_N$.
In view of the quantum numbers discussed in Appendix \ref{app:definitions} the initial and final states are
\begin{eqnarray}
\ket{aN'v'J'}&=&\ket{\gamma',S'|\Lambda'| s',\epsilon'_N,N'v',J'}\nonumber\\
\ket{bN''v''J''}&=&\ket{\gamma'',S''|\Lambda''| s'',\epsilon''_N,N''v'',J''}
\end{eqnarray}
where $\gamma',\gamma''$ stand for any electronic quantum numbers.

The spontaneous transition rate $A_{aN'v'J'}^{bN''v''J''}$ represents the partial decay rate of $aN'v'J'$ into $bN''v''J''$ and is given by
\begin{eqnarray}
    A_{aN'v'J'}^{bN''v''J''}
    &=&C_0\left(\tnu_{aN'v'}^{bN''v''}\right)^3\;\frac{S_{aN'v'J'}^{bN''v''J''}}{2J'+1}
\label{eq:arate}
\end{eqnarray}
where
\begin{equation}
    C_0=\frac{64\pi^4}{(4\pi\epsilon_0)3h}=
    2.02613\times10^{-6}\;\frac{\text{cm}^3\,\text{s}^{-1}}{(ea_0)^2},
    \label{eq:c0}
\end{equation}
$S_{aN'v'J'}^{bN''v''J''}$ is the line strength, and $\tnu_{aN'v'}^{bN''v''}$ is the transition wavenumber.
We follow Refs.\ \cite{hansson_comment_2005,watson_honllondon_2008} expressing the line strength through a squared effective dipole matrix element,
\begin{equation}
    S_{aN'v'J'}^{bN''v''J''}=|\bra{aN'v'J'}\mu_{\rm eff}\ket{bN''v''J''}|^2.
\label{eq:smueff}
\end{equation} 
When the initial and final states are expanded into parity eigenstates with definite values of the projection quantum number $\Omega$, the effective dipole moment can be written as
\begin{eqnarray}
    \bra{aN'v'J'}\mu_{\rm eff}\ket{bN''v''J''} =  \sum_{q,\Omega',\Omega''}  s^{J'J''}_{\Omega''q}\nonumber \\
    \times \bra{aJ'N'v'\Omega'}\mu_q\ket{bJ''N''v''\Omega''}
     \label{eq:mueff}
\end{eqnarray}
where
\begin{equation}
    s^{J'J''}_{\Omega''q} =\sqrt{[J',J'']}(-1)^{J'-\Omega'}\jjj{J'}{1}{J''}{-\Omega'}{q}{\Omega''}.
    \label{eq:sjj}
\end{equation}
We use the notation $(2J+1)(2J'+1)\ldots = [J,J',\ldots]$.
The parity eigenstates inside the sum of Eq.\ (\ref{eq:mueff}) are 
\begin{eqnarray}
    \ket{\alpha JNv\Omega}=X\left[r^{\Lambda J S}_{N\Omega}\ket{\alpha JNv\Lambda\Omega}
    \hspace{15mm}\right. \nonumber\\ \left.+\epsilon_N(-1)^s\,
   r^{-\Lambda\,J S}_{N\Omega}\ket{\alpha JNv{\;-\Lambda\,}\Omega}\right],
   \label{eq:mueff_elem}
\end{eqnarray}
which combine the Hund's case (a) basis states $\ket{\alpha JNv\Lambda\Omega}$. In this expression $\Lambda$ is assumed to be positive and the factor $X=1/\sqrt{2(1+\delta_{\Lambda,0})}$ ensures proper treatment of the exceptional case $\Lambda=0$ \cite{watson_honllondon_2008}.
When the upper and lower levels are described in Hund's case (b), the coefficients are given as
\begin{equation}
  r^{\Lambda J S}_{N\Omega} = \sqrt{[N]}\;(-1)^{J+\Omega}\jjj{S}{N}{J}{\Omega-\Lambda}{\Lambda}{-\Omega}
   \label{eq:rcoeff}.
\end{equation}
These coefficients
represent the expansion \cite{watson_honllondon_2008}
\begin{equation}
    \ket{\alpha JNv\Lambda} = \sum_{\Omega=\Lambda-S}^{\Lambda+S} r^{\Lambda J S}_{N\Omega} \ket{\alpha JNv\Lambda\Omega}
    \label{eq:rexp}
\end{equation}
of a Hund's case (b) state in the Hund's case (a) basis.
The Hund's-case basis states are not parity eigenstates (unless $\Lambda=0$) and use signed (i.e., positive or negative) $\Lambda$.

Via Eq.\ (\ref{eq:mueff_elem}), the sum in Eq.\ (\ref{eq:mueff}) is composed of ``channels'' 
with given $q$, $\Omega''$, $\Omega' =\Omega''+q$, $J''$, $J'$, and signed $\Lambda'',\Lambda'$.
A ``channel'' is represented by a matrix element between Hund's case (a) basis states.
Each of the basis states separates into a radial spatial and a spin part according to
\begin{equation}
    \ket{\alpha JNv\Lambda\Omega} = \ket{\alpha Nv\Lambda}\ket{S\,(\Omega-\Lambda)\,},
    \label{eq:mueff_spinsep}
\end{equation}
where the effect of the total angular momentum (quantum numbers $J$ and $\Omega$) is contained in the factors $s^{J'J''}_{\Omega''q}$ of Eq.\ (\ref{eq:mueff}).

When these states are applied in Eqs.\ (\ref{eq:mueff}) and (\ref{eq:mueff_elem}) and spin--orbit mixing is not considered, the spin parts of the transition matrix elements yield the spin selection rule ($S'=S''$) and the condition $\Lambda'-\Lambda''=\Omega'-\Omega''$, which effectively replaces $\Omega$ by $\Lambda$ in Eq.\ (\ref{eq:sjj}).
This leads to the familiar relative intensity factors of spin-allowed transitions.
However, Eq.\ (\ref{eq:mueff}) can also address spin--orbit mixed states and the resulting spin-forbidden transitions, for which results are less directly accessible in the literature.

\subsection{Doublet levels}
\label{app:rad_doub}
For the transitions X$^2\Sigma_g^+$-A$^2\Pi_u$ and A$^2\Pi_u$-X$^2\Sigma_g^+$ we apply Eq.\ (\ref{eq:mueff}) to the spin-allowed case.
After separation of the spin part according to Eq.\ (\ref{eq:mueff_spinsep}), the dipole matrix elements involving the states of Eq.\ (\ref{eq:mueff_elem}) take the forms
\begin{equation}
    \bra{aN'v'{}\,{\pm\Lambda'}}\mu_q\ket{bN''v''\,{\pm\Lambda''}}
\end{equation}
with all combinations of the signs.
Making use of symmetry properties, a factor corresponding to a single polarization component $q$ can \cite{watson_honllondon_2008} be pulled out of the sum in Eq.\ (\ref{eq:mueff}).
The line strength, Eq.\ (\ref{eq:smueff}), can be written (with unsigned, positive-only $\Lambda',\Lambda''$) as
\begin{equation}
    S_{aN'v'J'}^{bN''v''J''} =
    [\mu^{ab}_{\Lambda'-\Lambda''}\bm{(}N'v',N''v''\bm{)}]^2\;
    \hat{S}_{aN'J'}^{bN''J''}.
\label{eq:sall}
\end{equation}
It is thus separated into the squared, vibrationally averaged electronic dipole matrix element and the H\"onl--London factor $\hat{S}_{aN'J'}^{bN''J''}$, which reflects the terms remaining in the sum of Eq.\ (\ref{eq:mueff}).

The dipole matrix element in Eq.\ (\ref{eq:sall}),
\begin{equation}
    \mu^{ab}_{q}\bm{(}N'v',N''v''\bm{)} = \bra{aN'v'\Lambda'}\mu_{q}\ket{bN''v''\Lambda''},
    \label{eq:muab}
\end{equation}
with $q=\Lambda'-\Lambda''$ corresponds to the vibrational average
\begin{equation}
    \mu^{ab}_{q}\bm{(}N'v',N''v''\bm{)} = \bra{N'v'}\mu^{ab}_{q}(R) \ket{N''v''},
    \label{eq:mu_abinitio}
\end{equation}
where the dipole moment function
\begin{equation}
   \mu^{ab}_{q}(R)=
   \bra{a\Lambda'}\mu_{q}\ket{b\Lambda''}_R
\end{equation}
is obtained from ab-initio calculations of the X$^2\Sigma_g^+$ and A$^2\Pi_u$ states at several fixed $R$ (see Sec.\ \ref{sec:the_matelements}).
For either A$^2\Pi_u$-X$^2\Sigma_g^+$ or  X$^2\Sigma_g^+$-A$^2\Pi_u$ transitions, the same function $\mu^{\text{A-X}}(R)=\mu^{\text{A,X}}_{q\!=\!1}(R)$ can be used.

The H\"onl--London factor results from the angular factors, the definition of the parity eigenstates, Eq.\ (\ref{eq:mueff_elem}), and the summations in Eq.\ (\ref{eq:mueff}).
We can directly use Eq.\ (38) of Ref.\ \cite{watson_honllondon_2008}, taking into account that X$^2\Sigma_g^+$-A$^2\Pi_u$ and A$^2\Pi_u$-X$^2\Sigma_g^+$ both correspond to the case where exactly one of the two states has $\Lambda=0$.
This yields
\begin{eqnarray}
    &&\hat{S}^{\epsilon_N''N''J''} _{\epsilon_N'N'J'} = 2{P}_{\epsilon_N'\epsilon_N''}^{\Delta N} [J',J'',N',N''] \nonumber\\
    &&~\hspace{4mm}{}\times\sixj{S''}{N''}{J''}{1}{J'}{N'}^2 \jjj{N'}{1}{N''}{-\Lambda'}{q}{\Lambda''}^2
\label{eq:angall2}
\end{eqnarray}
with $q=\Lambda'-\Lambda''$, $S''=1/2$, and the parity selection factor ($\Delta N =N'-N''$)
\begin{equation}
      {P}_{\epsilon_N'\epsilon_N''}^{\Delta        N}=(1/2)\left[1-\epsilon_N'\epsilon_N''(-1)^{\Delta N}\right],
\label{eq:parity}
\end{equation}
which is consistent with a state parity of $\epsilon_N(-1)^N$ (see Appendix \ref{app:definitions}).
In the indices of $\hat{S}$ we maintained from the identifiers $a,b$ only the parity labels $\epsilon_N',\epsilon_N''$.
In the A$^2\Pi_u$ state, a sublevel with given $N,J$ further splits into two levels with $\epsilon_N=\pm1$, in contrast to X$^2\Sigma^+_g$ there is only one such sublevel with $\epsilon_N=1$.

We especially consider the partial decay rate from a state $\epsilon_N'N'v'J'$ to all sublevels $J'',\epsilon_N''$ of a final level $N''v''$.
In the required final-state sums over $\hat{S}$, the sum over $J''$ replaces the squared $6j$ symbol by $1/[J'',N']$ through its orthogonality relation.
In the $\epsilon_N''$ sum, the parity selection factor remains if the lower state is a $\Sigma$ state ($\Lambda''=0$).
If the lower state is a $\Pi$ state, a final level of fitting parity is always available.
Accordingly,
    \begin{eqnarray}
      \sum_{\epsilon_N'',J''}\hat{S}^{\epsilon_N''N''J''}_{\epsilon_N'N'J'}
      =2[J',N'']
          \jjj{N'}{1}{N''}{-\Lambda'}{q}{\Lambda''}^2\nonumber\\
      \times\left\lbrace
         \begin{array}{lr}{P}_{\epsilon_N',\epsilon_N''=+1}^{\Delta N}&\text{for
           $\Lambda''=0$ (A$^2\Pi_u$--X$^2\Sigma^+_g$)}\\1&\text{for
           $\Lambda''=1$ (X$^2\Sigma^+_g$--A$^2\Pi_u$).} \end{array}\right. 
         \label{eq:angall}
    \end{eqnarray}
The result illustrates (see also Sec.\ \ref{sec:the_potentials}) that levels of different nuclear exchange symmetry are not connected by radiative transitions. 
Hence, radiative transition probabilities and lifetimes are unaffected by the isotopical symmetry.
The expressions in this Appendix include all cases of nuclear exchange symmetry and the isotopical symmetry of the C$_2{}^-$ system is fully accounted for by choosing the proper initial levels.

With help of Eqs.\ (\ref{eq:arate}) and (\ref{eq:sall}), we then derive the following  partial decay rates.
For the A$^2\Pi_u$--X$^2\Sigma^+_g$ transitions,
\begin{eqnarray}
  A_{A,N'v'}^{X,N''v''}(\epsilon'_N)
  =2\,C_0 \left(\tnu_{A,N'v'}^{X,N''v''}\right)^3\;
      \left[\mu^{\text{A-X}}(N'v',N''v'')\right]^2\;
     \nonumber\\
    {}\times
         [N'']\jjj{N'}{1}{N''}{-1}{1}{0}^2
         \;\frac{1-\epsilon'_N(-1)^{\Delta N}}{2}.\hspace{10mm}
\label{eq:resultAX1}
\end{eqnarray}
For the X$^2\Sigma^+_g$--A$^2\Pi_u$ transitions,
\begin{eqnarray}
  A_{X,N'v'}^{A,N''v''}
  &=&2\,C_0 \left(\tnu_{X,N'v'}^{A,N''v''}\right)^3\;
      \left[\mu^{\text{A-X}}(N'v',N''v'')\right]^2\;
     \nonumber\\ &&
    {}\times
         [N'']\jjj{N'}{1}{N''}{0}{-1}{1}^2.\hspace{10mm}
\label{eq:resultXA1}
\end{eqnarray}

\subsection{Quartet levels}
\label{app:rad_quart}
In contrast to spin-allowed transitions, line strengths of forbidden transitions were discussed less systematically.  An early study \cite{budo_intensitatsverteilung_1940,kovacs_1969} was performed for the $^4\Sigma$\,--\,$^2\Sigma$ case relevant here.  Spin--orbit induced forbidden transitions were also discussed under the aspect of perturbations in molecular spectra.  As an instructive example, recent studies \cite{james_transition_1971,rostas_band_2000} include intercombination transitions connecting states in the X\,$^1\Sigma^+$ ground potential of CO to various excited triplet states, which is made possible by their perturbation by states in the A\,$^1\Pi$ potential.

Since no systematic studies are available, we set up the description based on the definition of the effective dipole matrix element in Eq.\ (\ref{eq:mueff}).
In the picture of molecular perturbations \cite{rostas_band_2000}, the mixing of quantum states produces a set of perturbed vibrational and rotational molecular states where, through the relatively small size of the spin--orbit interaction, the levels predominantly keep their a$^4\Sigma^+_u$ and A$^2\Pi_u$ character, respectively.
Such mixed states share the same total angular momentum ($J$ and $\Omega$).
Radiative transitions from the states with an a$^4\Sigma^+_u$ dominance to X$^2\Sigma^+_g$ are possible only via the small A\,$^2\Pi_u$ admixture; hence, they remain much longer lived than the states with an A$^2\Pi_u$ dominance (whose very small relative lifetime modifications are neglected).

We describe the modification of the ``channel'' transition matrix elements entering Eq.\ (\ref{eq:mueff}) via Eq.\ (\ref{eq:mueff_elem}) by the spin--orbit interaction as a perturbation by the first-order admixture.
Perturbing states are the full set of rotational and vibrational levels in the  A$^2\Pi_u$ electronic state, described by Hund's case (b) basis states $\ket{c\bar{J}\bar{N}\bar{v}\bar{\Lambda}}$ [Eqs.\ (\ref{eq:rexp}), (\ref{eq:rcoeff})], which have the energies $E_{c\bar{N}\bar{v}}$.
The ``channel'' matrix elements [see the discussion preceding Eq.\ (\ref{eq:mueff_spinsep})] then take the form
\begin{eqnarray}
\bra{aJ'N'v'\Lambda'\Omega'}\,{\mu_{q}}\,\ket{bJ''N''v''\Lambda''\Omega''}=
\hspace{15mm}\nonumber\\
\sum_{\bar{N}\bar{v}}
\left(r^{\bar{\Lambda} J' \bar{S}}_{\bar{N} \Omega'}\right)^2 \frac{\bra{aJ'N'v'\Lambda'\Omega'}\,{H_{\text{SO}}}\,\ket{cJ'\bar{N}\bar{v}\bar{\Lambda}\Omega'}}
{E_{aN'v'}-E_{c\bar{N}\bar{v}}}
\hspace{8mm}\nonumber\\{}\times
\bra{cJ'\bar{N}\bar{v}\bar{\Lambda}\Omega'}\,\mu_q\,\ket{bJ''N''v''\Lambda''\Omega''}
\hspace{5mm}
\end{eqnarray}
which accounts for $\bar{J}=J'$ and $\bar{\Omega}=\Omega'$.
The matrix elements of the spin--orbit coupling Hamiltonian $H_{\text{SO}}$ follow from the electronic interaction in the body-fixed frame (see below) and couple the wave functions of the $a$ and $c$ states with the same symmetry ($\bar{\Omega}=\Omega'$) while the spin projection quantum numbers are different.
Thus, coupling occurs between states with $\Sigma'=\Omega'-\Lambda'$ and
$\bar{\Sigma}=\Omega'-\bar{\Lambda}$, which differ as $\bar{\Lambda}\neq\Lambda'$.

To account for Eq.\ (\ref{eq:mueff_elem}) we use the Hund's case (b) expansions of the initial
level $a=\text{a$^4\Sigma^+_u$}$ ($S'=3/2$)
\begin{equation}
    \ket{aJ'N'v'}=\sum_{\Omega'} r^{0\,J'S'}_{N'\Omega'} \ket{aJ'N'v'\,0\,\Omega'}\nonumber
\end{equation}
and the final level $b=\text{X$^2\Sigma^+_g$}$ ($S''=1/2$)
\begin{equation}
    \ket{bJ''N'v''}=\sum_{\Omega''} r^{0\,J''S''}_{N''\Omega''} \ket{bJ''N''v''\,0\,\Omega''}.\nonumber
\end{equation}
For these two levels, the Hund's case basis states are already parity eigenstates, corresponding to the exceptional case $\Lambda=0$ in Eq.\ (\ref{eq:mueff_elem}). 
Hence, Eq.\ (\ref{eq:mueff}) yields the effective dipole matrix element
\begin{eqnarray}
\bra{aN'v'J'}\,{\mu_{\text{eff}}}\,\ket{bN''v''J''}=
\sum_{q,\Omega',\Omega''}
r^{0\,J'S'}_{N'\Omega'}
r^{0\,J''S''}_{N''\Omega''}
\nonumber\\
\times \sum_{\bar{N}\bar{v}}
\left(r^{\bar{\Lambda} J' \bar{S}}_{\bar{N} \Omega'}\right)^2 
\frac{
\bra{aJ'N'v'\,0\,\Omega'}\,{H_{\text{SO}}}\,\ket{cJ'\bar{N}\bar{v}\bar{\Lambda}\Omega'}}
{E_{aN'v'}-E_{c\bar{N}\bar{v}}}
\nonumber\\{}\times
\bra{cJ'\bar{N}\bar{v}\bar{\Lambda}\Omega'}\,\mu_q\,\ket{bJ''N''v''\Lambda''\,0\,\Omega''} 
s^{J'J''}_{\Omega''q} .
\hspace{11mm}
\label{eq:mueff_so}
\end{eqnarray}
The $\mu_q$ matrix element corresponds to a spin-allowed optical transition where the selection rules lead to $\bar{\Sigma}=\Sigma''$, $\Omega'=\Omega''+q$, and $\bar{\Lambda}=\Lambda''+q$.
Similar to Eq.\ (\ref{eq:muab}) it is written as $\mu^{cb}_q\bm{(}N'v',N''v''\bm{)}$.
For our $\Pi$-$\Sigma$ transition, $q=\pm1$ only.
Simultaneous change of signs for $\Omega',\Omega''$, and $q$ leaves the sign of the matrix element unchanged ($\Sigma^-$ states are not considered).  
Moreover, since $\Lambda''=0$, $\bar{\Lambda}=q$.

The spin--orbit matrix elements can be written as $\hat{A}^{ac}_{\Omega'}\bm{(}N'v',\bar{N}\bar{v}\bm{)}$ where for the states $a=\text{a$^4\Sigma^+_u$}$ and $c=\text{A$^2\Pi_u$}$ the spin--orbit coupling function is obtained from the ab-initio calculation of the matrix elements 
\begin{eqnarray}
\hat{A}^{ac}_{3/2}(R)&\!=\!&
\Braket{a,\Lambda'\!=\!0,\Sigma'\!=\!
{\textstyle\frac{3}{2}}
|H_{\text{SO}}|
c,\bar{\Lambda}\!=\!1,\bar{\Sigma}\!=\!
{\textstyle\frac{1}{2}}}_R,
\nonumber\\
\hat{A}^{ac}_{-1/2}(R)&\!=\!&
\Braket{a,\Lambda'\!=\!0,\Sigma'\!=\!
-{\textstyle\frac{1}{2}}
|H_{\text{SO}}|
c,\bar{\Lambda}\!=\!-1,\bar{\Sigma}\!=\!
{\textstyle\frac{1}{2}}}_R\nonumber\\
\label{eq:a_abinitio}
\end{eqnarray}
with the property $\hat{A}^{ac}_{-\Omega'}=-\hat{A}^{ac}_{\Omega'}$.

The sum of Eq.\ (\ref{eq:mueff_so}) effectively runs over $q$ and $\Omega''$ as $\Omega'=\Omega''+q$.
The terms $T(q,\Omega'')$ under the sum are grouped considering the sign-change symmetry
$$
T(-q,-\Omega'')=\hat{s}\,T(q,\Omega'')
$$
where $\hat{s}=(-1)^{N'+N''+1}$.
All terms 
$$
T(1,|\Omega''|)+T(-1,-|\Omega''|)+T(-1,|\Omega''|)+T(1,-|\Omega''|)
$$
occurring for each value of $|\Omega''|$ can then be written as
$$
(1+\hat{s})\left[\,T(1,|\Omega''|)+T(-1,|\Omega''|)\,\right]
$$
This procedure (which would need special consideration of the $\Omega''=0$ term if $\Omega''$ would not be half-integer) leaves the $q=\pm1$ sum over the expressions for $\Omega''=|\Omega''|$ and $\Omega'=\Omega''+q$, multiplied by $1+(-1)^{N'+N''+1}$.

In the present case, with $|\Omega''|=1/2$ as the only value, the effective dipole matrix element results as
\begin{eqnarray}
\bra{aN'v'J'}\,{\mu_{\text{eff}}}\,\ket{bN'v''J''}=
\hspace{30mm}\nonumber\\
2 P(\Delta N)\sum_{q}
Q_q \sum_{\bar{N}\bar{v}}
\lambda_{\bar{N},q}\frac{
\hat{A}^{ac}_{\Omega'}\bm{(}N'v',\bar{N}\bar{v}\bm{)}}
{E_{N'v'}-E_{\bar{N}\bar{v}}}
\hspace{15mm}\nonumber\\{}\times
\mu^{cb}_q\bm{(}
\bar{N}\bar{v}
,N''v''\bm{)}
\hspace{15mm}
\label{eq:mueff_forb}
\end{eqnarray}
with the angular coefficients
\begin{eqnarray} Q_q&=&
r^{0J''S''}_{N''\Omega''}
\,r^{0J'S'}_{N'\Omega'}
\,s^{J'J''}_{\Omega''q},
\label{eq:coef_q}\\
\lambda_{\bar{N},q}&=&
\left(r^{q J' \bar{S}}_{\bar{N} \Omega'}\right)^2,
\label{eq:coef_lambda}
\end{eqnarray}
and the parity selection factor
\begin{equation}
P(\Delta N)=
(1/2)\left[1-(-1)^{\Delta N}\right].
\label{eq:parity_spinforb}
\end{equation}
The appropriate replacements are
\begin{eqnarray}
\Omega''&=&1/2,\nonumber\\ 
\Omega'\,&=&1/2+q\;\;[= (3/2,-1/2)\;\;{\rm for}\;q=(1,-1)],\nonumber\\ 
S''&=&1/2\;(=\bar{S}),\nonumber\\
S'\,&=&3/2.
\label{eq:qnrepl}
\end{eqnarray}
for the spin-forbidden transition a$^4\Sigma^+_u$-X$^2\Sigma_g^+$.

The ab-initio results for the spin--orbit coupling functions $\hat{A}^{ac}_{+}(R)=\hat{A}^{ac}_{3/2}(R)$ and $\hat{A}^{ac}_{-}(R)=\hat{A}^{ac}_{-1/2}(R)$ are found to obey at all $R$ the relation $\hat{A}^{ac}_{-}(R)=-(1/\sqrt{3})\hat{A}^{ac}_{+}(R)$. 
Hence, vibrational averaging needs to be performed only for $\hat{A}^{ac}_{+}(R)$ to obtain $\hat{A}^{ac}_{+}\!\left(N'v', \bar{N}\bar{v}\right)$ as well as $\hat{A}^{ac}_{-}\!\left(N'v', \bar{N}\bar{v}\right)=
-(1/\sqrt{3})\hat{A}^{ac}_{+}\!\left(N'v', \bar{N}\bar{v}\right)$.

To sum the two $q$ terms of Eq.\ (\ref{eq:mueff_forb}) we introduce the coefficients
\begin{eqnarray}
    \hat{Q}_+=\sqrt{3}\, r^{0J''S''}_{N''\Omega''}
\,r^{0J'S'}_{N'(\Omega'\!=\!\frac{3}{2})}
\,s^{J'J''}_{\Omega''(q\!=\!+1)},
\label{eq:qfac}
\end{eqnarray}
and
\begin{eqnarray}
    \hat{r}=\frac{r^{0J'S'}_{N'(\Omega'\!=\!-\frac{1}{2})}
\,s^{J'J''}_{\Omega''(q\!=\!-1)}}
{\sqrt{3}\,r^{0J'S'}_{N'(\Omega'\!=\!\frac{3}{2})}
\,s^{J'J''}_{\Omega''(q\!=\!+1)}},
\label{eq:rfac}
\end{eqnarray}
which yields $Q_{+1}=\hat{Q}_+/\sqrt{3}$ and
$Q_{-1}=\hat{r}\hat{Q}_+$.  The factor of $\sqrt{3}$ in Eq.\ (\ref{eq:qfac}) is introduced in order to link to previous results on molecular doublet--quartet transitions.  
All the angular factors are given in terms of $J',J''$, and the fine structure labels of the involved upper and lower levels (see the Supplemental Material).

With these definitions, the relation $\mu_{-1}=\mu_-=\mu_{1}=\mu_+$, and that between $\hat{A}_{-}$ and $\hat{A}_{+}$, the $q$ summation in Eq.\ (\ref{eq:mueff_forb}) leads to
\begin{eqnarray}
\bra{aN'v'J'}\,{\mu_{\text{eff}}}\,\ket{bN''v''J''}=
\frac{2P(\Delta N)}{\sqrt{3}} \hat{Q}_+ 
\hspace{10mm}\nonumber\\
\times\sum_{\bar{N}}
\left(\lambda_{\bar{N},1}
-\hat{r}\lambda_{\bar{N},-1}
\right)I(\bar{N})
\hspace{15mm}
\label{eq:mueff_forb2}
\end{eqnarray}
with
\begin{eqnarray}
I(\bar{N}) =\sum_{\bar{v}} \frac{
\hat{A}^{ac}_+\!\left(N'v',
\bar{N}\bar{v}\right)
\mu^{cb}_+\!\left(
\bar{N}\bar{v},N''v''\right)
}{E_{N'v'}-E_{\bar{N}\bar{v}}}.
\hspace{5mm}
\label{eq:mueff_I}
\end{eqnarray}
Here, $I(\bar{N})$ describes the perturbation sum for the vibrational levels belonging to the angular momentum $\bar{N}$. 
Among the perturbing states, the energy difference $E_{N'v'}-E_{\bar{N}\bar{v}}$ is most often large compared to the vibrational average of the spin--orbit interaction, yielding correspondingly
small admixture amplitudes $\hat{A}^{ac}_+\!\left(N'v',\bar{N}\bar{v}\right)/(E_{N'v'}-E_{\bar{N}\bar{v}})$.
For the very rare cases that the perturbing level lies energetically very close, the admixture amplitude is computationally limited to $<1/\sqrt{2}$.

For the transition a$^4\Sigma^+_u$-X$^2\Sigma_g^+$ via A$^2\Pi_u$ admixture, the coefficients $\lambda_{\bar{N},q}$ according to Eqs.\ (\ref{eq:coef_lambda}), (\ref{eq:rcoeff}) are
\begin{eqnarray}
\lambda_{\bar{N},q}&=&
[N]\;
\jjj{\bar{S}}{\bar{N}}{J'}{\Omega'-q}{q}{-\Omega'}^2
\nonumber\\
&=&[N]\;
\jjj{1/2}{\bar{N}}{J'}{1/2}{q}{-1/2-\Omega'}^2
\nonumber\\
&=&\frac{1}{2}\mp\frac{q}{[J']}
\label{eq:lambdaq}
\end{eqnarray}
corresponding to the only two possible values of $\bar{N}=J'\pm1/2$. 
Hence, performing the sum over $\bar{N}$ in Eq.\ (\ref{eq:mueff_forb2}) with $\lambda_{\bar{N},1}$ and $\lambda_{\bar{N},-1}$ leads to
\begin{eqnarray}
\bra{aN'v'J'}\,{\mu_{\text{eff}}}\,\ket{bN''v''J''} = \frac{2P(\Delta N)}{\sqrt{3}} \hat{Q}_+ 
\hspace{10mm}\nonumber\\
\times\left[(1-\hat{r})\frac{I^{J'}_1+I^{J'}_2}{2}
-(1+\hat{r})\frac{I^{J'}_1-I^{J'}_2}{[J']}\right].
\hspace{5mm}
\label{eq:mueff_forb3}
\end{eqnarray}
Given $J'$ for a$^4\Sigma^+_u$, there are two groups of perturbing $v$ levels with $\bar{N}=J\pm1/2$, for which the vibrational sums lead to $I^{J'}_1=I(J'+1/2)$ and $I^{J'}_2=I(J'-1/2)$ with $I(\bar{N})$ according to Eq.\ (\ref{eq:mueff_I}).  
According to Eq.\ (\ref{eq:smueff}), the vibrationally averaged line strength of the spin-forbidden doublet--quartet transition then becomes
\begin{eqnarray}
S^{N''v''J''}_{N'v'J'}=
\frac{4P(\Delta N)}{3} \hat{Q}^2_+ 
\left[(1-\hat{r})\frac{I^{J'}_1+I^{J'}_2}{2}\right.
\hspace{10mm}\nonumber\\
\left.-(1+\hat{r})\frac{I^{J'}_1-I^{J'}_2}{[J']}\right]^2.
\hspace{3mm}
\label{eq:s_dq_p}
\end{eqnarray}
where all vibrational dependence is contained in the functions $I^{J'}_{1,2}$ according to Eq.\ (\ref{eq:mueff_I}).  

It is instructive to note that the details of the vibrational perturbation calculation essentially enter via the variation of the vibrational energy differences between the initial (quartet) and the perturbing (doublet) levels in the denominator of Eq.\ (\ref{eq:mueff_I}). 
If the energies of all relevant levels of the perturbing state were strongly different from the initial-state energy, the energy denominators in  Eq.\ (\ref{eq:mueff_I}) could be well approximated by the electronic energy difference $\Delta E_e^{ac}$ and taken out of the sum as a $\bar{v}$-independent factor.  
Considering the completeness of the vibrational wave functions, the sum over
$\bar{v}$ including only the product of the two vibrational integrals could be replaced by a single vibrational integral $I=\tilde{\mu}^{a(c)b}_+\!\left(
N'v',N''v''\right)$ over the product function
\begin{eqnarray}
    \tilde{\mu}^{a(c)b}_+(R)=\frac{\hat{A}^{ac}_+(R)\,\mu^{cb}_+(R)}{\Delta E_e^{ac}}. 
\end{eqnarray}
In addition, this leads to $I^{J'}_1=I^{J'}_2=I$ (independent of $J'$).  Then, the line strength could be given as
\begin{equation}
\tilde{S}^{N''J''}_{N'J'}=
\left[\frac{4P(\Delta N)}{3} \hat{Q}^2_+ 
(1-\hat{r})^2\right]\;
\left[\tilde{\mu}^{a(c)b}_+\!\left(
N'v',N''v''\right)\right]^2,
\label{eq:s_dq_ee}
\end{equation}
where a H\"onl--London-like angular factor (first brackets) is multiplied by a Franck--Condon-type integral.  In particular, the angular factor gives the invariant set of relative intensities of the fine-structure components, while the absolute line intensities are given by the $\tilde{\mu}^{a(c)b}_+$ integral only.  Published relative fine-structure intensities of doublet--quartet transitions
\cite{kovacs_1969} obtained with this approximation agree with the values of $\hat{Q}^2_+(1-\hat{r})^2$ used in Eq.\ (\ref{eq:s_dq_ee}) \cite{suppl}.

In our case, however, considering the a$^4\Sigma^+_u$ and A$^2\Pi_u$ vibrational energy levels in C$_2{}^-$, the energy denominator of Eq.\ (\ref{eq:mueff_I}) varies strongly with $\bar{v}$ and a full perturbation calculation is required. 
For this calculation, all vibrationally bound levels $N'v'$ of a$^4\Sigma^+_u$ with $N'\geq155$ (odd $N'$) are included as initial levels, requiring their energies and vibrational wave functions. 
Levels with lower $N'$ should rapidly autodetach into the a$^3\Pi_u$ state of C$_2$. 
The fine-structure terms F$i$ have $i=i'=1,2,3,4$ and $J'=N'+5/2-i'$.  
Considering perturbation by A$^2\Pi_u$ levels, the relevant $\bar{N}\bar{v}$ for given $N'$, see Eq.\ (\ref{eq:lambdaq}), have $\bar{N}=N'-2,\ldots,N'+2$ in steps of $1$.
The fine-structure terms F$i$ of levels $N''v''$ in the final X$^2\Sigma^+_g$ state have $i=i''=1,2$ and $J''=N''+3/2-i''$.
Considering the allowed transitions with $\Delta J=J'-J''=-1,0,1$, see Eq.\ (\ref{eq:sjj}), values of $N''$ for which at least one fine-structure transition is possible are $N''=N'-3,\ldots,N'+3$, taken in steps of 2 because of the parity selection rule, see Eq.\ (\ref{eq:parity_spinforb}).
This implies $\Delta N=N'-N''=-3,-1,1,3$.
Accordingly, energies and vibrational wave functions are included for all vibrationally bound levels $\bar{N}\bar{v}$ of A$^2\Pi_u$ with $\bar{N}\geq153$ (step of 1) and for all vibrationally bound levels $N''v''$ in the final X$^2\Sigma^+_g$ state with $N''\geq152$ (step of 2).  

Using the ab-initio functions, Eqs.\ (\ref{eq:mu_abinitio}) and (\ref{eq:a_abinitio}), we combined all relevant pairs of individual-level vibrational wave functions, calculating the vibrational integrals $\mu^{cb}_+\!\left(\bar{N}\bar{v},N'',v''\right)$ and $\hat{A}^{ac}_+\!\left(N'v',\bar{N}\bar{v}\right)$. 
Together with the level energy differences, we evaluated Eq.\ (\ref{eq:mueff_I}) and then combined the resulting $\bar{v}$ sums to find line strengths $S^{N''v''i''}_{N'v'i'}$ from Eq.\ (\ref{eq:s_dq_p}).
At the $N'$ ($J'$) considered, the main contribution in Eq.\ (\ref{eq:s_dq_p}) is seen to arise from the positive-sign interference of $I^{J'}_{1,2}$, while the effect of the negative-sign interference is reduced through a factor $<10^{-2}$.  The negative-sign contribution was nevertheless included as the additional computational effort was negligible compared to the evaluation of the large set of vibrational integrals.

\begin{figure}[!t]
    \centering
    \includegraphics[width=\columnwidth]{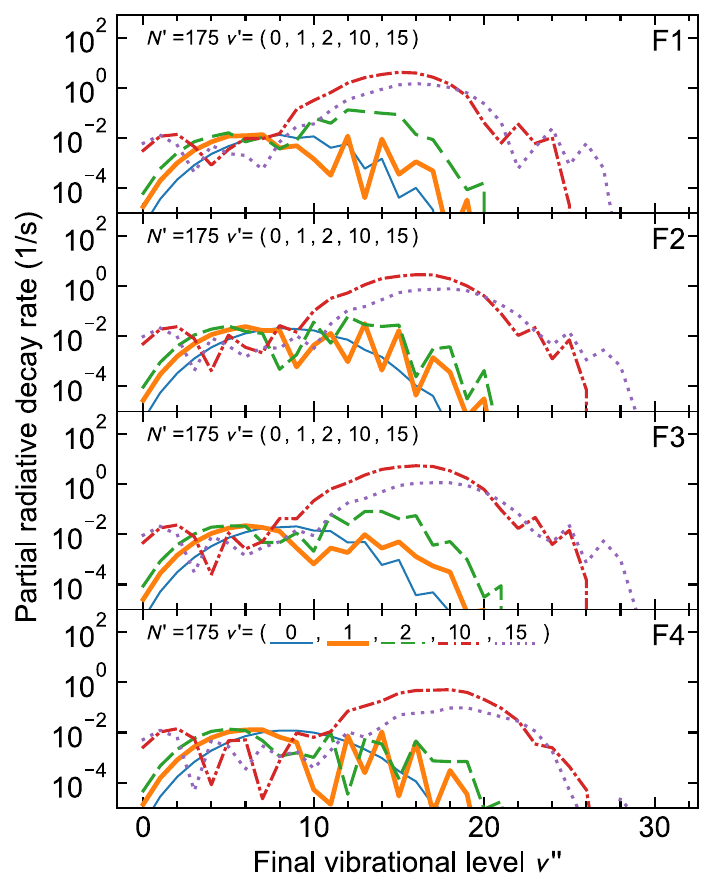}
    \caption{Partial rates $A^{v''}_{N'v'i'}$ [Eq.\ (\ref{eq:apartial})] of spin-forbidden radiative decay for fine-structure levels F$i$ ($i=i'=1,\ldots,4$) in the a$^4\Sigma^+_u$  electronic state with $N'=175$ and $v'$ as labeled into final levels $v''$ of X$^2\Sigma^+_g$. Colored lines as marked in the lowest window are used for the selected initial vibrational levels $v'$.}
    \label{fig:quartet_partial}
\end{figure}

Applying Eq.\ (\ref{eq:arate}), transition rates $A^{N''v''i''}_{N'v'i'}$ (where $E_{N''v''} < E_{N'v'}$) were obtained for the individual fine structure components of the $(N'v')$-$(N''v'')$ transition arrays. 
This fine-structure is included in the $J'$-dependence of the squared bracket in Eq.\ (\ref{eq:s_dq_p}) as well as in the dependence of the parameters $\hat{Q}_+$ and $\hat{r}$ on the rotational quantum numbers. 
Considering the decay from a fine-structure level $N'v'i'$, these angular parameters can be written as matrices $(\hat{Q}_+^2/[J'])^{\Delta N}_{i'i''}$ and $(\hat{r})^{\Delta N}_{i'i''}$, which can be expressed using $N'$ and are given in the Supplemental Material \cite{suppl}.

In Fig.\ \ref{fig:quartet_partial} we show as an example the partial decay rates
\begin{equation}
   A^{v''}_{N'v'i'}=\sum_{N''}\sum_{i''=1,2}
   A^{N''v''i''}_{N'v'i'},
   \label{eq:apartial}
\end{equation}
where $N''=N'-\Delta N$ with $\Delta N=-3,-1,1,3$, for some levels $v'$ with $N'=175$.
As expected because of the energy denominator in Eq.\ (\ref{eq:mueff_I}) and the variation of the vibrational overlaps, the perturbation leads to significant fluctuation and also to a substantial enhancement for a range of final levels where the overlap of the A$^2\Pi_u$ and X$^2\Sigma^+_g$ vibrational wavefunctions is favorable.

Total radiative decay rates are obtained as
\begin{equation}
    A_{N'v'i'}=\sum_{v''}
    A^{v''}_{N'v'i'}.
    \label{eq:atotal}
\end{equation}
They scatter quite irregularly between the levels as shown in Fig.\ \ref{fig:rdec_quartet} (Sec.\ \ref{sec:calculation_rad_quartet}).

We are aware that also continuum vibrational levels of the A$^2\Pi_u$ perturber state might contribute to the spin--orbit mixing.
However, as confirmed by inspecting the contributions of individual $\bar{v}$ perturber levels in a number of cases, the mixing effect is strongly dominated by the perturber vibrational states energetically close to the initial level.
Moreover, considering the a$^4\Sigma^+_u$ and  A$^2\Pi_u$ states near their rotational barriers ($R\sim3$\,\AA\ for $N$ near 171, see Fig.\ \ref{fig:rearrange}), the A$^2\Pi_u$ potential supports bound vibrational levels up to higher energies than the a$^4\Sigma^+_u$ potential, so that the  A$^2\Pi_u$ perturber levels energetically closest to the bound a$^4\Sigma^+_u$ vibrational levels will always be bound. 
Hence, by omitting the  A$^2\Pi_u$ vibrational continuum, we only neglect perturber states with larger (more unfavorable) energy denominators.
In addition, the vibrational overlap of the A$^2\Pi_u$-X$^2\Sigma^+_g$ radiative matrix element entering Eq.\ (\ref{eq:mueff_I}) should also become less favorable for the A$^2\Pi_u$ continuum vibrational states.
Therefore, we expect the overall pattern of a$^4\Sigma^+_u$ spin-forbidden decay rates to be well represented even without including the A$^2\Pi_u$ vibrational continua.

\newpage\null\thispagestyle{empty}\newpage



\section*{Supplementary material}

**Supplemental Material**

# Unimolecular processes in diatomic carbon anions at high rotational excitation

Viviane C. Schmidt[1], Roman Čurík[2], Milan Ončák[3], Klaus Blaum[1], Sebastian George[1],

Jürgen Göck[1], Manfred Grieser[1], Florian Grussie[1] Robert von Hahn[1], Claude Krantz[1],

Holger Kreckel[1], Oldřich Novotný[1], Kaija Spruck[1], Andreas Wolf[1]

[1] *Max-Planck-Institut für Kernphysik, 69117 Heidelberg, Germany*

[2] *J. Heyrovský Institute of Physical Chemistry, Academy of Sciences of the Czech Republic,v.v.i, Dolejškova 3, 18223 Prague 8, Czech Republic*

[3] *Institut für Ionenphysik und Angewandte Physik, Leopold-Franzens-Universität Innsbruck, Technikerstraße 25/32, Innsbruck 6020, Austria*

---

The Material is addressed in the sequence as referenced in the main paper.

## List of Supplemental Material

## 1. Low-rotation limit of the ab-initio model

Results from the model potentials for $C_2^-$ and $C_2$ are listed in Table S1 and compared to results of recent more sophisticated calculations for $C_2^-$ only [1] and with available theoretical results. This includes the electronic excitation energies (energy differences between potential curve minima) $T_e$, the internuclear distace $R_e$ at the potential minimum, the vibrational constant $\omega_e$, and the lowest anharmonicity coefficient $\omega_e x_e$. The $A^2\Pi_u$ term energy is overestimated by the model by about $253\,\mathrm{cm}^{-1}(0.031\ \mathrm{eV})$. The difference between the $v = 0$ energy levels in the ab-initio potentials of the $X^2\Sigma_g^+$ state of $C_2^-$ and the $X^1\Sigma_g^+$ of $C_2$ underestimates the electron affinity of $C_2$ by $\sim 0.22$ eV compared to the experimental value [2]. Therefore, the potentials applied in the autodetachment rate calculations were modified as described in Sec. IVC of the main paper.

TABLE S1: Results for various quantities extracted from the present model potentials for $C_2^-$ and $C_2$ (correlation-consistent basis set cc-pVQZ, valence polarization only) with recent calculations [1] for $C_2^-$ (basis set cc-pCV5Z, including core-valence polarization) and with experimental results. The electron affinity $E_\mathrm{EA}$ of $C_2$ is included.

| Quantity (unit) | Molecule | Level | cc-pVQZ[a] | cc-pCV5Z[b] | exp |
|---|---|---|---|---|---|
| $T_e(\mathrm{cm}^{-1})$ | $C_2^-$ | $A^2\Pi_u$ | 4182.1 | 4009.9 | 3928.660(17)[c] |
| | | $B^2\Sigma_u^+$ | 19441.2 | 18981.3 | 18390.723(15))[d] |
| | | $C^4\Sigma_u^+$ | 33762.3 | 33227.0 | |
| | $C_2$ | $a^3\Pi_u$ | 745.7 | | 720.0083(21)[e] |
| $R_e(\text{Å})$ | $C_2^-$ | $X^2\Sigma_g^+$ | 1.2744 | 1.2689 | 1.26831(13)[c] |
| | | $A^2\Pi_u$ | 1.3127 | 1.3075 | 1.30768(13)[c] |
| | | $B^2\Sigma_u^+$ | 1.2301 | 1.2238 | $(1.2234)^{d,f}$ |
| | | $C^4\Sigma_u^+$ | 1.4797 | 1.4641 | |
| | $C_2$ | $X^1\Sigma_g^+$ | 1.2466 | | $(1.2394)^{e,f}$ |
| | | $a^3\Pi_u$ | 1.3180 | | $(1.3087)^{e,f}$ |
| $(E_\mathrm{EA})_\mathrm{el}$ $(\mathrm{cm}^{-1})$ [g] | | | | 24540.8 | |
| $E_\mathrm{EA}$ $(\mathrm{cm}^{-1})$ [h] | | | | 24587.1 | |
| $E_\mathrm{EA}$ (eV) [g] | | | | 3.048 | 3.269(6)[i] |
| $\omega_e/\omega_e x_e$ $(\mathrm{cm}^{-1})$ | $C_2^-$ | $X^2\Sigma_g^+$ | 1770/12.4 | 1781.6/11.5 | 1781.189(18)/11.6717(48)[c] |
| | | $A^2\Pi_u$ | 1648/9.7 | 1668.7/10.8 | 1666.7(10)/10.8(26)[c] |
| | | $C^4\Sigma_u^+$ | 1161/12.7 | 1129.0/11.3 | |
| | $C_2$ | $X^1\Sigma_g^+$ | 1866/19.3 | | $(1855/13.6)^{e,f}$ |

(continued on next page)

| Quantity (unit) | Molecule | Level | cc-pVQZ[a] | cc-pCV5Z[b] | exp |
|---|---|---|---|---|---|
| | | | (continued from previous page) | | |
| $G_0$ (cm$^{-1}$) [j] | | a$^3\Pi_u$ | 1629/12.5 | | (1641/11.6)[e,f] |
| | C$_2^-$ | X$^2\Sigma_g^+$ | 881.9 | | (887.677)[c,f] |
| | C$_2$ | X$^1\Sigma_g^+$ | 928.2 | | (924.125)[e,f] |

[a] Model potentials, this work
[b] Shi *et al.* [1]
[c] Rehfuss *et al.* [3]
[d] Mead *et al.* [4]
[e] Chen *et al.* [5]
[f] Derived value(s)
[g] Energy difference of potential minima
[h] $E_{\mathrm{EA}}$ estimated as $(E_{\mathrm{EA}})_{\mathrm{el}} + G_0(\mathrm{X}^1\Sigma_g^+) - G_0(\mathrm{X}^2\Sigma_g^+)$
[i] Ervin and Lineberger [2]
[j] Vibrational zero-point energy $G_0$ estimated as $\omega_e/2 - \omega_e x_e/4$

## 2. Extrapolation of vibrational energies

We consider the vibrational energy eigenvalues in the rotational-electronic potentials $E_{\gamma N}(R)$ for a given electronic state $\gamma$. At high and intermediate $N$, these energies $E(N, v)$ were obtained by numerical eigenvalue calculations for all possible bound levels $v$, while at lower $N$ these explicit eigenvalues were obtained only for a subset of bound levels with quantum numbers up to $v_x$ and energies up to $E_x = E(v_x, N)$. We describe a method to estimate the energies of bound levels missing in the low-$N$ calculations from the available set of eigenvalues at high and intemediate $N$.

From the available energy eigenvalues, functions $f_N\big(E(N, v)\big)$ can be determined which give the energy difference from a state $v$ to the next-higher state $v + 1$:

$$f_N\big(E(N, v)\big) = \Delta E = E(N, v + 1) - E(N, v). \tag{S1}$$

These functions are shown in Fig. S1(a) for the example of the $A^2\Pi_u$ levels and for some intermediate values of $N$ – where $f_N\big(E(N, v)\big)$ was determined for all bound levels – as well as for lower $N$ where the eigenvalue calculation stops at $v_x(N)$ before reaching the highest vibrational eigenvalue. For the high $N$ where all bound level were calculated, the clustering of high-$v$ levels close to the maximum of the rotational-electronic potential is visible in $f_N\big(E(N, v)\big)$. For a given $N$, this occurs before reaching the maximum $E_m(N)$ of the potential-energy function $E_{\gamma N}(R)$, which can be easily derived.

Motivated by the similarity between the curves for all $N$ revealed by Fig. S1(a), we obtain the parts of the $f_N\big(E(N, v)\big)$ function missing for small $N$ from the linear transformation of a 'reference' level-spacing function

$$f_{N_r}(E) = E(N_r, v + 1) - E(N_r, v), \tag{S2}$$

where the reference angular momentum quantum number $N_r$ is chosen as the smallest $N$ for which the eigenvalues were obtained completely ($N_r = 118$ for the $X^2\Sigma_g^+$ state and $N_r = 116$ for the $A^2\Pi_u$ state). In order to estimate the unavailable eigenvalues at a lower $N_x < N_r$, where the highest calculated eigenvalue has the energy $E_x$, the value of the transformed level-spacing function is scaled by the ratio $f_{N_x}(E_x)/f_{N_r}(E_x)$. Moreover, the energy variable of the reference function, denoted as $E'$, is linearly transformed for such that for the given $N_x$, $E'$ runs from $E_x$ to $E_m(N_r)$ as $E$ varies from $E_x$ to $E_m(N_x)$:

$$E' = E_x + (E - E_x)\frac{E_m(N_r) - E_x}{E_m(N_x) - E_x}. \tag{S3}$$

Altogether, the level-spacing function (defined for $E > E_x$ and a given $N_x$) after this linear transform can be expressed as

$$f_{N_x}(E) = f_{N_r}(E')\frac{f_{N_x}(E_x)}{f_{N_r}(E_x)} \tag{S4}$$

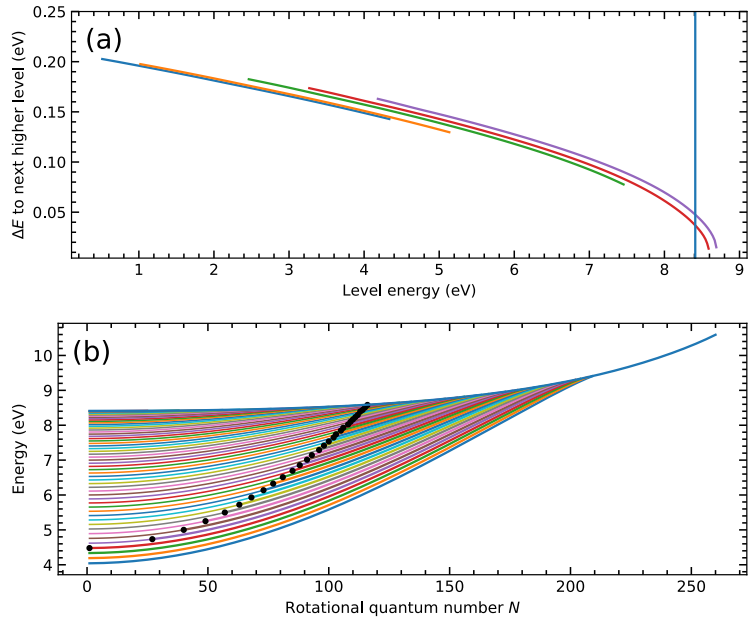

FIG. S1. (a) Energy difference functions $\Delta E = f_N\big(E(N,v)\big)$ as defined in Eq. (S1) from the calculated vibrational eigenvalues, plotted as functions of the lower-level energy $E(N,v)$ for the example of the $A\,^2\Pi_u$ electronic state and $N = 1, 50, 100, 120$, and 140. For the three lower $N$, the eigenvalues were obtained only for a subset of lower $v$ levels, as seen by the line ends which correspond to $E_x$ for a given $N$. Vertical line: potential-energy maximum $E_m(N=0)$ without rotation. (b) Vibrational level energies as functions of $N$ for $A\,^2\Pi_u$ shown by the colored curves for individual $v$ for $v \leq 20$. Blue (upper) curve: potential-energy maximum $E_m(N)$. For the higher $v$, black dots indicate the $N$-value above which eigenvalues were explicitly determined. Above the last explicitly calculated vibrational energy for a given $N$, the higher vibrational energies (thin curves) were obtained by the described approximate method, setting $N_x = N$ and $v_x$ to the quantum number of the highest explicitly calculated level at this $N$.

with $E'$ from Eq. (S3). The missing vibrational energies are then obtained iteratively by

$$E(N_x, v+1) = E(N_x, v) + f_{N_x}\big(E(N_x, v)\big) \tag{S5}$$

starting with $v = v_x$ while remaining in the range $E(N_x, v) < E_m(N_x)$ [corresponding to $E' < E_m(N_r)$]. Estimated high vibrational level energies filling the gap due to the unavailablity of explicitly calculated eigenvalues are illustrated by Fig. S1(b). The only application of these estimated vibrational energies in this work is for the normalization of the thermal energy distributions where, moreover, the high vibrations affected by the approximation are expected to become of a certain relevance only at the discussed high-temperature limit of $30000\,\mathrm{K}$.

## 3. Angular factors of the doublet–quartet line strengths

The radiative decay rates of the $C_2^-$ quartet levels (Sec. IVA2 of the main paper) follow from Eq. (36) using the line strengths $S_{N'v'J'}^{N''v''J''}$ between the upper (quartet) term F($i'$) (i.e., F1...F4) with $J' = N' + 5/2 - i'$ and the lower doublet term F($i''$) (i.e., F1...F2) with $J'' = N'' + 3/2 - i''$. The full angular factor in the decay rate $A_{C,N'v'i'}^{X,N''v''i''}$ of Eq. (36) is given by $S_{N'v'J'}^{N''v''J''}/(2J' + 1)$, where the line strength $S_{N'v'J'}^{N''v''J''}$ is given by Eq. (B36). Taken together, the angular factors required for the decay rate calculation are $\hat{Q}_+^2/(2J' + 1)$ and $\hat{r}$, where $\hat{Q}_+^2$ follows from Eq. (B30) and $\hat{r}$ is given by Eq. (B31).

The definitions for $s_{\Omega''q}^{J'J''}$ in Eq. (B6) and $r_{N\Omega}^{\Lambda JS}$ in Eq. (B8), with $[J, J', \ldots] = (2J+1)(2J'+1)\ldots$, yield

$$\frac{\hat{Q}_+^2}{[J']} = 3\,[J'', N', N''] \begin{pmatrix} \frac{1}{2} & N'' & J'' \\ \frac{1}{2} & 0 & -\frac{1}{2} \end{pmatrix}^2 \begin{pmatrix} \frac{3}{2} & N' & J' \\ \frac{3}{2} & 0 & -\frac{3}{2} \end{pmatrix}^2 \begin{pmatrix} J' & 1 & J'' \\ -\frac{3}{2} & 1 & \frac{1}{2} \end{pmatrix}^2,$$

$$\hat{r} = \begin{pmatrix} \frac{3}{2} & N' & J' \\ -\frac{1}{2} & 0 & \frac{1}{2} \end{pmatrix} \begin{pmatrix} J' & 1 & J'' \\ \frac{1}{2} & -1 & \frac{1}{2} \end{pmatrix} \left[ \sqrt{3} \begin{pmatrix} \frac{3}{2} & N' & J' \\ \frac{3}{2} & 0 & -\frac{3}{2} \end{pmatrix} \begin{pmatrix} J' & 1 & J'' \\ -\frac{3}{2} & 1 & \frac{1}{2} \end{pmatrix} \right]^{-1}. \quad (S6)$$

For use in the present model we express these angular factors using only the angular momentum quantum $N'$ of the upper state, taking into account the fine-structure labels. The results for particular fine-structure transitions are presented by using matrices with two columns (for the lower-states $i'' = 1, 2$) and four rows (for the upper states $i' = 1, 2, 3, 4$). Moreover, enclosed in brackets as $[(\,), (\,), (\,), (\,)]$, four separate matrices are given for $\Delta N = -3, -1, +1$, and $+3$, i.e., $N'' = N' + 3, N' + 1, N' - 1$, and $N' - 3$. The factorial expressions for the Wigner symbols yield

$$\left(\frac{\hat{Q}_+^2}{[J']}\right)_{i''i'} = \left[ \begin{pmatrix} 0 & \frac{3(N'+1)(N'+2)(N'+3)}{8(2N'+3)(2N'+4)(2N'+5)} \\ 0 & 0 \\ 0 & 0 \\ 0 & 0 \end{pmatrix}, \begin{pmatrix} \frac{3(N'+1)(N'+3)^2}{4(2N'+3)(2N'+4)(2N'+5)} & \frac{3(N'+3)^2(2N'+1)}{16(2N'+2)(2N'+3)^2} \\ \frac{9N'}{32(2N'+3)} & \frac{9N'(N'+2)^2}{4(2N'+2)(2N'+3)^2} \\ 0 & \frac{9(N'-1)^2}{16(2N'-1)(2N'+2)} \\ 0 & 0 \end{pmatrix}, \right.$$

$$\left. \begin{pmatrix} 0 & 0 \\ \frac{9(N'+2)^2(2N'-1)}{32N'(2N'+1)(2N'+3)} & 0 \\ \frac{9(N'-1)^2(N'+1)}{8N'(2N'-1)(2N'+1)} & \frac{9(N'+1)(2N'-3)}{32(2N'-1)^2} \\ \frac{3(N'-1)^2}{32N'(2N'-1)} & \frac{3N'(N'-2)^2}{4(2N'-2)(2N'-1)^2} \end{pmatrix}, \begin{pmatrix} 0 & 0 \\ 0 & 0 \\ 0 & 0 \\ \frac{3N'(N'-2)}{16(2N'-3)(2N'-1)} & 0 \end{pmatrix} \right] \quad (S7)$$

with the high-$N$ limit

$$\left(\frac{\hat{Q}_+^2}{[J']}\right)_{i''i'} \xrightarrow{N'\gg1} \left[\begin{pmatrix} 0 & \frac{3}{64} \\ 0 & 0 \\ 0 & 0 \\ 0 & 0 \end{pmatrix}, \begin{pmatrix} \frac{3}{32} & \frac{3}{64} \\ \frac{9}{64} & \frac{9}{32} \\ 0 & \frac{9}{64} \\ 0 & 0 \end{pmatrix}, \begin{pmatrix} 0 & 0 \\ \frac{9}{64} & 0 \\ \frac{9}{32} & \frac{9}{64} \\ \frac{3}{64} & \frac{3}{32} \end{pmatrix}, \begin{pmatrix} 0 & 0 \\ 0 & 0 \\ 0 & 0 \\ \frac{3}{64} & 0 \end{pmatrix}, \right] \tag{S8}$$

and

$$(\hat{r})_{i''i'} = \left[\begin{pmatrix} 0 & 1 \\ 0 & 0 \\ 0 & 0 \\ 0 & 0 \end{pmatrix}, \begin{pmatrix} -\frac{N'+2}{N'+3} & \frac{N'+1}{N'+3} \\ -\frac{1}{3} & \frac{N'+1}{3(N'+2)} \\ 0 & -\frac{N'+1}{3(N'-1)} \\ 0 & 0 \end{pmatrix}, \begin{pmatrix} 0 & 0 \\ -\frac{N'}{3(N'+2)} & 0 \\ \frac{N'}{3(N'-1)} & -\frac{1}{3} \\ \frac{N'}{N'-2} & -\frac{N'-1}{N'-2} \end{pmatrix}, \begin{pmatrix} 0 & 0 \\ 0 & 0 \\ 0 & 0 \\ 1 & 0 \end{pmatrix}\right] \tag{S9}$$

with the high-$N$ limit

$$(\hat{r})_{i''i'} \xrightarrow{N'\gg1} \left[\begin{pmatrix} 0 & 1 \\ 0 & 0 \\ 0 & 0 \\ 0 & 0 \end{pmatrix}, \begin{pmatrix} -1 & 1 \\ -\frac{1}{3} & \frac{1}{3} \\ 0 & -\frac{1}{3} \\ 0 & 0 \end{pmatrix}, \begin{pmatrix} 0 & 0 \\ -\frac{1}{3} & 0 \\ \frac{1}{3} & -\frac{1}{3} \\ 1 & -1 \end{pmatrix}, \begin{pmatrix} 0 & 0 \\ 0 & 0 \\ 0 & 0 \\ 1 & 0 \end{pmatrix}\right]. \tag{S10}$$

With these expressions the full $N'$ dependence can be easily taken into account, while it is also instructive to consider the given limits for $N' \gg 1$.

Given $N'$, $\Delta N$, as well as the initial and final fine structure labels, these matrices determine the angular factors for calculating the spontaneous transition rate with Eqs. (36) and (B35) of the main paper.

## 4. Spectroscopic doublet–quartet angular factors

For spectroscopic work, the angular factors in the line strength are conventionally expressed using the angular momentum $J''$ of the lower spectroscopic term [6, 7] and grouped according to the fine structure levels $F(i'')$ of the lower term. Starting in a specified lower level, transitions are then distinguished by the change $\Delta J = J' - J''$ of the total angular momentum and the fine structure level $F(i')$ within the upper $J'$ term. For the $^2\Sigma$–$^4\Sigma$ transitions studied here, line strength angular factors based on an approximate separation into a Franck–Condon-type electronic integral and Hönl–London-type angular factors [see the discussion in Appendix B around Eqs. (B37), (B38)] were obtained in earlier work [6, 8]. These earlier angular factors are combinations of the results in our work, where we go beyond the earlier approximation.

To link to the spectroscopic convention, we represent quantities related to the fine-structure lines by matrices with three columns for $\Delta J = -1, 0, +1$ (i.e., $J' = J'' - 1, J'', J'' + 1$) and four rows for the upper states $i' = 1, 2, 3, 4$. Matrices giving $\Delta N$ and the selection-rule factor $\bar{P}(\Delta N)$ [Eq. (B28)] for the spectroscopic line array are

$$
(\Delta N)^{i''=1}_{\Delta J, i'} = \begin{pmatrix} -2 & -1 & 0 \\ -1 & 0 & 1 \\ 0 & 1 & 2 \\ 1 & 2 & 3 \end{pmatrix} \quad \Rightarrow \quad (\bar{P})^{i''=1}_{\Delta J, i'} = \begin{pmatrix} 0 & 1 & 0 \\ 1 & 0 & 1 \\ 0 & 1 & 0 \\ 1 & 0 & 1 \end{pmatrix} \quad \Rightarrow \quad \begin{pmatrix} 0 & Q_{12} & 0 \\ P_2 & 0 & R_2 \\ 0 & Q_{32} & 0 \\ P_{42} & 0 & R_{42} \end{pmatrix}
$$

$$
(\Delta N)^{i''=2}_{\Delta J, i'} = \begin{pmatrix} -3 & -2 & -1 \\ -2 & -1 & 0 \\ -1 & 0 & 1 \\ 0 & 1 & 2 \end{pmatrix} \quad \Rightarrow \quad (\bar{P})^{i''=2}_{\Delta J, i'} = \begin{pmatrix} 1 & 0 & 1 \\ 0 & 1 & 0 \\ 1 & 0 & 1 \\ 0 & 1 & 0 \end{pmatrix} \quad \Rightarrow \quad \begin{pmatrix} P_{13} & 0 & R_{13} \\ 0 & Q_{23} & 0 \\ P_3 & 0 & R_3 \\ 0 & Q_{43} & 0 \end{pmatrix} \tag{S11}
$$

The right-hand sides give the spectroscopic labels of the allowed fine-structure lines in the notation used by the authors of Refs. [6, 8], where the conventional letters P, R, and Q are used to denote $\Delta J$ and the indices of the letters are $i', (i'' + 1)$.

For the angular factors, we again use Eq. (S6) now expressing the Wigner symbols only by $J''$. The results are

$$
\left( P\, \hat{Q}_+^2 \right)^{i''=2}_{\Delta J, i'} = \begin{pmatrix} 0 & \frac{3\left(J''-\frac{1}{2}\right)\left(J''+\frac{1}{2}\right)\left(J''+\frac{3}{2}\right)^2}{8J''^2(2J''+2)} & 0 \\ \frac{9\left(J''-\frac{3}{2}\right)\left(J''-\frac{1}{2}\right)\left(J''+\frac{1}{2}\right)}{32J''^2} & 0 & \frac{9\left(J''+\frac{3}{2}\right)\left(J''+\frac{5}{2}\right)^2}{8(2J''+2)(2J''+4)} \\ 0 & \frac{9\left(J''-\frac{1}{2}\right)^2\left(J''+\frac{1}{2}\right)\left(J''+\frac{3}{2}\right)}{8J''^2(2J''+2)} & 0 \\ \frac{3\left(J''-\frac{3}{2}\right)^2\left(J''-\frac{1}{2}\right)}{32J''^2} & 0 & \frac{3\left(J''+\frac{1}{2}\right)\left(J''+\frac{3}{2}\right)\left(J''+\frac{5}{2}\right)}{8(2J''+2)(2J''+4)} \end{pmatrix} \tag{S12}
$$

and

$$\left(P\,\hat{Q}_+^2\right)_{\Delta J, i'}^{i''=3} = \begin{pmatrix} \frac{3\left(J''-\frac{3}{2}\right)\left(J''-\frac{1}{2}\right)\left(J''+\frac{1}{2}\right)}{32J''(J''-1)} & 0 & \frac{3\left(J''+\frac{3}{2}\right)\left(J''+\frac{5}{2}\right)^2}{16(J''+1)(2J''+2)} \\ 0 & \frac{9\left(J''-\frac{1}{2}\right)\left(J''+\frac{1}{2}\right)\left(J''+\frac{3}{2}\right)^2}{4J''(2J''+2)^2} & 0 \\ \frac{9\left(J''-\frac{3}{2}\right)^2\left(J''-\frac{1}{2}\right)}{32J''(J''-1)} & 0 & \frac{9\left(J''+\frac{1}{2}\right)\left(J''+\frac{3}{2}\right)\left(J''+\frac{5}{2}\right)}{16(J''+1)(2J''+2)} \\ 0 & \frac{3\left(J''-\frac{1}{2}\right)^2\left(J''+\frac{1}{2}\right)\left(J''+\frac{3}{2}\right)}{4J''(2J''+2)^2} & 0 \end{pmatrix} \quad \text{(S13)}$$

as well as

$$(\hat{r})_{\Delta J, i'} = \begin{pmatrix} 1 & -\frac{J''+\frac{1}{2}}{J''+\frac{3}{2}} & \frac{J''+\frac{1}{2}}{J''+\frac{5}{2}} \\ -\frac{1}{3} & \frac{J''+\frac{1}{2}}{3\left(J''+\frac{3}{2}\right)} & -\frac{J''+\frac{1}{2}}{3\left(J''+\frac{5}{2}\right)} \\ -\frac{J''+\frac{1}{2}}{3\left(J''-\frac{3}{2}\right)} & \frac{J''+\frac{1}{2}}{3\left(J''-\frac{1}{2}\right)} & -\frac{1}{3} \\ \frac{J''+\frac{1}{2}}{J''-\frac{3}{2}} & -\frac{J''+\frac{1}{2}}{J''-\frac{1}{2}} & 1 \end{pmatrix} \quad \text{(S14)}$$

independent of $i''$. In the approximation used by the authors of Refs. [6, 8], the Franck–Condon-type pre-factors (intensity factors) are proportional to $\tilde{S}^2 = P\,\hat{Q}_+^2(1-\hat{r})^2$. When the results given separately in Eqs. (S12)–(S14) are combined, evaluating the expression for $\tilde{S}^2$ element per element, the intensity factors are

$$\left(\tilde{S}^2\right)_{\Delta J, i'}^{i''=2} = \begin{pmatrix} 0 & \frac{3\left(J''+\frac{1}{2}\right)(J''+1)(2J''-1)}{8J''^2} & 0 \\ \frac{\left(J''+\frac{1}{2}\right)(2J''-3)(2J''-1)}{8J''^2} & 0 & \frac{(J''+2)(2J''+3)}{4(J''+1)} \\ 0 & \frac{(J''-1)^2\left(J''+\frac{1}{2}\right)(2J''+3)}{8J''^2(J''+1)} & 0 \\ \frac{3(2J''-1)}{16J''^2} & 0 & 0 \end{pmatrix} \quad \text{(S15)}$$

and

$$\left(\tilde{S}^2\right)_{\Delta J, i'}^{i''=3} = \begin{pmatrix} 0 & 0 & \frac{3(2J''+3)}{16(J''+1)^2} \\ 0 & \frac{\left(J''+\frac{1}{2}\right)(J''+2)^2(2J''-1)}{8J''(J''+1)^2} & 0 \\ \frac{(J''-1)(2J''-1)}{4J''} & 0 & \frac{\left(J''+\frac{1}{2}\right)(2J''+3)(2J''+5)}{8(J''+1)^2} \\ 0 & \frac{3J''\left(J''+\frac{1}{2}\right)(2J''+3)}{8(J''+1)^2} & 0 \end{pmatrix}. \quad \text{(S16)}$$

The matrix elements for $\tilde{S}^2$ agree with the expressions for the relative intensities given in Table 3.26 of Ref. [6], using column 3 of this table together with the replacement for $J$ implied by column 2 and putting $J = J''$. While the results in Ref. [6] are quoted only up to a common constant factor, this factor turns out to be unity in the described comparison. Note that the intensity of transitions with $\Delta N = \mp 3$ (lines labeled P$_{13}$ and R$_{42}$) vanishes in this approximation, while it remains finite (albeit small) in the results of Sec. IV A 2.

The limits for high $J''$ are:

$$\left(\tilde{S}^2\right)^{i''=2}_{\Delta J, i'} \xrightarrow{N' \gg 1} [J'] \begin{pmatrix} 0 & \frac{3}{8} & 0 \\ \frac{1}{4} & 0 & \frac{1}{4} \\ 0 & \frac{1}{8} & 0 \\ 0 & 0 & 0 \end{pmatrix} \quad , \quad \left(\tilde{S}^2\right)^{i''=3}_{\Delta J, i'} \xrightarrow{N' \gg 1} [J'] \begin{pmatrix} 0 & 0 & 0 \\ 0 & \frac{1}{8} & 0 \\ \frac{1}{4} & 0 & \frac{1}{4} \\ 0 & \frac{3}{8} & 0 \end{pmatrix}. \tag{S17}$$

The intensities of the lines labeled $R_{13}$ and $P_{42}$ vanish in the high-$N$ limit, on top of the intensities for $P_{13}$ and $R_{42}$ ($\Delta N = \mp 3$) which, as mentioned, always vanish in this approximation. For the four fine-structure sublevels within an upper level of given $N'$, all radial matrix elements are set equal in the approximation of Eqs. (B37), (B38) in the main paper. The relative total decay rates of the four sublevels, and hence their relative radiative lifetimes, follow by summing in a given row (representing the fine structure component $i'$) over all matrix elements for the different $\Delta J$ and $i''$. In the high-$N$ limit, it is seen that this sum is 3/8 for $i' = 1$ and 4 and 5/8 for $i' = 2$ and 3. Hence, as was already implied by the earlier calculations [6], the levels $i' = 2$ and 3 decay faster by a factor of 5/3 than the levels $i' = 1$ and 4 when the high-$N$ limit is considered. Different radiative lifetimes of fine-structure components are often found when spin-forbidden decay is considered in atomic systems, such as, e.g., in Ref. [9].

## 5. Rotationally assisted autodetachment rates

Table S2 complements the data presented in Fig. 11 of the main paper. The properties of those levels in the lowest quartet state $C^4\Sigma_u^+$ of $C_2^-$ are given which undergo rotationally assisted autodetachment (AD) into $A^3\Pi_u$ levels of $C_2$ with a rotational quantum number $N_n$ in the range of $N-6 \leq N_n < N-4$ (see Sec. IVC1 of the main paper).

TABLE S2: List of $C_2^-$ quartet levels with caluclated rotationally assisted AD rates [10] in the range of $10^2$–$10^4\,\mathrm{s}^{-1}$ for the three quartet potentials Q0, Qp, and Qm lying within the uncertainty of the ab-initio calculations. Given are the rotational and vibrational quantum numbers $N$ and $v$, the calculated AD rate $k_d$, and the modeled energy $E_{\gamma N v}$ ($\gamma = C^4\Sigma_u^+$) of the anionic quartet level above the $X^2\Sigma_g^+$, $N = v = 0$ ground state of $C_2^-$. The numbers given as $a\,(b)$ are $a \times 10^b$.

| Potential Q0 | | | | Potential Qp | | | | Potential Qm | | | |
|---|---|---|---|---|---|---|---|---|---|---|---|
| $N$ | $v$ | $k_d$ (s$^{-1}$) | $E_{\gamma N v}$ (eV) | $N$ | $v$ | $k_d$ (s$^{-1}$) | $E_{\gamma N v}$ (eV) | $N$ | $v$ | $k_d$ (s$^{-1}$) | $E_{\gamma N v}$ (eV) |
| 165 | 0 | 1.282 (+3) | 7.799 | 171 | 0 | 1.340 (+3) | 8.106 | 159 | 0 | 7.263 (+2) | 7.485 |
| 167 | 0 | 6.381 (+2) | 7.870 | 173 | 0 | 6.453 (+2) | 8.178 | 161 | 0 | 4.112 (+2) | 7.553 |
| 169 | 0 | 2.591 (+2) | 7.940 | 175 | 0 | 2.473 (+2) | 8.250 | 163 | 0 | 1.488 (+2) | 7.621 |
|  | 1 | 2.461 (+3) | 7.994 |  | 1 | 2.378 (+3) | 8.302 |  | 1 | 1.704 (+3) | 7.677 |
| 171 | 0 | 1.374 (+2) | 8.010 | 177 | 1 | 8.805 (+2) | 8.371 | 165 | 1 | 7.004 (+2) | 7.742 |
|  | 1 | 1.121 (+3) | 8.061 |  | 2 | 4.447 (+3) | 8.418 |  | 2 | 3.918 (+3) | 7.793 |
|  | 2 | 4.902 (+3) | 8.109 | 179 | 1 | 3.473 (+2) | 8.439 | 167 | 1 | 3.105 (+2) | 7.807 |
| 173 | 1 | 3.791 (+2) | 8.128 |  | 2 | 1.810 (+3) | 8.483 |  | 2 | 1.761 (+3) | 7.856 |
|  | 2 | 2.077 (+3) | 8.173 |  | 3 | 5.613 (+3) | 8.525 |  | 3 | 6.175 (+3) | 7.903 |
| 175 | 2 | 6.629 (+2) | 8.236 | 181 | 2 | 5.784 (+2) | 8.547 | 169 | 2 | 6.281 (+2) | 7.918 |
|  | 3 | 2.575 (+3) | 8.277 |  | 3 | 1.815 (+3) | 8.586 |  | 3 | 2.643 (+3) | 7.963 |
|  | 4 | 6.712 (+3) | 8.315 |  | 4 | 5.325 (+3) | 8.623 |  | 4 | 7.466 (+3) | 8.005 |
| 177 | 3 | 8.099 (+2) | 8.336 | 183 | 4 | 6.290 (+2) | 8.647 | 171 | 3 | 8.919 (+2) | 8.022 |
|  | 4 | 2.090 (+3) | 8.372 |  | 5 | 1.594 (+3) | 8.681 |  | 4 | 3.039 (+3) | 8.062 |
|  | 5 | 5.554 (+3) | 8.407 |  | 6 | 3.979 (+3) | 8.713 |  | 5 | 7.336 (+3) | 8.100 |
| 179 | 4 | 7.000 (+2) | 8.429 |  | 7 | 7.572 (+3) | 8.744 | 173 | 4 | 9.869 (+2) | 8.119 |
|  | 5 | 1.612 (+3) | 8.461 |  |  |  |  |  | 5 | 2.398 (+3) | 8.155 |
|  | 6 | 3.882 (+3) | 8.492 |  |  |  |  |  | 6 | 6.150 (+3) | 8.190 |
|  | 7 | 7.295 (+3) | 8.521 |  |  |  |  | 175 | 6 | 1.880 (+3) | 8.242 |
| 181 | 6 | 1.035 (+3) | 8.543 |  |  |  |  |  | 7 | 4.497 (+3) | 8.274 |

| | Potential Q0 | | | | Potential Qp | | | | Potential Qm | | |
|---|---|---|---|---|---|---|---|---|---|---|---|
| $N$ | $v$ | $k_d$ (s$^{-1}$) | $E_{\gamma Nv}$ (eV) | $N$ | $v$ | $k_d$ (s$^{-1}$) | $E_{\gamma Nv}$ (eV) | $N$ | $v$ | $k_d$ (s$^{-1}$) | $E_{\gamma Nv}$ (eV) |

(continued from previous page)

| | Potential Q0 | | | | Potential Qp | | | | Potential Qm | | |
|---|---|---|---|---|---|---|---|---|---|---|---|
| | 7 | 1.987 (+3) | 8.571 | | | | | | 8 | 8.456 (+3) | 8.305 |
| | 8 | 4.060 (+3) | 8.597 | | | | | 177 | 7 | 1.276 (+3) | 8.324 |
| | 10 | 6.732 (+3) | 8.623 | | | | | | 8 | 2.471 (+3) | 8.353 |
| 183 | 9 | 9.684 (+2) | 8.644 | | | | | | 9 | 5.220 (+3) | 8.382 |
| | 12 | 1.646 (+3) | 8.668 | | | | | | 10 | 8.862 (+3) | 8.409 |
| | 14 | 2.567 (+3) | 8.690 | | | | | 179 | 9 | 1.426 (+3) | 8.428 |
| | 17 | 4.268 (+3) | 8.710 | | | | | | 10 | 2.440 (+3) | 8.454 |
| | 19 | 5.599 (+3) | 8.729 | | | | | | 11 | 4.032 (+3) | 8.479 |
| | 22 | 6.398 (+3) | 8.746 | | | | | | 12 | 7.348 (+3) | 8.504 |
| | | | | | | | | | 13 | 1.100 (+4) | 8.527 |
| | | | | | | | | 181 | 12 | 1.896 (+3) | 8.545 |
| | | | | | | | | | 13 | 2.860 (+3) | 8.567 |
| | | | | | | | | | 14 | 4.792 (+3) | 8.588 |
| | | | | | | | | | 15 | 6.494 (+3) | 8.608 |
| | | | | | | | | | 16 | 8.388 (+3) | 8.627 |
| | | | | | | | | 183 | 17 | 1.458 (+3) | 8.644 |
| | | | | | | | | | 18 | 1.815 (+3) | 8.660 |
| | | | | | | | | | 19 | 1.975 (+3) | 8.674 |
| | | | | | | | | | 20 | 1.269 (+3) | 8.685 |
| | | | | | | | | | 21 | 1.512 (+3) | 8.695 |


\end{document}